\documentclass{aa}
\usepackage{graphicx,amsmath}
\usepackage{epsf}
\usepackage{txfonts}
\usepackage{geometry}
\usepackage{tablefootnote}
\usepackage{longtable}
\usepackage{lscape}
\usepackage[hyperindex,breaklinks=true, colorlinks, citecolor=blue]{hyperref}  
\usepackage{comment}
\usepackage{natbib}
\usepackage{upgreek}
\usepackage{float}
\usepackage[switch,pagewise]{lineno}
\usepackage{multirow}

\newcommand{\postreport}[1]{{\bf #1}}
\renewcommand{\postreport}[1]{#1}

\newcommand{\postRVcorr}[1]{{\bf #1}}
\renewcommand{\postRVcorr}[1]{#1}

\newcommand{\kps}{km\,s$^{-1}$}
\newcommand{\vLOS}{$v_{\rm LOS}$}

\newcommand{\sigmavHI}{\postreport{10.8}}
\newcommand{\sigmavCO}{\postreport{6.6}}
\newcommand{\meanEkHI}{\postRVcorr{0.11}}
\newcommand{\meanEkCO}{\postRVcorr{0.04}}

\providecommand{\sorthelp}[1]{}

\def\deg{\ifmmode^\circ\else$^\circ$\fi}
\def\pdeg{\ifmmode $\setbox0=\hbox{$^{\circ}$}\rlap{\hskip.11\wd0 .}$^{\circ} \else \setbox0=\hbox{$^{\circ}$}\rlap{\hskip.11\wd0 .}$^{\circ}$\fi}
\def\arcs{\ifmmode {^{\scriptstyle\prime\prime}} \else $^{\scriptstyle\prime\prime}$\fi}
\def\arcm{\ifmmode {^{\scriptstyle\prime}} \else $^{\scriptstyle\prime}$\fi}
\newdimen\saa  \newdimen\sbb
\def\parcs{\saa=.07em \sbb=.03em
     \ifmmode \hbox{\rlap{.}}^{\scriptstyle\prime\kern -\sbb\prime}\hbox{\kern -\sa}
     \else \rlap{.}$^{\scriptstyle\prime\kern -\sbb\prime}$\kern -\sa\fi}
\def\parcm{\saa=.08em \sbb=.03em
     \ifmmode \hbox{\rlap{.}\kern\saa}^{\scriptstyle\prime}\hbox{\kern-\sbb}
     \else \rlap{.}\kern\saa$^{\scriptstyle\prime}$\kern-\sbb\fi}

\defcitealias{tchernyshyov2017}{TP17} 
\defcitealias{edenhofer2024}{E24}

\begin{document}

\title{Kinetic tomography of the Galactic plane within 1.25 kiloparsecs from the Sun}
\subtitle{The interstellar flows revealed by H{\sc i} and CO line emission and 3D dust}

\titlerunning{Kinetic tomography of the Milky Way's disk}
    \author{
       J.~D.~Soler$^{1}$\thanks{Corresponding author; \email{juandiegosolerp@gmail.com}},
       S.~Molinari$^{1}$,
       S.~C.~O.~Glover$^{2}$,
       R.~J.~Smith$^{3}$,
       R.~S.~Klessen$^{2,4,5,6}$, 
       R.~A.~Benjamin$^{7}$, 
       P.~Hennebelle$^{8}$,
       J.~E.~G.~Peek$^{9}$,
       H.~Beuther$^{10}$,
       G.~Edenhofer$^{5}$,
       E.~Zari$^{11}$,
       C.~Swiggum$^{12}$,
       C.~Zucker$^{5}$
} 
\institute{
1. Istituto di Astrofisica e Planetologia Spaziali (IAPS). INAF. Via Fosso del Cavaliere 100, 00133 Roma, Italy\\
2. Universit\"{a}t Heidelberg, Zentrum f\"{u}r Astronomie, Institut f\"{u}r Theoretische Astrophysik, Albert-Ueberle-Str. 2, 69120, Heidelberg, Germany\\ 
3. SUPA, School of Physics and Astronomy, University of St Andrews, North Haugh, St Andrews, KY16 9SS, UK\\
4. Universit\"{a}t Heidelberg, Interdisziplin\"{a}res Zentrum f\"{u}r Wissenschaftliches Rechnen, Im Neuenheimer Feld 225, 69120 Heidelberg, Germany\\
5. Center for Astrophysics $|$ Harvard \& Smithsonian, 60 Garden St., Cambridge, MA 02138, USA\\
6. Elizabeth S. and Richard M. Cashin Fellow at the Radcliffe Institute for Advanced Studies at Harvard University, 10 Garden Street, Cambridge, MA 02138, USA\\
7. Department of Physics, University of Wisconsin-Whitewater, Whitewater, WI, USA\\
8. Laboratoire AIM, Paris-Saclay, CEA/IRFU/SAp - CNRS - Universit\'{e} Paris Diderot. 91191, Gif-sur-Yvette Cedex, France\\
9. Space Telescope Science Institute, 3700 San Martin Drive, Baltimore, MD 21218, USA\\
10. Max-Planck Institute for Astronomy, K{\"o}nigstuhl 17, 69117, Heidelberg, Germany\\
11. Dipartimento di Fisica e Astronomia, Universit{\`a} degli Studi di Firenze, Via G. Sansone 1, I-50019, Sesto F.no (Firenze), Italy\\
12. University of Vienna, Department of Astrophysics, T{\"u}rkenschanzstra{\ss}e 17, 1180 Wien, Austria
}
\authorrunning{Soler,\,J.D. et al.}

\date{Received 15 November 2024 / Accepted 4 February 2025}

\abstract{
We present a reconstruction of the line-of-sight motions of the local interstellar medium (ISM) based on the combination of a model of the three-dimensional dust density distribution within 1.25\,kpc from the Sun and the H{\sc i} and CO line emission within Galactic latitudes $|b|$\,$\leq$\,5\deg.
We used the histogram of oriented gradient (HOG) method, a computer vision technique for evaluating the morphological correlation between images, to match the plane-of-the-sky dust distribution across distances with the atomic and molecular line emission.
We identified a significant correlation between the 3D dust model and the line emission.
We employed this correlation to assign line-of-sight velocities to the dust across density channels and produce a face-on map of the local ISM radial motions with respect to the local standard of rest (LSR).
We find that most of the material in the 3D dust model follows the large-scale pattern of Galactic rotation; however, we also report local departures from the rotation pattern with standard deviations of \sigmavHI\ and \sigmavCO\,\kps\ for the H{\sc i} and CO line emission, respectively.
The mean kinetic energy densities \postreport{corresponding} to these streaming motions are around \meanEkHI\ and \meanEkCO\,eV/cm$^{3}$ from either gas tracer.
Assuming homogeneity and isotropy in the velocity field, these values are within a factor of a few of the total kinetic energy density.
These kinetic energy values are roughly comparable to other energy densities, thus confirming the near-equipartition in the local ISM.
Yet, we identify energy and momentum overdensities of around a factor of ten concentrated in the Radcliffe Wave, the Split, and other local density structures.
Although we do not find evidence of the local spiral arm's impact on these energy overdensities, their distribution suggests the influence of large-scale effects that, in addition to supernova feedback, shape the energy distribution and dynamics in the solar neighborhood.
}

\maketitle

\section{Introduction}

Tracing the flows of matter and energy in the interstellar medium (ISM) is instrumental for understanding the structure and evolution of the Milky Way and other galaxies \citep[see, for example,][]{ballesteros-paredes2020,tacconi2020,chevance2023}.
Reconstructing the flow of gas in the ISM enables the identification of the mechanisms that make gas available for star formation and allows us to quantify the impact of galactic dynamics, magnetic fields, stellar winds, outflows, and supernovae (SNe) in the ISM \cite[e.g.,][]{fieldANDsaslaw1965,tomisaka1984,koyama2000,blitz2007,hennebelle2008,klessen2010, dobbs2014,krumholz2014,schmidt2016,hennebelleANDinutsuka2019}.
In this work, we present a reconstruction of the ISM line-of-sight (LOS) motions in the Solar neighborhood using state-of-the-art models of the three-dimensional dust density distribution, which we associate with the velocity information from the neutral atomic hydrogen (H{\sc i}) and carbon monoxide (CO) line emission using a machine vision tool to quantify the morphological similarities between distance and velocity channels \citep{soler2019}.

Classic studies of the ISM distribution and kinematics rely on the assumption of circular motions around the Galactic center to compute what is usually called ``kinematic distances'' \citep[see, for example,][]{oort1958,roman-duval2009,sofue2011,wenger2018,hunter2024}.
Kinematic distances depend on the Galactic rotation model and are affected by the kinematic distance ambiguity \citep[see, for example,][]{eilers2020,reid2022}.
Moreover, the motion of the Galactic ISM is not purely rotational; non-circular streaming motions are induced by the Galactic bar, the spiral arms, and stellar feedback \citep{burton1971,gomez2006,moises2011}.
The latter effect is expected to be prominent in the Solar neighborhood, where the motions introduced by supernovae (SNe)-blown bubbles and other local effects are potentially dominant over Galactic rotation as seen from the local standard of rest \citep[LSR; see, for example,][and references therein]{zucker2023}.

In recent years, the simultaneous rise of {\it Gaia} and deep wide-field photometric surveys, such as the Panoramic Survey Telescope and Rapid Response System \citep[Pan-STARRS;][]{chambers2016} and the Apache Point Observatory Galactic Evolution Experiment \citep[APOGEE;][]{allende-prieto2008}, has sparked a revolution in our understanding the Galactic ISM in 3D.
The remarkable stellar parallax and reddening observations provided by these surveys have triggered unprecedented reconstructions of the interstellar dust distribution in the three spatial dimensions \citep[see, for example,][]{green2019,leike2020,lallement2022}.
These ``3D dust'' maps have enabled a variety of ISM studies, which include the characterization of the Local Bubble \citep[][]{pelgrims2020,zucker2022,oneill2024}, SNe-blown cavities \citep[][]{bracco2023,liu2024}, and molecular cloud (MC) envelopes
\citep[][]{zucker2021,mullens2024}, among others.

In this work, we associated the state-of-the-art 3D dust models presented by \cite{edenhofer2024} with the LOS velocity information provided by H{\sc i} and CO surveys toward the Galactic plane.
Following the pioneering work in \citet[][]{tchernyshyov2017, tchernyshyov2018}, we adopted the term ``kinetic tomography'' to describe our reconstruction of the density and LOS velocity distribution.
Our approach is based on the morphological similarity quantified by the histogram of oriented gradients (HOG) method \citep{soler2019}.
Using HOG, we linked the position-position-distance (PPD) 3D dust density cubes and the position-position-velocity (PPV) H{\sc i} and CO line emission cubes.
The result is a position-position-distance-velocity (PPDV) hypercube, which we employ to characterize the local ISM motions along the line of sight.
Line emission from ISM tracers and 3D dust have previously been combined to investigate the ISM in four dimensions \citep[see, for example,][]{ivanova2021,duchene2023}, but not with the level of reconstruction in the \cite{edenhofer2024} maps, and never before with the HOG method.

We used the H{\sc i} line emission at 21-cm wavelength, which traces both the cold, pre-molecular state before star formation and the warm, diffuse ISM before and after star formation \citep[see,][for a recent review]{mcclure-griffiths2023}.
Since its discovery, H{\sc i} has been instrumental in studying the diffuse ISM in the Milky Way \citep{ewenANDpurcell1951,mullerANDoort1951,pawsey1951}.
Early observations of the H{\sc i} absorption against radio continuum sources revealed the presence of narrow, few-\kps-wide, spectral features \citep[e.g.,][]{hagen1955,clark1965}.
In emission, these narrow features appear on top of broader 10 to 20-\kps-wide features \citep[e.g.,][]{matthews1957}.
These observations inspired the formulation of a ``two phase'' H{\sc i} model \citep{field1966,mckeeANDostriker1977}, in which at the pressure of the ISM, the heating and cooling processes naturally lead to two thermally stable states: a dense cold neutral medium (CNM; $T$\,$\approx$\,50\,K and $n$\,$\approx$\,50\,cm$^{-3}$) immersed in a diffuse warm neutral medium (WNM; $T$\,$\approx$\,8\,000\,K and $n$\,$\approx$\,0.3\,cm$^{-3}$).

Part of our understanding of the H{\sc i} multiphase structure comes from studying absorption toward continuum sources. \citep[see, for example,][]{strasser2007,stanimirovic2014,murray2018}.
Comparisons between 21-cm H{\sc i} emission and absorption measurements indicate that, in the vicinity of the Sun, the WNM has roughly the same column density as the CNM \citep{falgaronANDlequeux1973,liszt1983}.
Crucial additional information about the distribution of the CNM in and around MCs comes from the portion of the CNM sampled by the extended absorption of background H{\sc i} emission by cold foreground H{\sc i}, which is generically known as H{\sc i} self-absorption \citep[HISA;][]{heeschen1955,gibson2000,seifried2020}.
Observations indicate that the HISA distribution is related to that of the CO emission, although the CNM it reveals corresponds to less than 15\% than the total gas mass \citep[see, for example,][]{gibson2005,krco2008,wang2020hisa,syed2020}.

We also used the CO~$(J$\,$=$\,$1$\,$\rightarrow$\,$0)$ emission at 2.6\,mm wavelength, which is the archetypal tracer of the cooler and denser ISM \citep{wilson1970}.
With its low excitation energy and critical density, CO provides an irreplaceable proxy for H$_{2}$, which is the most abundant molecule in the Galaxy but is much harder to observe \citep{combes1991,bolatto2013,heyerANDdame2015}.
The focus on the molecular material traced by CO in star formation (SF) studies was cemented by the observed correlation between the SF and CO surface densities in nearby galaxies, which contrasts with the lack of correlation between SF and H{\sc i} at low surface densities in those objects \citep[see, for example,][]{kennicutt1998,wong2002,leroy2008,schinnererANDleroy2024}.

\cite{soler2023a} presented a pilot study of the HOG-based kinetic tomography using the 3D dust density reconstruction from \cite{leike2020} and archival H{\sc i} and CO line emission toward the Taurus molecular cloud (MC).
The authors found anti-correlation between the dust density and the H{\sc i} emission, which uncovers the CNM associated with the MC.
They also found a pattern in LOS velocities and distances consistent with converging gas motions in the Taurus MC, with the cloud's near side moving at higher velocities than the far side.
This result is consistent with the kinematic imprint of the MC location at the intersection of two bubble surfaces, the Local Bubble \citep{pelgrims2020, zucker2022} and the Per-Tau shell \citep{bialy2021}.

In this paper, we applied the HOG method to study the atomic and molecular gas motions toward the nearby Milky Way's disk, defined as the Galactic latitude range $|b|$\,$\leq$\,5\deg\ within the 1.25-kpc extent of the \cite{edenhofer2024} 3D dust reconstruction.
The presentation of the data, analysis methods, and results is organized as follows. 
In Sec.~\ref{sec:data}, we present the 3D dust models and the H{\sc i} and CO observations used in the analysis.
Section~\ref{sec:methods} summarizes the main aspects of the HOG method implementation for analyzing the Galactic plane.
We describe the morphological correlation between the 3D dust models and the line emission in Sec.~\ref{sec:results}.
In Sec.~\ref{sec:physics}, we detail the streaming motions obtained with the HOG method and the energy and momentum densities derived from them. 
Section~\ref{sec:discussion} discusses our results and their implications for understanding the ISM dynamics in the Solar neighborhood.
We present our conclusions of this work in Sec.~\ref{sec:conclusions}.
We complement the main results of this work with the analysis shown in a set of appendices.
Appendix~\ref{app:hog} presents details on the HOG method's error propagation and selection of parameters.
We consider the effect of HISA in our results and the potential of the HOG method to identify these features in App.~\ref{app:hisa}.
In App.~\ref{app:physics}, we evaluate the effects of the fixed angular resolution and distance in the HOG results.
Appendix~\ref{app:synthobs} presents the HOG analysis of synthetic line emission and 3D density from a multiphase magnetohydrodynamic (MHD) MC simulation.
Appendix~\ref{app:masers} compares our results with the distances and line-of-sight velocities for the five maser sources within the studied volume. 
Finally, App.~\ref{app:anticenter} presents the distance-velocity mapping for a few regions of interest, including areas toward the Galactic center and anticenter, where uncertainties in the kinematic distance estimates are considerable.

\begin{figure*}[ht!]
\centerline{\includegraphics[width=0.99\textwidth,angle=0,origin=c]{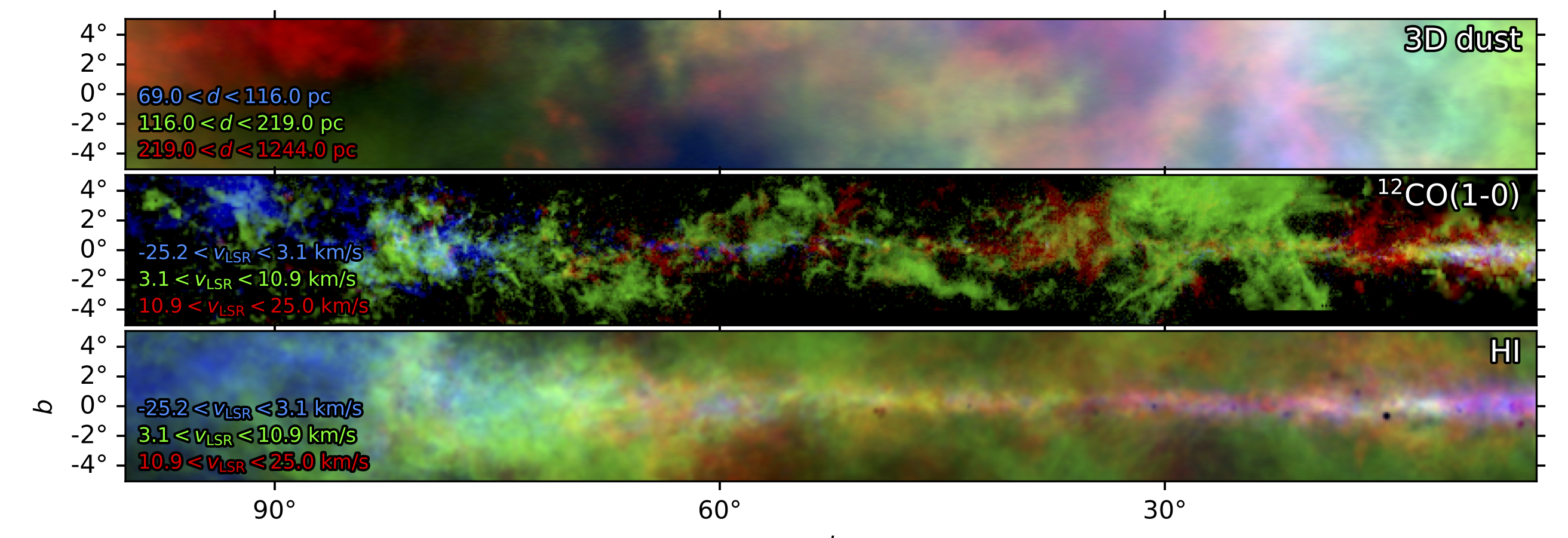}}
\caption{Examples of the 3D dust, $^{12}$CO, and H{\sc i} data combined in this paper.
{\it Top}. Nucleon density derived from the 3D extinction model presented by \cite{edenhofer2024} for three distance bins indicated in the figure.
{\it Middle}. $^{12}$CO line emission from the \cite{dame2001} survey for the three LOS velocity intervals indicated in the figure. 
{\it Bottom}. H{\sc i} 21-cm line emission observations presented by \cite{hi4pi2016}. 
}
\label{fig:RGBmaps}
\end{figure*}

\section{Data}\label{section:observations}\label{sec:data}

We limit our analysis to the band of sky within $|b|$\,$<$\,5\deg, which is the area covered by the most extensive spectroscopic survey of CO emission for the same energy transition and the same isotopologue \citep{dame2001}.
This region is covered by the most wide-ranging MC studies \citep{rice2016,miville-deschenes2017}.
As our results later indicate, the proximity to the Galactic midplane also facilitates the interpretation of the reconstructed LOS velocity pattern.
We complement the CO observations with the whole-sky 3D dust models from \cite{edenhofer2024} and H{\sc i} emission observations introduced in \cite{hi4pi2016}.
We show an example of these three data sets in Fig.~\ref{fig:RGBmaps}.

In this paper, we use \vLOS\ to refer to the radial velocity inferred from the emission line Doppler shift relative to the LSR, usually identified as $v_{\rm LSR}$ in the H{\sc i} and CO observations.
We introduce this convention to distinguish it from the general motion relative to the LSR, for which the three components of the velocity vector can be inferred in the Solar vicinity from the measurement of stellar motions \citep[see, for example,][]{miret-roig2022,ratzenbock2023}.

\begin{figure}[ht]
\centerline{
\includegraphics[width=0.5\textwidth,angle=0,origin=c]{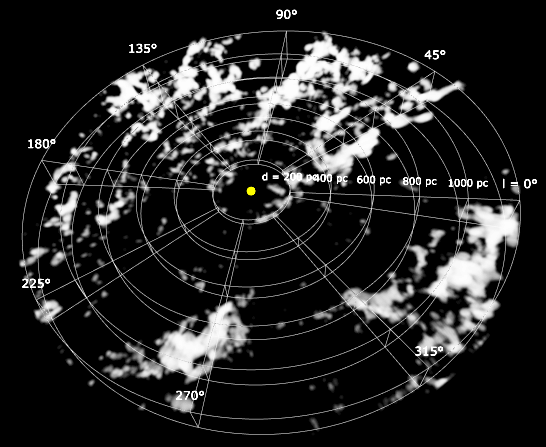}}
\caption{Rendering of the 3D dust density distribution from the extinction models presented in \cite{edenhofer2024} for the region $|b|$\,$<$\,5\deg\ considered in this paper.
The yellow sphere represents the position of the Sun.
The associated movie is available online.
}
\label{fig:3DdustDiagram}
\end{figure}

The selected $b$ range implies that we are studying the volume covered by the revolution of a trapezoid with base sizes 12.16 and 220.4\,pc located at 69 and 1250\,pc from the LSR, as illustrated in Fig.~\ref{fig:3DdustDiagram}.
The limits along the LOS are set by the boundaries of the \cite{edenhofer2024} 3D dust model.
At the furthest point in this analysis, we cover a region roughly within the characteristic CO scale height around the Solar circle \citep[$\sim$\,100\,pc,][]{heyerANDdame2015}.

We consider the line emission in the range $-25$\,$<$\,\vLOS\,$<$\,$25$\,\kps, which corresponds to the expected amplitude of LOS velocities for a Galactic rotation model for heliocentric distances within 1250\,pc\ plus or minus a few kilometers per second.
\postreport{This restricted input range mitigates the spurious chance correlations produced by the limited angular resolution in the 3D dust reconstruction, as detailed in App~\ref{app:hog}.
However, it also introduces a generic limitation in our analysis, as some local ISM structures can have significant morphological correlations beyond the input \vLOS\ range.
Thus, in practice, the restricted \vLOS\ input range limits the amplitude of reconstructed LOS motions for the local density structures.
Until future higher-resolution 3D dust reconstructions break the ambiguities that produce chance correlations and expand the reliable input \vLOS\ range, our HOG-based reconstruction should be considered lower limits for the streaming motions and other derived quantities.
}

\subsection{3D dust density distribution}

The primary dataset enabling our study is the reconstruction of the 3D dust density between 69 and 1250\,pc from the Sun presented in \cite{edenhofer2024}.
This model uses the extinction estimates from \cite{zhang2023}, which are primarily based on the {\it Gaia} satellite's BP/RP spectra, with a spectral resolution ($R$) of roughly between 30 and 100 \citep{gaia2023}.
\cite{zhang2023} forward-modeled the extinction, distance, and intrinsic parameters of each star given the combination of the {\it Gaia} spectra \citep{carrasco2021,deangeli2023,montegriffo2023} and infrared photometry from the Two Micron All Sky Survey \citep[2MASS;][]{skrutskie2006}, and unWISE, a processed catalog based on the observations from NASA's Wide-field Infrared Survey Explorer (WISE) mission \citep{wright2010,schlafly2019}.
The model was trained using a subset of stars observed with higher spectral resolution, $R$\,$\sim$\,1800, available from the Large Sky Area Multi-object Fibre Spectroscopic Telescope \citep[LAMOST;][]{wang2022,xiang2022}.

The \cite{edenhofer2024} 3D dust maps achieve a compromise between angular resolution and volume coverage, alleviating the limitations of previous 3D extinction models, which can be roughly divided into two groups.  
First, there are Cartesian reconstructions, which commonly feature fewer artifacts produced by the smearing of extinction structures along the line of sight (``fingers of god'') but which are either limited to covering a small volume at high resolution \citep[see, for example,][]{leike2019,leike2020} or a large volume at a low resolution \citep[see, for example,][]{capitanio2017,lallement2019,vergely2022}.
Second, there are spherical reconstructions, which have a much higher angular resolution and probe large volumes of the Galaxy but feature more pronounced finger-of-god artifacts \citep[see, for example,][]{chen2019,green2019}.
Other approaches, such as using many small reconstructions \citep[e.g.,][]{leike2022}, an analytical approach \citep[e.g.,][]{rezaeikh2017}, or inducing-point methods \citep[e.g.,][]{dharmawardena2022}, have so far been unsuccessful in modeling dust distributions at high resolution over large volumes without artifacts.

\cite{edenhofer2024} obtains a spherical-coordinate reconstruction beyond 1\,kpc while still resolving nearby dust clouds at parsec-scale resolution by implementing a new Gaussian process (GP) methodology to incorporate smoothness in a spherical coordinate system, mitigating fingers-of-god artifacts.
The authors modeled the 3D distribution of differential extinction for stars in the \cite{zhang2023} catalog, assuming that the dust extinction distribution is spatially smooth.
The posterior of their extinction model was reconstructed using variational inference and Gaussian variational inference \citep[MGVI,][]{knollmuller2019,leike2019}.
We refer to \cite{edenhofer2024} for further data modeling and reconstruction details.

The result of the \cite{edenhofer2024} model is a set of 12 samples drawn from the variational posterior of the 3D dust extinction distribution; that is, a set of 12 possible distributions consistent with the observations within the uncertainties.
Each sample gives the value of the dust density in voxels arranged as a series of HEALPix\footnote{HEALPix is the Hierarchical Equal Area isoLatitude Pixelization; see \url{http://healpix.sf.net} and \citet{gorski2005}.} spheres. 
There are 516 of these spheres, each corresponding to a distance logarithmically spaced grid between 69 to 1250\,pc. 
Each sphere has a resolution parameter $N_{\rm side}$\,$=$\,256, corresponding to an angular size of around 13\arcmin7 for each voxel.
We analyzed each of the 3D dust extinction samples, considering them independent realizations of 3D dust extinction, and reported the mean trends, as described in detail in App.~\ref{app:hog}.

Using the tools in the Python {\tt healpy} package, we produced a Cartesian projection of the 3D dust extinction covering the range $-180$\deg\,$<$\,$l$\,$<$\,180\deg\ and $-5$\deg\,$<$\,$b$\,$<$\,5\deg\ with a pixel size $\delta l$\,$=$\,$\delta b$\,$=$\,7\parcm5.
The reconstruction was made in terms of the unitless extinction defined by \cite{zhang2023}, which we transformed into V-band extinction ($A_{\rm V}$) by multiplying with the 2.8 factor derived from their extinction curve.
We then transformed $A_{\rm V}$ into hydrogen nucleon column density ($N_{\rm H}$) by multiplying by the mean extinction per H nucleon factor 5.8\,$\times$\,$10^{21}$\,cm$^{-2}$/mag from \cite{bohlin1978} \postRVcorr{and dividing by the total-to-selective extinction ratio, $R_{\rm V}$\,$\approx$\,3.1 \citep{savageANDmathis1979}.}
Finally, we computed the hydrogen nucleon density ($n_{\rm H}$) by dividing by the width of each distance channel.

Figure~\ref{fig:polarDens} shows a face-on view of the 3D dust distribution across the studied regions.
The pixelization in this representation corresponds to the distance cells in which we divide the studied volume for the analysis presented in Sec.~\ref{sec:methods}.
For reference, we indicate in the figure the position of the large-scale features identified in the Solar neighborhood \citep{zucker2023}.
They are the Sagittarius spur \citep{kuhn2021}, the Cepheus Spur \citep{pantaleoni-gonzalez2021}, the Split \citep{lallement2019}, the Radcliffe Wave \citep{alves2020}, and the Local Spiral Arm, as defined in \cite{reid2019}.
Although our analysis does not assume any prior information on the presence of these features, they are relevant for interpreting our results.

\subsection{Carbon monoxide (CO) emission}

We employed the $^{12}$CO$(J$\,$=$\,$1$\,$\rightarrow$\,$0)$ emission maps presented by \cite{dame2001}, which cover the whole Galactic plane and have an angular resolution comparable to the 3D dust data.
This dataset is a combination of the observations obtained over two decades with two 1.2-meter-aperture telescopes: one at Columbia University in New York City, and later in Cambridge,
Massachusetts, and one at the Cerro Tololo Inter-American Observatory in Chile.
These observations have an angular resolution of 8\parcm5 at 115\,GHz, the frequency of the $^{12}$CO$(J$\,$=$\,$1$\,$\rightarrow$\,$0)$ line.
This study used the dataset covering the whole Galactic plane within the Galactic latitude range $|b|$\,$\leq$\,5\deg\ with 1.3-\kps-wide spectral channels.
We used the raw dataset, which is not interpolated in the spatial coordinates or the spectral axis. 

The noise level throughout the \cite{dame2001} data is not uniform, as it comprises the combination of surveys with different instruments acquired at various times.
This is discussed in appendix~A of \cite{miville-deschenes2017}, where the authors identified three peaks in the noise distribution at 0.06, 0.10, and 0.19\,K.
We conservatively adopt the latter as the global noise level for this dataset.

We used the {\tt astropy} {\tt reproject} package to project this data into the same spatial grid of the 3D dust \citep[][]{astropy2018}.
We also applied the {\tt astropy} {\tt spectral-cube}\footnote{\url{http://spectral-cube.readthedocs.io}} package to project the spectral axis of these observations into that of the H{\sc i} observations introduced below.
This spectral reprojection does not affect the results of our analysis; it was performed for convenience in treating this large dataset. 
An example of the resulting CO emission data is shown in the middle panel of Fig.~\ref{fig:RGBmaps}.

\subsection{Neutral atomic hydrogen (H{\sc i}) emission}

We employed the publicly available H{\sc i} 21-cm-wavelength line observations in the H{\sc i} 4$\pi$ (H{\sc i}4PI) survey \citep[][]{hi4pi2016}.
This survey is based on data from the Effelsberg-Bonn H{\sc i} Survey \citep[EBHIS,][]{kerp2011} and the Galactic All-Sky Survey \cite[GASS,][]{McClure-Griffiths2009,kalberla2010}.
It comprises observations over the whole sky in the range $-600$\,$<$\,\vLOS\,$<$\,$600$\,\kps\ for declination $\delta$\,$>$\,0\deg\ and $-470$\,$<$\,\vLOS\,$<$\,470\,\kps\ for $\delta$\,$<$\,0\,\deg, as observed with the Effelsberg 100-m radio telescope in Bad M{\"u}nstereifel, Germany and the 64-m radio telescope at Parkes, New South Wales, Australia.

The HI4PI observations have complete spatial sampling, thus overcoming the central issue of pioneering whole-sky H{\sc i} observations in the Leiden/Argentine/Bonn (LAB) survey \citep[][]{kalberla2005}.
The final H{\sc i}4PI data product is a set of whole-sky H{\sc i} maps with an angular resolution of 16\parcm2 and sensitivity of 43\,mK per 1.29-\kps\ velocity channel.

We used the data distributed in FITS-format binary tables containing lists of spectra sampled on a {\tt HEALPix} grid with $N_{\rm side}$\,=\,$1024$, which corresponds to a pixel angular size of 3\parcm44 \citep{gorski2005}.
We arranged these spectra in all-sky HEALPix maps corresponding to the 1.29-\kps-wide velocity channels.
Using the tools in the Python {\tt healpy} package, we produced a Cartesian projection for each one of these velocity channels into the same spatial grid used for the 3D dust data.
An example of the resulting H{\sc i} emission data is shown on the bottom panel of Fig.~\ref{fig:RGBmaps}.

\begin{figure}[ht]
\centerline{
\includegraphics[width=0.49\textwidth,angle=0,origin=c]{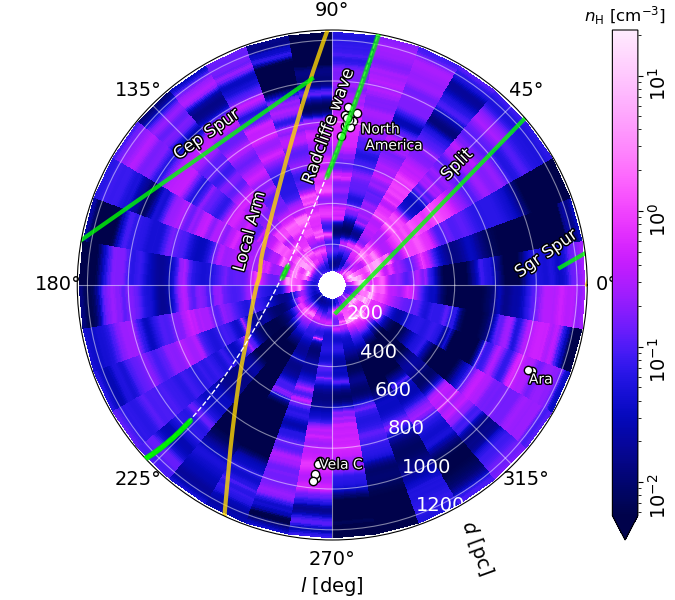}}
\caption{Heliocentric distribution of the mean nucleon density derived from the 3D dust extinction models presented in \cite{edenhofer2024} for the sky region $|b|$\,$\leq$\,5\deg.
The superimposed curves indicate large-scale features in the Solar neighborhood, as reported in \cite{zucker2023}.
For the Radcliffe Wave, we indicate the full extent of the structure with the dashed white line and the segments within $|b|$\,$\leq$\,5\deg\ in green.
The white disks indicate the locations of MC complexes referenced in the text.
}
\label{fig:polarDens}
\end{figure}

\section{Methods}\label{sec:methods}

\subsection{Gradient comparison}

We matched the H{\sc i} and CO emission across velocity channels and the 3D dust cube using the histogram of oriented gradients (HOG) method introduced in \cite{soler2019}.
This method is based on characterizing the similarities in the emission distribution using the orientation of its gradients.
In the HOG method, a LOS-velocity channel map from the line emission and a distance channel from the 3D dust are morphologically similar if their gradients are mainly parallel and dissimilar if they are randomly oriented.
The distribution of angles between gradient vectors is evaluated using the projected Rayleigh statistic ($V$), a statistical test of non-uniformity in an angle distribution around a particular direction \citep[see][and references therein]{jow2018}.

We implemented HOG as follows.
We split the sky band within $|b|$\,$<$\,5\deg\ into 36 10\deg\,$\times$\,10\deg\ regions, identified by the index $k$.
For each region, we used the PPV cube derived from the line emission by tracer $X$, $I_{ijkp}^{\rm X}$, and the nucleon density PPD cube derived from the 3D dust reconstruction $n_{ijkq}$.
The $i$ and $j$ indexes run over the pixels in the sky coordinates, which in our case are Galactic longitude ($l$) and latitude ($b$), and the indexes $p$ and $q$ run over LOS velocity and distance channels, respectively.
We calculated the relative orientation angles between the emission and density gradients,
\begin{equation}\label{eq:phi}
\theta^{X}_{ijkpq}=\arctan\left(\frac{\left\Vert\nabla I^{\rm X}_{ijkp} \times \nabla n_{ijkq}\right\Vert}{\nabla I^{\rm X}_{ijkp} \cdot \nabla n_{ijkq}}\right),
\end{equation}
where $\nabla$ is the differential operator corresponding to the gradient in the spatial sky coordinates $l$ and $b$ \postreport{and $\Vert\ldots\Vert$ indicates the vector norm}.

We computed the gradients using Gaussian derivatives resulting from the image's convolution with the spatial derivative of a two-dimensional Gaussian function.
The width of the two-dimensional Gaussian determines the area of the vicinity over which the gradient is calculated.
Varying the width of the Gaussian derivative kernel enables the sampling of different scales and reduces the effect of noise in the pixels \citep[see,][and references therein]{soler2013}.

\subsection{Quantifying the morphological correlation}

We used Eq.~\eqref{eq:phi} to calculate the relative orientation angles in the range $(-\pi,\pi]$, thus accounting for the direction of the gradients.
The values of $\theta^{X}_{ijkpq}$ are only meaningful in regions where both $|\nabla I^{\rm X}_{ijkp}|$ and $|\nabla n_{ijkq}|$ are greater than zero or above thresholds that are estimated according to the noise properties of each emission and 3D dust cube, as further discussed in App.~\ref{app:hog}.
We synthesized the information contained in $\theta^{X}_{ijkpq}$ by summing over the spatial coordinates, indexes $i$ and $j$, using the projected Rayleigh statistic \citep[][]{jow2018}, which we defined as
\begin{equation}\label{eq:Vd}
(V_{\rm d})_{kpq} = \frac{\sum_{ij}w_{ijkpq}\cos(\theta_{ijkpq})}{\left(\sum_{ij}w_{ijkpq}/2\right)^{1/2}}.
\end{equation}
This definition differs from that used by \cite{soler2019}, where the gradients' orientation (and not their direction) was considered.
To distinguish our definition from that provided by \cite{soler2019}, we introduced the subscript ``d''
, for direction. 
This modification improves the significance when comparing 3D dust and CO, where only parallel gradients are meaningful in the column density range considered in the 3D dust reconstruction. 
A comparison between the two metrics is presented in App.~\ref{app:VandVd}.

The projected Rayleigh statistic (either $V$ or $V_{\rm d}$) tests non-uniformity in a distribution of angles around a particular direction.
In Eq.~\ref{eq:Vd}, the angles of interest are $\theta_{0}$\,$=$\,$0\deg$ and $180\deg$, such that $V_{\rm d}$\,$>$\,0 or $V_{\rm d}$\,$<$\,0 correspond to clustering around those angles \citep{durandANDgreenwood1958}.
Values of $V_{\rm d}$\,$>$\,0, which imply that the gradients are primarily parallel, quantify the significance of the morphological similarity between the line emission, $I^{X}$, and the density, $n$, for a pair distance-\vLOS\ channels.
Values of $V_{\rm d}$\,$<$\,0, which imply that the gradients are mostly antiparallel, are potentially relevant toward HISA features, which are characterized by an anti-correlation between the dust density and the H{\sc i} emission intensity.

The projected Rayleigh statistic's null hypothesis is that the angle distribution is uniform.
In the particular case of independent and uniformly distributed angles and for a large number of samples, values of $V_{\rm d}$\,$\approx$\,$1.64$ and $2.57$ correspond to the rejection of the null hypothesis with a probability of 5\% and 0.5\%, respectively \citep{batschelet1972}.
Thus, a value of $V_{\rm d}$\,$\approx$\,2.87 is roughly equivalent to a 3$\sigma$ confidence interval.
Similarly to the $\chi^2$-test probabilities, $V_{\rm d}$ and its corresponding null hypothesis rejection probability are reported in the classical circular statistics literature as tables of ``critical values'', as, for example, in \cite{batschelet1972}, or computed in the circular statistics packages, such as {\tt circstats} in {\tt astropy} \citep{astropy2018}.

\begin{figure}[ht]
\centerline{\includegraphics[width=0.5\textwidth,angle=0,origin=c]{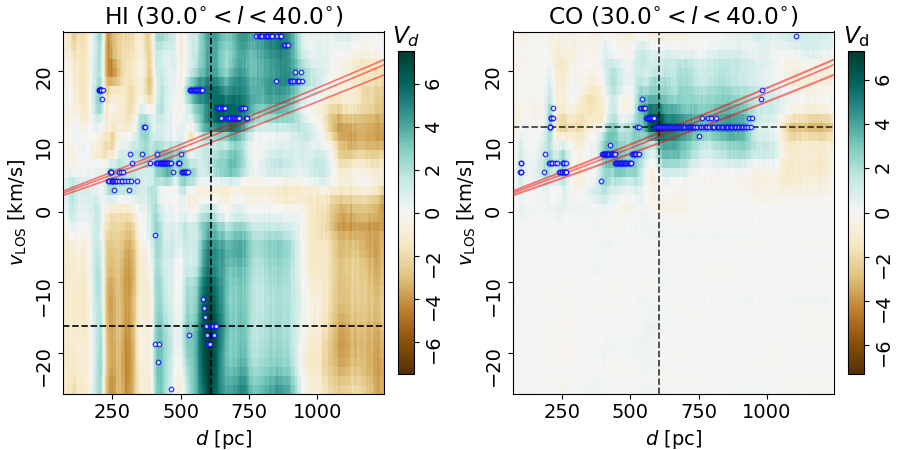}}
\caption{Example of the HOG method morphological correlation between the 3D dust density reconstruction and the line emission observations for the region 30\deg\,$<$\,$l$\,$<$\,40\deg\ and $|b|$\,$<$\,5\deg.
Each panel corresponds to the correlation between distance channels, from the 3D dust density, and velocity channels, from the H{\sc i} and CO line emission, shown in the left and right panels, respectively.
The correlation metric is the direction-sensitive projected Rayleigh statistic ($V_{\rm d}$, Eq.~\ref{eq:Vd}).
Values of $V_{\rm d}$\,$\approx$\,0 correspond to a random orientation between the gradients of the two tracers, thus indicating low morphological correlation. 
Values of $V_{\rm d}$\,$>$\,2.87 correspond to mostly parallel gradients, thus indicating a significant morphological correlation.
Values of $V_{\rm d}$\,$<$\,2.87 correspond to mostly antiparallel gradients.
The white markers indicate the \vLOS\ with the highest $V_{\rm d}$ for each distance channel.
The green dashed lines correspond to the distance and \vLOS\ with the highest $V_{\rm d}$ for the whole velocity and distance ranges.
The three cyan lines represent the kinematic distances for the Galactic longitude at the center and the region's edges, according to the \cite{reid2019} Galactic rotation model.
}
\label{fig:Vplanes}
\end{figure}

\subsubsection{Selection of HOG parameters}

There are two main parameters in HOG calculation.
The first is the pixel vicinity for the finite-differences gradient calculation, that is, the angular scale at which the gradients are calculated.
The second is the statistical weight in Eq.~\ref{eq:Vd}, which accounts for the statistical dependence of the gradients.

The derivative kernel size ($\Delta$) sets the area over which the finite differences derivative is calculated.
Varying the size of the derivative kernel enables the sampling of different angular scales and reduces the effect of noise in the pixels.
For this particular application, we chose $\Delta$\,$=$\,30\arcmin, which roughly Nyqvist-samples the observations and provides 40\,$\times$\,40 independent relative orientation measurements per tile.
The results of different kernel size selections are discussed in App.~\ref{app:VandVd}.
The size of the derivative kernel sets a common angular scale for comparing the line emission and 3D dust channels.
Thus, we did not smooth the input data to the same angular resolution for the HOG computation.

We accounted for the spatial correlations introduced by the scale of the derivative kernel by choosing the statistical weights $w_{ijkpq}$\,$=$\,$(\delta x/\Delta)^{2}$, where $\delta x$ is the angular size of the pixel and $\Delta$ is the derivative kernel's FWHM.
For pixels where the gradient norm is negligible or can be confused with the signal produced by noise, we set $w_{ijkpq}$\,$=$\,0.
If all the gradients in an image pair are negligible, Eq.~\ref{eq:Vd} has an indeterminate form that is treated numerically as Not a Number ({\tt NaN}).

\subsubsection{Handling of errors in the {\tt HOG} morphological correlation and LOS velocity estimation}

We estimated the errors in $V_{\rm d}$ by applying the HOG method to the line emission cubes and the twelve 3D dust distribution realizations produced by \cite{edenhofer2024} from the variational posterior.
The result is twelve $V_{\rm d}$ values for each line emission and 3D dust channel map.
Additionally, we analyzed 100 realizations of the line emission maps produced with Monte Carlo sampling, assuming a Gaussian distribution centered on the observed value in each channel and a standard deviation equal to the noise level of each data set.
\postreport{We reported the mean value of $V_{\rm d}$ for the twelve 3D dust cubes and the 100 realizations.}

\postreport{The $V_{\rm d}$ standard deviation estimated from the 12\,$\times$\,100 samples, $\sigma_{V_{\rm d}}$, is a measurement of variance in the morphological correlation.
In App.~\ref{app:errors}, we show that $\sigma_{V_{\rm d}}$ is led by the variance among the twelve 3D dust realizations rather than by the noise in the line emission observations.
Given that each of the twelve realizations is equally valid, large $\sigma_{V_{\rm d}}$ value reflects an ambiguity in the 3D dust morphology that limits the significance of the morphological matching with a line emission channel.
Thus, we exclude distance-\vLOS\ pairs with $|V_{\rm d}/\sigma_{V_{\rm d}}|$\,$<$\,3.0 from the calculations of the representative \vLOS\ in each distance channel, as described in Sec.~\ref{sec:dustvlsr}.
We also exclude distance-\vLOS\ pairs with $V_{d}$ below the chance correlation thresholds obtained when flipping one of the input maps in the vertical, horizontal, and diagonal directions, as described in App.~\ref{app:chancecorrelation}.
}

\subsection{Dissecting the nearby Galactic plane}\label{sec:dustvlsr}

We carried out the Galactic plane analysis by splitting the observations within the $|b|$\,$<$\,5\deg\ range into 36 contiguous 10\deg\,$\times$\,10\deg\ tiles centered on $b$\,$=$\,0\deg.
For each tile, we computed Eq.~\eqref{eq:Vd} using the {\tt HOGcorr\_ima} routine in the publicly available {\tt astroHOG} package\footnote{\url{https://github.com/solerjuan/astroHOG}}.
The result is a $V_{\rm d}$ matrix representing the correlation between each distance and velocity-channel map toward a tile, as illustrated in Fig.~\ref{fig:Vplanes}.

The $V_{\rm d}$ matrix is a map of the relationship between $d$ and \vLOS\ for each studied region, something which is usually estimated by assuming circular motions around the Galactic center to compute kinematic distances \citep[see, for example,][]{oort1958,sofue2011,wenger2018,hunter2024}.
In the HOG method, this mapping is based on the morphological similarity between the line emission and the 3D dust density distribution for each $d$ and \vLOS\ pair.
Thus, $d$ and \vLOS\ are not linked by a one-to-one relation but rather by a distribution of $V_{\rm d}$, which accounts for the correlation across LOS velocities introduced by the emission linewidth and the coherence expected in 3D dust structure for contiguous distance channels. 

We employed the $V_{\rm d}$ matrices to assign a \vLOS\ to each distance channel in a tile.
We used the critical value 2.87 and the $V_{\rm d}$ standard deviation, $(\sigma_{V_{\rm d}})_{vk}$, as significance thresholds for the LOS velocity assignment to each distance channel.
If the maximum value of $V_{\rm d}$ is below either of these values, no \vLOS\ is assigned to a distance channel.
If more than one velocity channel has $V_{\rm d}$ above these thresholds, we assign the \vLOS\ corresponding to the maximum $V_{\rm d}$.
\postreport{The selection by maximum $V_{\rm d}$ excludes distance channels potentially dominated by HISA from the \vLOS\ reconstruction.
This choice is motivated by analysis in Appendix~\ref{app:hisa}, which shows that a selection based on $|V_{\rm d}|$ instead of $V_{\rm d}$ only favors negative values in a small number of the distance channels and that the interpretation of $V_{\rm d}$\,$<$\,0 exclusively as the product of HISA is not always correct.
We reserve the specific study of HISA using {\tt HOG} for a subsequent publication.}

The simplification of the velocity field in our {\tt HOG} application aims to identify a representative LOS velocity for the bulk of the dust in each distance slice along a 10\deg\,$\times$\,10\deg\ area.
Thus, we estimated the average LOS motions of volumes between 12\,$\times$\,12\,$\times$\,0.4\,pc$^{3}$ and 220\,$\times$\,220\,\,$\times$\,7.0\,pc$^{3}$.
We acknowledge that this selection extracts just a segment of the turbulent energy cascade, which connects galactic motions to the smallest scales in the ISM.
However, our goal is reconstructing the bulk motions of density structures revealed by the 3D dust model rather than describing the full complexity of the ISM velocity field across scales, which has been considered in other works \citep[see, for example,][and references therein]{elmegreenANDscalo2004,hennebelleANDfalgarone2012}.

Figure~\ref{fig:Vplanes} illustrates an example of the LOS velocity selection from a $V_{\rm d}$ matrix, where the markers represent the assigned \vLOS\ for each distance channel.
In this example, the representative LOS motions for CO roughly follow the expected \vLOS\ from Galactic rotation, indicated by the solid lines.
In contrast, the H{\sc i} \vLOS\ shows deviations up to tens of \kps\ from the CO \vLOS\ and the LOS velocities expected from the Galactic rotation model.
In the following sections, we discuss this and other dynamical effects on the reconstructed velocity field after presenting the global HOG results.

\section{Results}\label{sec:results}

\begin{figure*}[ht]
\centerline{
\includegraphics[width=0.5\textwidth,angle=0,origin=c]{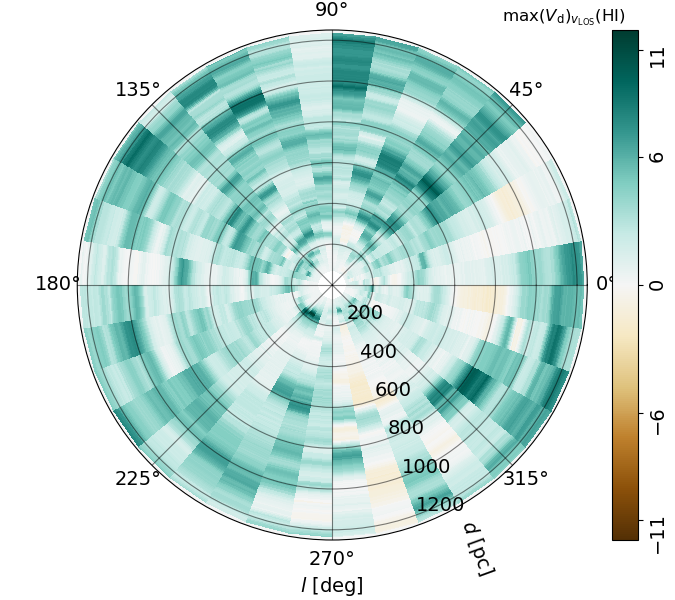}
\includegraphics[width=0.5\textwidth,angle=0,origin=c]{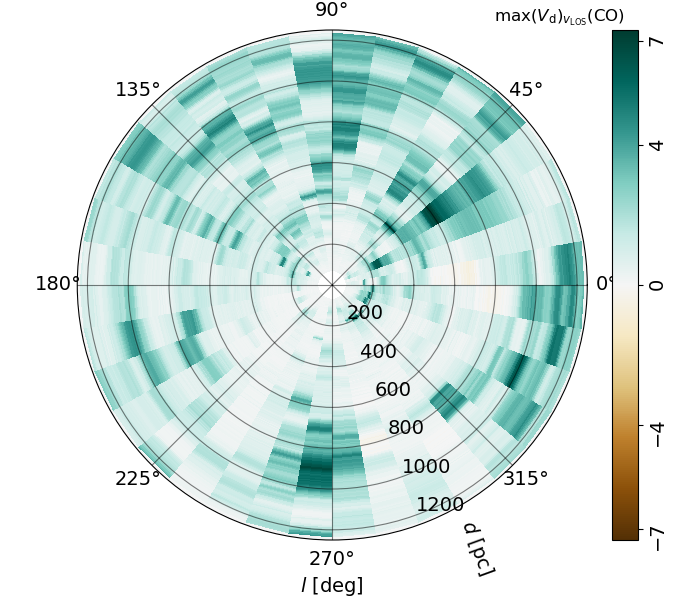}
}
\caption{Maximum morphological correlation between the 10\deg\,$\times$\,10\deg\ distance channels and the H{\sc i} ({\it left}) and CO ({\it right}) line emission in the range $-25$\,$<$\,\vLOS\,$<$\,25\,\kps, as quantified by the direction-sensitive projected Rayleigh statistic ($V_{\rm d}$; Eq.~\ref{eq:Vd}).
}
\label{fig:polarPRS}
\end{figure*}

\begin{figure*}[ht]
\centerline{
\includegraphics[width=0.5\textwidth,angle=0,origin=c]{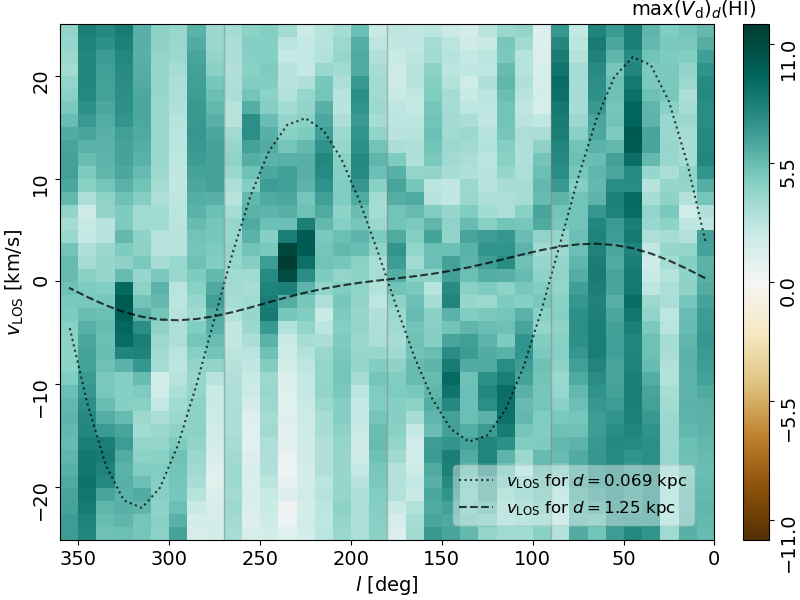}
\includegraphics[width=0.5\textwidth,angle=0,origin=c]{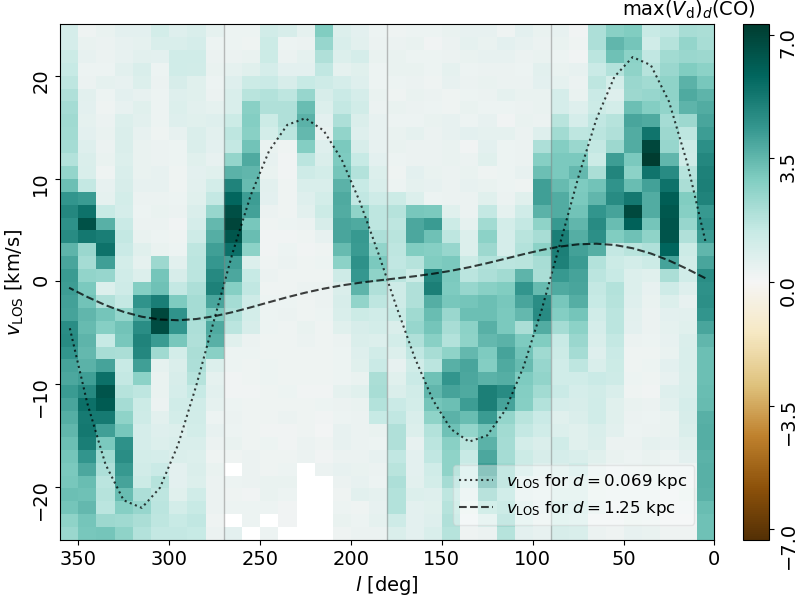}
}
\caption{Maximum morphological correlation between the 10\deg\,$\times$\,10\deg\ LOS velocity channels and the dust density in the range $69$\,$<$\,$d$\,$<$\,1250\,pc, as quantified by the direction-sensitive projected Rayleigh statistic ($V_{\rm d}$; Eq.~\ref{eq:Vd}).
The cyan curves represent the \vLOS\ expected for $d$\,$=$\,69 and 1250\,pc according to the \cite{reid2019} Galactic rotation model, shown by dotted and dashed lines, respectively.
}
\label{fig:lvPRS}
\end{figure*}

\subsection{Correlation between 3D dust and line emission morphology}

The fundamental hallmark of the HOG analysis is quantifying the similarity in the plane-of-the-sky distribution of the 3D dust extinction and the H{\sc i} and CO line emission across distance and velocity channels.
Figure~\ref{fig:polarPRS} shows the maximum values of the correlation metric across \vLOS, $\max(V_{\rm d})_{v}$, for the distance channels in the 36 tiles in which we divided the $|b|$\,$<$\,5\deg\ sky region.
That is, each pixel in Fig.~\ref{fig:polarPRS} corresponds to the value of Eq.~\eqref{eq:Vd} for the particular case of $p$\,$=$\,$p^{*}$, where $p^{*}$ is the velocity channel with the maximum value of $V_{\rm d}$.

The values of $\max(V_{\rm d})_{v}$ reported in Fig.~\ref{fig:polarPRS} indicate whether or not a dust parcel has a match in the H{\sc i} or CO emission in the range $-25$\,$<$\,\vLOS\,$<$\,25\,\kps.
By construction, $\max(V_{\rm d})_{v}$ highlights configurations where H{\sc i} and 3D dust density gradients are preferentially parallel.
Although this choice is biased against regions potentially dominated by HISA features, where $V_{\rm d}$\,$<$\,0, it minimizes the spurious signal produced by chance correlation between voids in the 3D dust and H{\sc i} emission, as further discussed in App.~\ref{app:hisa}.

Figure~\ref{fig:polarPRS} shows that most of the distance channels display a significant correlation ($V_{\rm d}$\,$>$\,2.87) between the 3D dust morphology and the line emission.
This result is remarkable given that it comes from two pairs of independent datasets, H{\sc i} and 3D dust and CO and 3D dust.
Although the association of 3D dust and line emission components has been previously studied toward particular regions in the Solar vicinity \citep[see, for example,][]{piecka2024,rybarczyk2024}, this is the first time this relation is reported using a quantitative measure of the morphological similarity and covering such an extensive region.

An alternative representation of the correlation between the line emission and the 3D dust is presented in Fig.~\ref{fig:lvPRS}, where we show the maximum values of $V_{\rm d}$ across distances, $\max(V_{\rm d})_{d}$, for the \vLOS\ channels in the 36 tiles in which we divided the $|b|$\,$<$\,5\deg\ sky region.
That is, each pixel in Fig.~\ref{fig:lvPRS} corresponds to the value of Eq.~\eqref{eq:Vd} for the particular case of $q$\,$=$\,$q^{*}$, where $q^{*}$ is the distance channel with the maximum value of $V_{\rm d}$.
Figure~\ref{fig:lvPRS} can be understood as a different projection of the PPDV space mapped with the HOG method, although some of the line emission signal in the $|v_{\rm LOS}|$\,$\leq$\,25\,\kps\ may come from density structures beyond the 3D dust reconstruction distance range.
Therefore, Fig.~\ref{fig:lvPRS} should be understood as a visual representation of whether a line emission channel has a morphological match in the 3D dust distribution between 69 and 1250\,pc, rather than as a full reconstruction of the PPV space from PPD information.

It is evident on the right panel of Fig.~\ref{fig:lvPRS} that the CO emission has the most considerable morphological correlation with the 3D dust within the velocity ranges expected from the circular motion around the Galactic center, which we estimated using the \cite{reid2019} model.
In contrast, the H{\sc i} emission is highly correlated with the 3D dust beyond the LOS velocity limits expected from pure circular motions.
We further discuss the prevalence of these streaming motions in Sec.~\ref{sec:driftmotions}.

\begin{figure*}[ht]
\centerline{\includegraphics[width=0.9\textwidth,angle=0,origin=c]{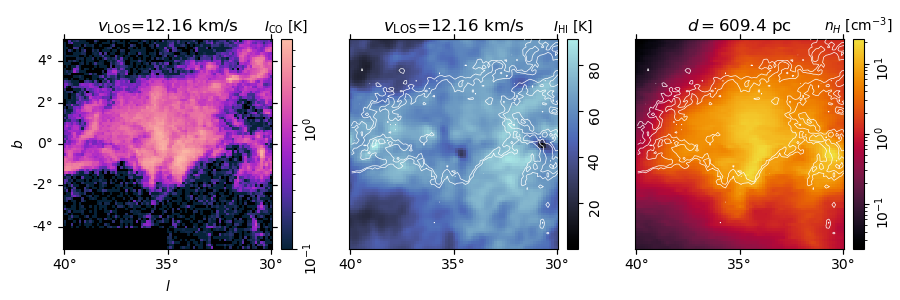}}
\caption{Example of the distance and CO velocity channel with high morphological correlation.
It corresponds to the highest $V_{\rm d}$ in the comparison between the CO line emission and the 3D dust presented in the right panel of Fig.~\ref{fig:polarPRS}.
From left to right, the panels show the CO and H{\sc i} line emission at the indicated \vLOS\ and the nucleon density inferred from the 3D dust model at the corresponding distance.}
\label{fig:maxVtilesCO}
\end{figure*}
\begin{figure*}[ht]
\centerline{\includegraphics[width=0.9\textwidth,angle=0,origin=c]{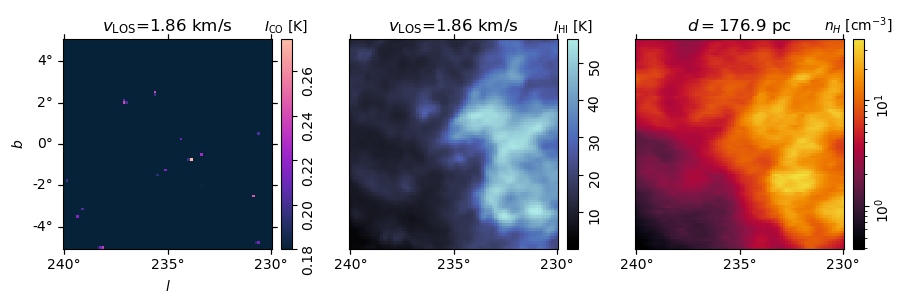}}
\caption{Same as Fig.~\ref{fig:maxVtilesCO}, but for high morphological correlation between H{\sc i} and 3D dust.
This example corresponds to the highest $V_{\rm d}$ in the comparison between the H{\sc i} line emission and the 3D dust density presented in the left panel of Fig.~\ref{fig:polarPRS}.}
\label{fig:maxVtilesHI}
\end{figure*}

Figure~\ref{fig:maxVtilesCO} shows an example of the H{\sc i} and CO line emission and the density distribution for the tile with the highest morphological correlation between CO and the 3D dust, as identified by the highest $V_{\rm d}$ in the right-hand-side panel of Figure~\ref{fig:polarPRS}.
The noticeable similarity between the CO emission and the density distribution visually confirms the HOG results.
Note that the fact we see CO emission associated with gas with a density $n_{\rm H}\,\sim\,10$--20$ \, {\rm cm^{-3}}$, which we would typically expect to be atomic, is likely an indication that the gas is clumped on scales smaller than the resolution of the 3D dust map at this distance, as confirmed by higher-angular-resolution CO surveys \citep[see, for example,][]{jackson2006,benedettini2020,ma2021}.

The H{\sc i} emission in Fig.~\ref{fig:maxVtilesCO} shows the typical shadows produced by the cold H{\sc i} observed on top of the warm background emission, which is a characteristic signature of a HISA feature \citep[see, for example,][]{gibson2000}.
We presume that the lower H{\sc i} emission within the CO contours in the central panel of Fig.~\ref{fig:maxVtilesCO} is a HISA based on the morphological similarity between the CO emission and HISA identified in other observations of the Galactic plane \citep[see, for example,][]{gibson2005,soler2019,wang2020hisa}.
However, a confirmation of the HISA for this example and throughout the Galactic plane requires further analysis of the H{\sc i} spectra, which is beyond the scope of this paper \citep[see, for example][]{syed2023}.
The presumed HISA in this particular example does not produce a prominent $V_{\rm d}$\,$<$\,0 in the HOG comparison between H{\sc i} and 3D dust, as shown in Fig.~\ref{fig:Vplanes}.
This is most likely due to the combination of antiparallel gradients in the HISA feature and the parallel gradients outside it, resulting in $V_{\rm}$\,$>$\,0.

Figure~\ref{fig:maxVtilesHI} shows the H{\sc i} and CO line emission and density distribution for the tiles with the highest $V_{\rm d}$ in the H{\sc i} and 3D dust comparison, as identified in the left-hand-side panel of Fig.~\ref{fig:polarPRS}.
In contrast with the region in Fig.~\ref{fig:maxVtilesCO}, there is no CO counterpart to the matching structures in H{\sc i} and 3D dust.
This example illustrates that the morphological correlation traced by H{\sc i} and CO is not necessarily identical because of the different structures each tracer shows.

\begin{figure}[ht]
\centerline{\includegraphics[width=0.5\textwidth,angle=0,origin=c]{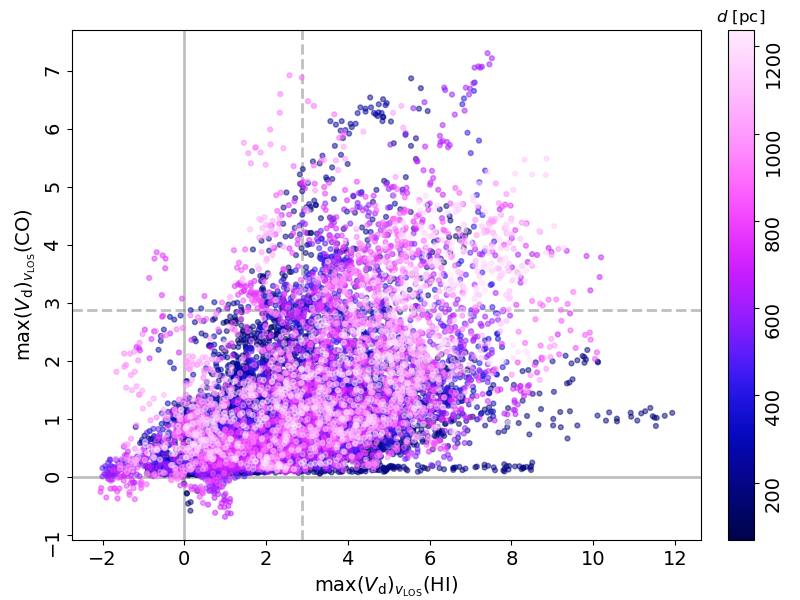}}
\caption{Comparison of the maximum $V_{\rm d}$ values for H{\sc i} and CO reported in Fig.~\ref{fig:polarPRS}.
}
\label{fig:scatterVforHIandCO}
\end{figure}

We further compared the morphological correlations of the 3D dust and the two gas tracers in Fig.~\ref{fig:scatterVforHIandCO}.
The scatter plot contrasts the values of $\max(V_{\rm d})_{v}$ for H{\sc i} and CO reported in Fig.~\ref{fig:polarPRS}.
There are some tiles where $\max(V_{\rm d})$ is high for the two gas tracers.
However, the correlation between the 3D dust and one gas tracer does not indicate its correlation with the other.

We found that the highest $\max(V_{\rm d})_{v}$ are not concentrated around near or far distances.
We interpret this observation as an indication that no significant bias is introduced by distance in the morphological correlation, as further considered in App.~\ref{app:physics}.
The highest $\max(V_{\rm d})_{v}$ is found for H{\sc i}, as expected from the largest angular extend of H{\sc i} emission.

Figure~\ref{fig:scatterVforHIandCO} shows a few tiles where $\max(V_{\rm d})_{v}$ is negative, most of them in H{\sc i}.
This suggests that despite our selection of the maximum $V_{\rm d}$ in each tile, there are regions where the H{\sc i} emission gradients are mostly antiparallel to the 3D dust gradients.
However, their significance is low, $V_{\rm d}$\,$<$\,$-2.87$, and they are related to chance correlation rather than HISA, as discussed in App.~\ref{app:hisa}.

\subsection{Local ISM motions}\label{sec:driftmotions}

\begin{figure}[ht]
\centerline{
\includegraphics[width=0.49\textwidth,angle=0,origin=c]{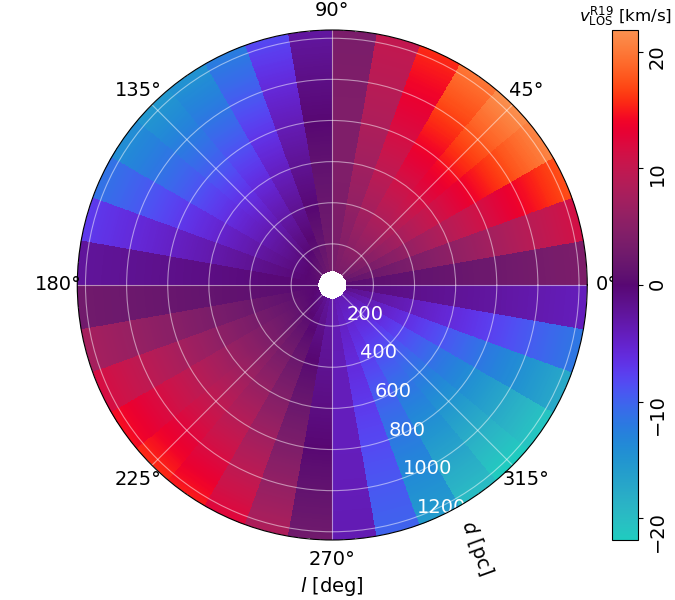}
}
\caption{Expected line-of-sight velocity (\emph{right}) from the \cite{reid2019} Galactic rotation model across the cells in our analysis. 
}
\label{fig:polarReid2019vLSR}
\end{figure}

We used the regions with high morphological correlation ($V_{\rm d}$ \,$>$\,2.87) to assign a prevalent velocity to the dust parcels within the 3D dust reconstruction, following the procedure described in Sec.~\ref{sec:dustvlsr}.
Assuming circular rotation around the Galactic center implies a likely LOS velocity pattern in our studied region.
For example, Fig.~\ref{fig:polarReid2019vLSR} shows the expected \vLOS\ pattern derived from the state-of-the-art Galactic rotation model presented in \cite{reid2019}.

\begin{figure*}[ht]
\centerline{
\includegraphics[width=0.49\textwidth,angle=0,origin=c]{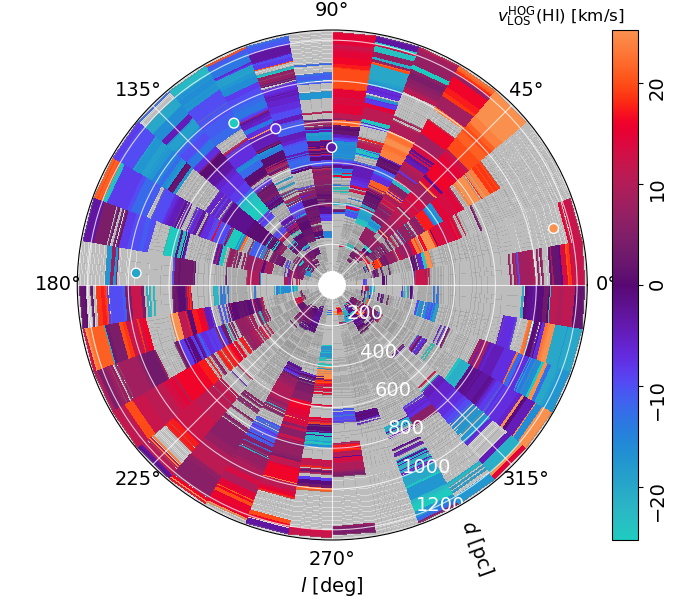}
\includegraphics[width=0.49\textwidth,angle=0,origin=c]{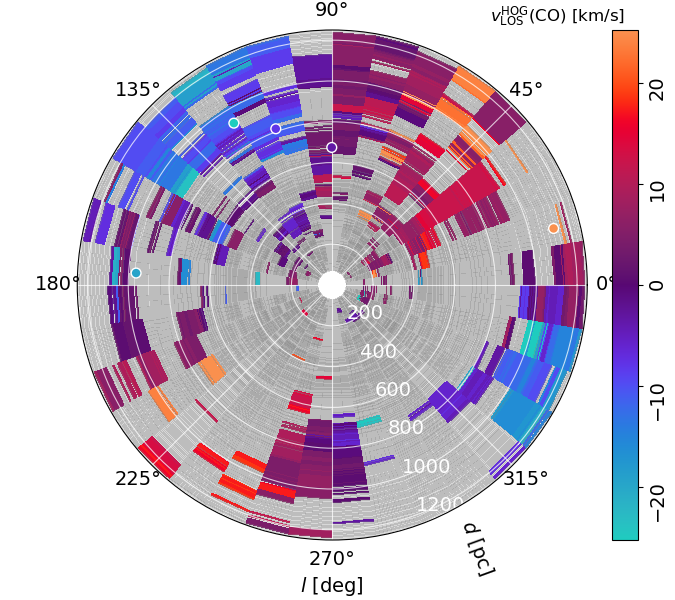}
}
\caption{Line-of-sight velocity derived from the H{\sc i} ({\it left}) and CO ({\it right}) line emission associated to the distance channels in the 3D dust reconstruction using the HOG method ($v^{\rm HOG}_{\rm LOS}$).
The colored circles correspond to the position and \vLOS\ for the five high-mass star-forming regions with VLBI parallax observations within the studied volume.
}
\label{fig:polarvLSRfromHIandCO}
\end{figure*}

Figure~\ref{fig:polarvLSRfromHIandCO} shows the velocity field reconstructed from 3D dust and line emission correlation identified using the HOG method.
The regions without assigned \vLOS\ correspond to distance channels with $V_{\rm d}$\,$<$\,2.87 for all velocity channels in either tracer.
The first compelling result from the velocity reconstruction is the large-scale similarity between the velocity fields in Fig.~\ref{fig:polarvLSRfromHIandCO} and the theoretical expectation shown in Fig.~\ref{fig:polarReid2019vLSR}, that is, the correspondence between the motions away from the Sun in the first and third Galactic quadrants and toward the Sun in the second and fourth quadrants.
The correspondence of this quadrupolar pattern in the model and our reconstruction is evident in both gas tracers, despite the large portion of the first quadrant for which the morphological correlation with H{\sc i} does not result in an unambiguous \vLOS\ assignment.

Appendix~\ref{app:synthobs} presents the HOG analysis of synthetic line emission from a simulated MC, demonstrating that our approach correctly reproduces the mean \vLOS\ of the numerical cloud.
Dispersions around the central \vLOS\ are dominated by the emission linewidth. 
Larger velocity excursions registered with the HOG are produced by the MC's complex dynamic structure rather than by a systematic effect introduced by the method.

We quantified the departures from the LOS motions produced by pure Galactic rotation by subtracting the expected LOS velocity from the \citet{reid2019} model, $v^{\rm R19}_{\rm LOS}$, from the reconstructions presented in Fig.~\ref{fig:polarvLSRfromHIandCO}, $v^{\rm HOG}_{\rm LOS}$.
In what follows, we refer to the quantity $v^{\rm R19}_{\rm LOS}-v^{\rm HOG}_{\rm LOS}$ as streaming motions.

\begin{figure}[ht]
\centerline{\includegraphics[width=0.49\textwidth,angle=0,origin=c]{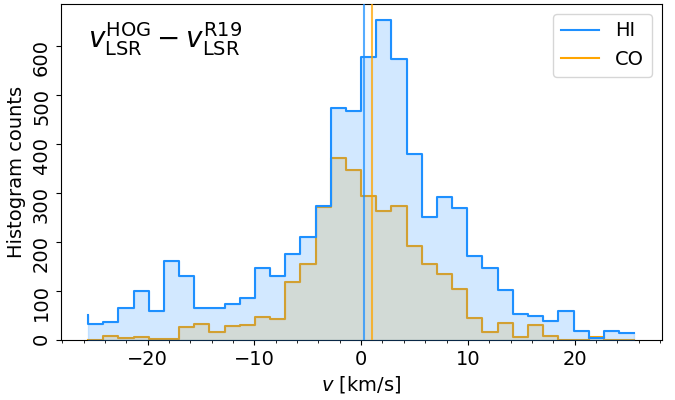}
}
\caption{Histograms for the differences between the LOS velocity estimated from the H{\sc i} and CO line emission, $v^{\rm HOG}_{\rm LOS}$, as reported in Fig.~\ref{fig:polarvLSRfromHIandCO}, and the expected LOS velocity according to the Galactic rotation model presented by \cite{reid2019}, $v^{\rm R19}_{\rm LOS}$, presented in Fig.~\ref{fig:polarReid2019vLSR}.
}
\label{fig:histdiffv}
\end{figure}

The streaming motions distribution, shown in Fig.~\ref{fig:histdiffv}, is not strictly symmetric but is centered around zero, indicating that, on average, the departures from pure Galactic rotation are relatively small.
The mean values streaming motions are roughly \postreport{0.2} and \postreport{1.1}\,\kps\ for H{\sc i} and CO, respectively. 
These values are below the 1.3-\kps\ spectral resolution of the line emission data, so they are within the LOS velocity determination uncertainties.
The characteristic amplitude of the streaming motions, quantified by the standard deviation of $v^{\rm R19}_{\rm LOS}-v^{\rm HOG}_{\rm LOS}$, is roughly \sigmavHI\ and \sigmavCO\,\kps, for H{\sc i} and CO, respectively.
\postreport{Given the $|v_{\rm LOS}|$\,$<$\,25\,\kps\ input range limitation, large excursions from Galactic rotation are currently excluded from the reconstruction, so these dispersions may represent lower limits of the actual streaming motions.
}

\begin{figure*}[ht]
\centerline{
\includegraphics[width=0.49\textwidth,angle=0,origin=c]{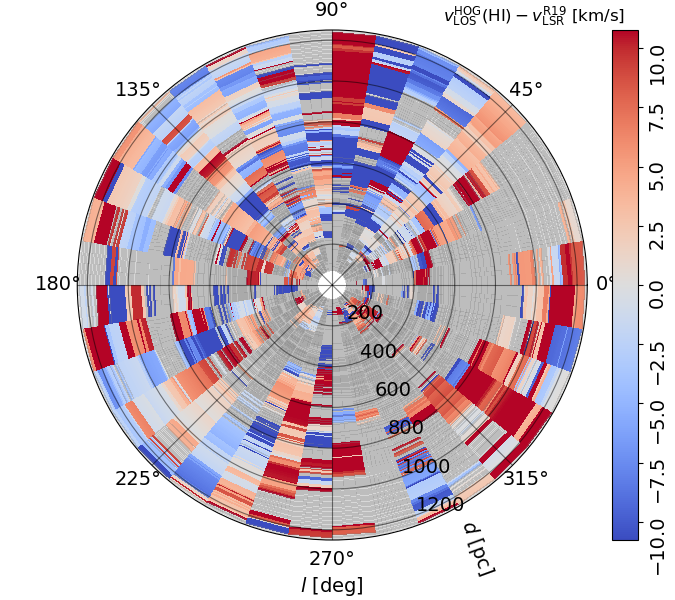}
\includegraphics[width=0.49\textwidth,angle=0,origin=c]{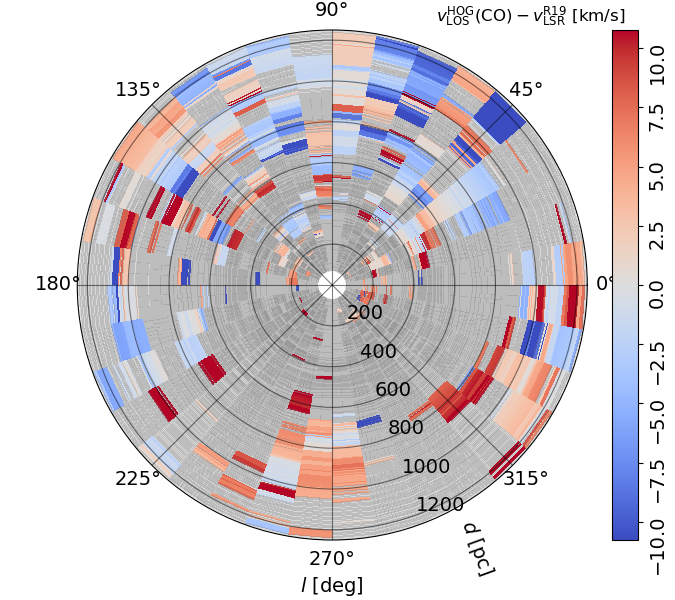}
}
\caption{Differences between the line-of-sight velocity derived from the H{\sc i} ({\it left}) and CO ({\it right}) line emission associated with the distance channels in the 3D dust reconstruction using the HOG method, $v^{\rm HOG}_{\rm LOS}$, and the expected LOS velocities according to the Galactic rotation model presented by \cite{reid2019}, $v^{\rm R19}_{\rm LOS}$.}
\label{fig:diffv}
\end{figure*}

The streaming motions map, presented in Fig.~\ref{fig:diffv}, does not show a prevalent large-scale pattern of streaming motions that may indicate the effect of compression toward the Local Arm.
Nor does it reveal global agreement in the radial velocity pattern sampled by the H{\sc i} and CO line emission, indicating that the residual motions due to Galactic rotation are relatively small.
In some positions, for example, at $d$\,$\sim$\,800\,pc for 80\,$<$\,$l$\,$<$\,90\deg, the H{\sc i} shows diverging motions that the CO does not match.
However, in general, the radial velocities do not reveal the concentration of converging or diverging motions but rather a succession of compressions and rarefactions along the line of sight, as expected from the behavior of a turbulent medium. 
\postreport{Given the scales considered in this study, averaging of cells that range between 12\,pc\,$\times$\,12\,pc\,$\times$\,0.4\,pc and 218\,pc\,$\times$\,218\,pc\,$\times$\,7\,pc in size, it is plausible that the reported motions result from a turbulent energy cascade introduced at kiloparsec scales by global galactic motions \citep[see, for example,][and references therein]{colman2022}.}

\section{Physical quantities derived from the local ISM motions}\label{sec:physics}

\begin{figure*}[ht]
\centerline{
\includegraphics[width=0.49\textwidth,angle=0,origin=c]{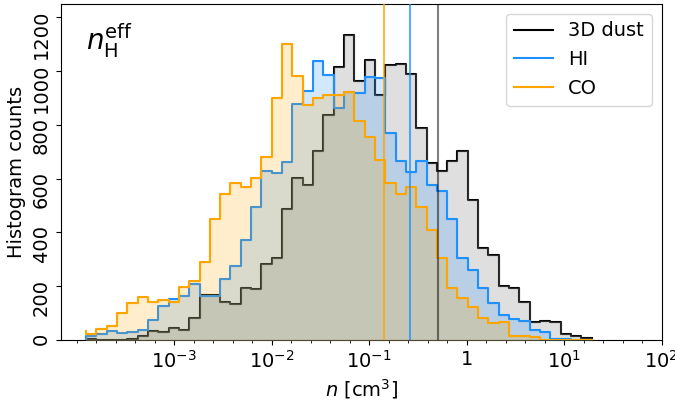}
\includegraphics[width=0.49\textwidth,angle=0,origin=c]{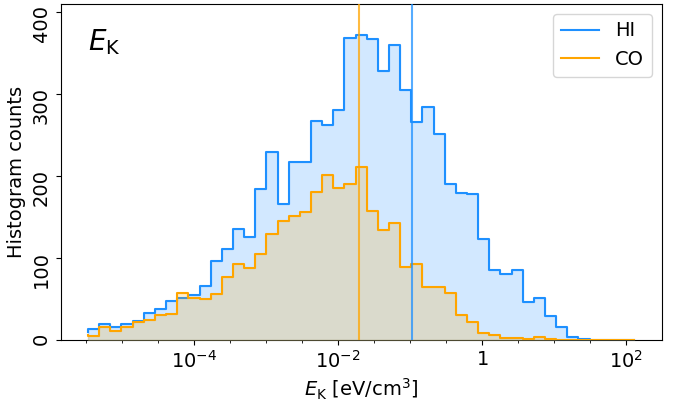}
}
\centerline{
\includegraphics[width=0.49\textwidth,angle=0,origin=c]{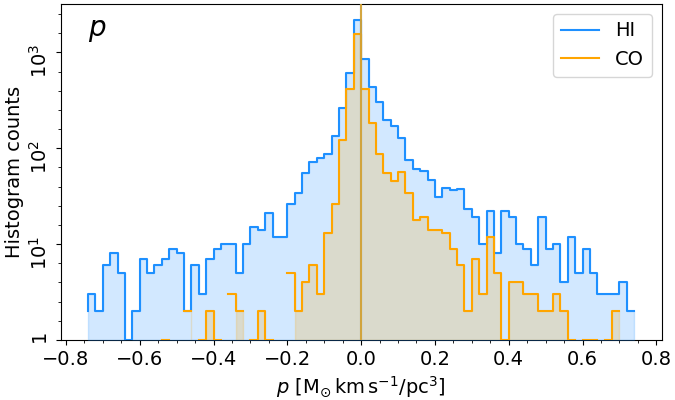}
\includegraphics[width=0.49\textwidth,angle=0,origin=c]{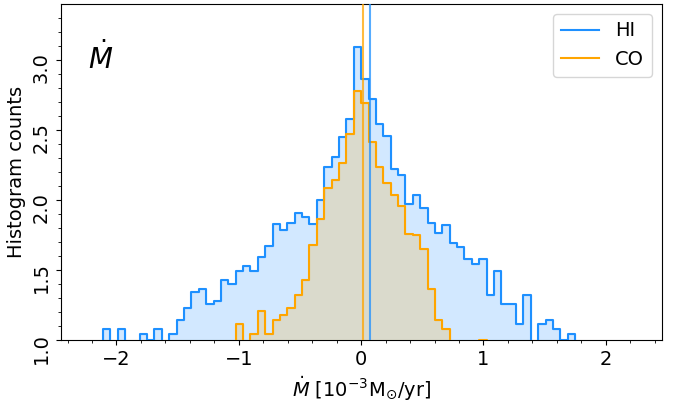}
}
\caption{Histograms of the physical quantities derived from the HOG analysis.
{\it Top left}. Effective energy density, $n_{\rm eff}$, that is assigned to each gas tracer, as defined in Eq.~\eqref{eq:neff}.
{\it Top right}. Kinetic energy density, $E_{\rm k}$, calculated with Eq.~\eqref{eq:Ek}.
{\it Bottom left}. Radial momentum, $p$, calculated with Eq.~\eqref{eq:p}.
{\it Bottom right}. Mass flow rate, $\dot{M}$, calculated with Eq.~\eqref{eq:mdot}.
}
\label{fig:histPhysicalQuantities}
\end{figure*}

The results presented in Sec~\ref{sec:results} indicate a significant correlation between the density traced by the 3D extinction models and the line emission.
We used this fact to estimate the distribution of departures from circular motions, also known as streaming motions.
Using these streaming motions, we calculated the distributions of 
kinetic energy density, momentum density, and mass flow rates associated with each gas tracer, as reported in  Fig.~\ref{fig:histPhysicalQuantities}.
We calculated these physical quantities as follows.

\subsection{Effective densities}

\postreport{In the standard, line-emission-based PPV approach, the density and LOS velocity of an object are obtained by isolating a line component, identifying its centroid, and integrating the emission \citep[see, for example,][]{miville-deschenes2017}.
Thus, the density and the LOS motion are defined for the same PPV object.
Our analysis defines the object in PPD space as a slice of 3D dust density for a particular sky area.
We assigned a \vLOS\ to that object using the HOG method through morphological matching.
However, not all the 3D dust slice is represented by the \vLOS\ from one tracer.
For example, the HI and CO emission can match different portions of a 3D dust slice, and each portion can have different velocities, as illustrated in Fig.~\ref{fig:Vplanes}.
Thus, we introduce an ``effective density'' corresponding to the portions of the 3D dust slice with a significant morphological match with either gas tracer.
This approach has the advantage of separating the H{\sc i} and the CO motions and assigning them a portion of the 3D dust instead of using the whole slice, which would lead to an overestimation of the energy and momentum densities.}

We employed the map of relative orientation angles between gradients for each \postreport{3D density slice and \vLOS-channel pair}.
We split this map into 9\,$\times$\,9 blocks, computed $V_{\rm d}$ for each block, and calculated the mass contained in the blocks with $V_{\rm d}$\,$>$\,1.0, which roughly corresponds to a 1-$\sigma$ significance for $V_{\rm d}$.
Given the limited number of independent gradient vectors in each block, the threshold on $V_{\rm d}$ is necessarily less rigid than that we used for the whole tile.
\postreport{Currently, the HOG analysis is limited by the angular resolution in the gas observations and the 3D dust reconstruction, which ultimately define the number of independent gradient vectors used in the morphological comparison.}
More stringent values lead to null values of the effective density for most of the distance channels.

The effective density \postreport{in each density slice} is determined by dividing the total mass in the blocks with \postreport{$V_{\rm d}$\,$>$\,1.0} by the volume of the \postreport{slice}.
Explicitly, for each distance channel $q$ in the portion of the Galactic plane identified with the subindex $k$, we identified a velocity channel $p^{*}$ that has the highest $V_{\rm d}$ for the gas tracer $X$.
For that distance channel, we compute the effective nucleon density,
\begin{equation}\label{eq:neff}
n^{{\rm eff},X}_{kp^{*}q}=\frac{\sum^{N_{A},N_{B}}_{ab}\left(\omega_{ab}\sum^{N_{i^\prime},N_{j^\prime}}_{i^\prime j^\prime} n_{i^\prime j^\prime}\mathcal{V}_{i^\prime j^\prime kq}\right)}{\mathcal{V}_{kq}},
\end{equation}
where $n$ is the nucleon density derived from the 3D dust distribution model, 
the indices $a$ and $b$ run over the blocks, $i^\prime$ and $j^\prime$ run over the pixels within the block, and $\mathcal{V}_{i^\prime j^\prime kq}$ is the volume occupied by the $i^\prime$-th and $j^\prime$-th pixel at the distance $d^{2}_{kq}$.
These definitions imply that the total volume in the distance channel $q$ is
\begin{equation}
\mathcal{V}_{kq}=\sum^{N_i,N_j}_{i,j}\mathcal{V}_{ijkq}\,=\,\sum^{N_i,N_j}_{i,j}d^{2}_{kq}\tan(\Delta l)\tan(\Delta b)\Delta d_q, 
\end{equation}
where $\Delta l$ and $\Delta b$ are the angular sizes of the pixel and $\Delta d_q$ is the distance channel width along the LOS. 
The statistical weight $\omega_{ab}$ is equal to one if $V_{\rm d}$\,$>$\,1.0 in the block identified by the indices $a$ and $b$ and zero otherwise.

The top-left panel of Fig.~\ref{fig:histPhysicalQuantities} shows the effective density distribution across distance cells.
By construction, the effective densities for the gas tracers are below the total dust mean.
While the mean density from all the dust is around \postRVcorr{0.51}\,cm$^{-3}$, the effective densities are around \postRVcorr{0.27} and \postRVcorr{0.14}\,cm$^{-3}$ for H{\sc i} and CO, respectively.
The lower value for CO reflects that tracer's lower volume filling factor, which implies a smaller contribution than that from H{\sc i} to the kinetic energy and momentum for each volume element in our reconstruction.

We note that the effective densities calculated using Eq.~\eqref{eq:neff} are not the specific density of a gas tracer but a combination of its volume filling factor and a selection based on its correlation with the 3D dust density.
For example, CO is associated with densities roughly above 100\,cm$^{-3}$, but the \cite{edenhofer2024} 3D dust models do not have the spatial resolution to detail the regions where these densities are prevalent.
The value of $n^{\rm eff}$ \postreport{does not aim to estimate the specific volumetric density of the gas but rather identify the portion of the 3D dust slice associated with the motion of the gas tracer.}

The fundamental uncertainties in estimating the densities come from the limitation of the 3D extinction reconstruction, described in detail in section~6 of \cite{edenhofer2024}.
We converted the extinction into nucleon density using the value from \cite{bohlin1978}, which recent observations of H{\sc i} emission and interstellar reddening indicate is too low by a factor of 1.5 \cite[][]{lenz2017}.
Variations from the diffuse environment at high Galactic latitude to the denser environment in the plane are also relevant, adding a factor of a few to the uncertainty.
\postRVcorr{Recent observations also reveal spatial variability of $R_{V}$ characterized by a standard a standard deviation of around 0.25 \citep{zhang2023rv}.}

\postreport{The HOG analysis of synthetic line emission from a simulated MC presented in App.~\ref{app:synthobs} indicates that the reconstructed H{\sc i} densities are roughly within a factor of two from the actual values.
The spatial averaging in Eq.~\eqref{eq:neff} results in smearing the density peaks across the area of a block, effectively limiting the maximum reconstructed density values.
For CO, the comparison with the numerical model is limited by the extent of the synthetic diffuse CO emission, which is a complex chemical modeling problem beyond this work's scope.
However, the HOG method reproduces the CO density profile, although it over-estimates the CO density by a factor of a few.
In either case, the limitations in the HOG reconstruction are below the variance across the studied region.}

\subsection{Kinetic energy}

We estimated the average kinetic energy from streaming motions for each distance channel $q$ as
\begin{equation}\label{eq:Ek}
\left<E^{X}_{\rm K}\right>_{kq}=\frac{1}{2}\,\rho^{{\rm eff},X}_{kp^{*}q}\left[(v^{X}_{\rm LOS})_{kp^{*}q}-(v^{R19}_{\rm LOS})_{kq}\right]^{2},
\end{equation}
where $X$ corresponds to the velocity tracer, either H{\sc i} or CO, 
$(v^{R19}_{\rm LOS})_{kq}$ is the expected LOS velocity from the \cite{reid2019} Galactic rotation model for the Galactic plane position $k$ and the distance channel $q$, and
$(v^{X}_{\rm LOS})_{kp^{*}q}$ is the LOS velocity from the line emission channel $p^{*}$ with the highest morphological correlation ($V_{\rm d}$) with the distance channel $q$.
We calculated the effective volumetric density, 
\begin{equation}\label{eq:rhoeff}
\rho^{{\rm eff},X}_{kp^{*}q}=\mu\,m_{p}n^{{\rm eff},X}_{kp^{*}q},
\end{equation}
using Eq.~\eqref{eq:neff}, the proton mass, $m_{p}$, and $\mu$ is the mean nucleon mass factor, which for solar metallicity is 1.402 \citep{asplund2009,draine2011}.

The $E_{\rm K}$ distribution is shown in the top-right panel of Fig.~\ref{fig:histPhysicalQuantities}.
We found that the average $E_{\rm K}$ is around \meanEkHI\ and \meanEkCO\,eV/cm$^{3}$ for H{\sc i} and CO, respectively.
These estimates are comparable to the 0.22\,eV/cm$^{3}$ obtained for the turbulent kinetic energy in the local ISM, obtained with a mean nucleon density $n$\,$=$\,30\,cm$^{-3}$ and velocity $v$\,$=$\,1\,\kps, or mean values $\left<n_{\rm H}\right>$\,$=$\,1\,cm$^{-3}$ and $\sigma_v$\,$=$\,5.5\,\kps\ \citep{draine2011}.
Given that we are only sampling one component of the velocity field, the actual $E_{\rm K}$ value can be up to a factor of three higher.

The $E_{\rm K}$ standard deviations are around \postreport{1.95} and \postreport{0.53}\,eV/cm$^{3}$ for H{\sc i} and CO, respectively.
These values, roughly a factor of four above the mean, suggest significant fluctuations of this quantity within the 2.5-kpc-diameter ISM parcel considered in this paper.
The maximum $E_{\rm K}$ values are around 100 and 30\,eV/cm$^{3}$ for H{\sc i} 
In general, the highest $E_{\rm K}$ are associated with the highest $n^{\rm eff}$, suggesting that density is the dominant factor in our reconstruction of the kinetic energy distribution.

\begin{figure*}[ht]
\centerline{
\includegraphics[width=0.49\textwidth,angle=0,origin=c]{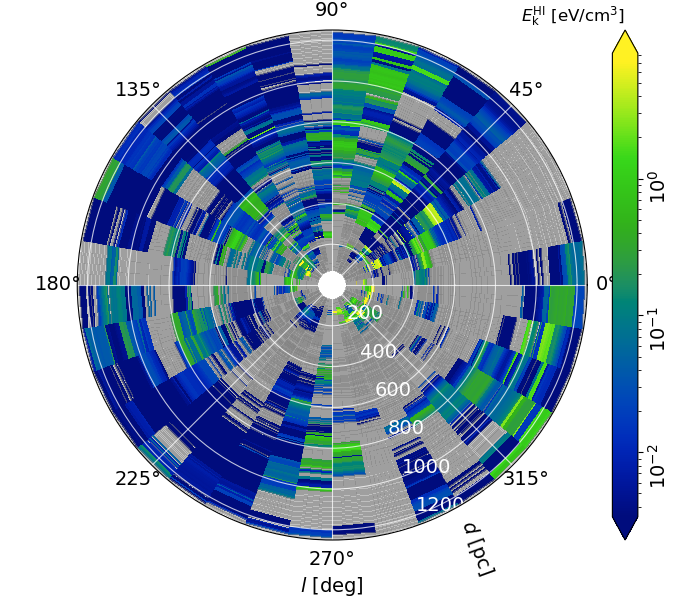}
\includegraphics[width=0.49\textwidth,angle=0,origin=c]{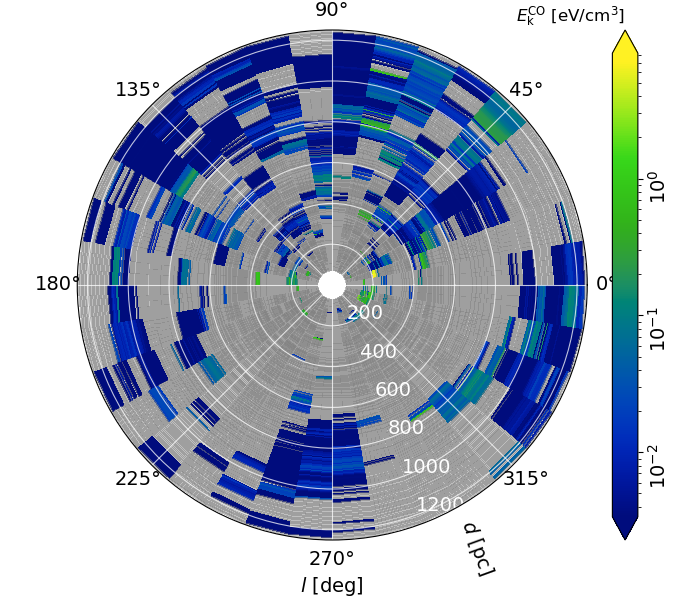}
}
\caption{Kinetic energy density ($E_{\rm k}$) derived from the velocity field reconstruction in Fig.~\ref{fig:diffv} and $n$ derived from the 3D dust extinction modeling, as detailed in Eq.~\eqref{eq:Ek}.
}
\label{fig:polarEk}
\end{figure*}

Figure~\ref{fig:polarEk} \postreport{presents} the $E_{\rm K}$ maps obtained from the H{\sc i} and CO emission.
We found no evident similarity in the $E_{\rm K}$ distribution in the two gas tracers.
However, for both tracers the highest $E_{\rm K}$ are found in the regions at $d$\,$<$\,300\,pc.

\begin{figure}[ht]
\centerline{\includegraphics[width=0.5\textwidth,angle=0,origin=c]{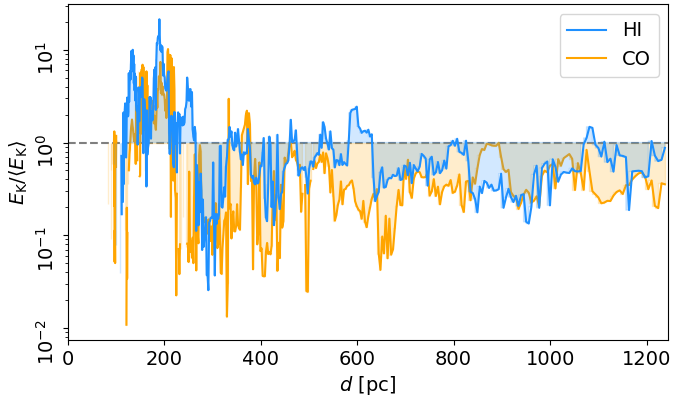}}
\centerline{\includegraphics[width=0.5\textwidth,angle=0,origin=c]{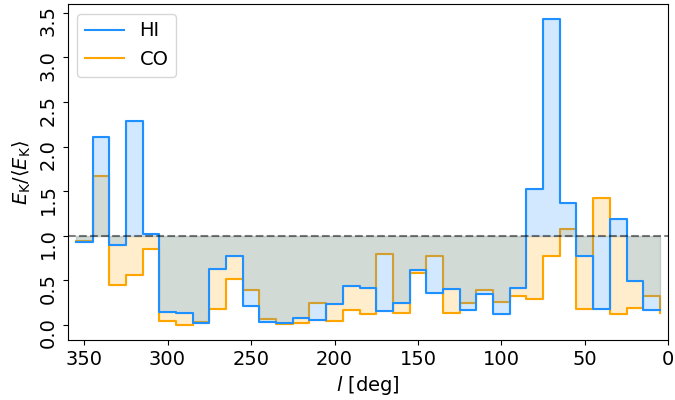}}
\caption{Kinetic energy density, $E_{\rm K}$, radial and azimuthal profiles normalized to the mean values over the studied region.  
}
\label{fig:profilesEk}
\end{figure}

The $E_{\rm K}$ radial profile, presented in the top panel of Fig.~\ref{fig:profilesEk}, shows values up to \postreport{ten} times over the mean for regions outside $d$\,$\approx$\,300\,pc and up to 20 times for regions inside that range.
The $E_{\rm K}$ peaks at $d$\,$<$\,300\,pc do not seem to result from a concentration of high $V_{\rm d}$ due to a bias in the morphological correlation toward near distances, as discussed in App.~\ref{app:physics}.
However, the angular resolution in the 3D dust reconstruction may limit the mapping of $E_{\rm K}$ enhancements to further distances. 
One potential explanation for the $E_{\rm K}$ peak emerging at $d$\,$\approx$\,100\,pc is the LSR location within the Local Bubble.
We are observing the gas motions from a location near the center of that cavity so that the peak is at a similar radius in all directions.
This is less likely to be the case for material that is further away from us.

The $E_{\rm K}$ azimuthal profile, obtained by adding the energy along the \postreport{LOS}, is presented in the bottom panel of Fig.~\ref{fig:profilesEk}, 
It shows peaks toward $l$\,$\approx$\,40 to 90\deg\ for H{\sc i} and CO.
This range covers the general directions along the axis of two large-scale structures identified in 3D dust reconstructions: the Split \citep{lallement2019} and the Radcliffe wave \citep{alves2020}, shown in Fig.~\ref{fig:polarDens}.
Additional overdensities are found toward the fourth quadrant, roughly between $l$\,$\approx$\,300 to 340\deg.

\subsection{Momentum density and momentum}

We estimated the LOS momentum density from the streaming motions for each distance channel $q$ as
\begin{equation}\label{eq:p}
(p^{X}_{\rm K})_{kq}=\rho^{{\rm eff},X}_{kp^{*}q}\left[(v^{X}_{\rm LOS})_{kp^{*}q}-(v^{R19}_{\rm LOS})_{kq}\right],
\end{equation}
where the LOS velocities $(v^{X}_{\rm LOS})_{kp^{*}q}$ and $(v^{R19}_{\rm LOS})_{kq}$ are defined in the same way as in Eq.~\eqref{eq:Ek}.
The effective density for tracer $X$, $\rho^{{\rm eff},X}_{kp^{*}q}$, is estimated using Eq.~\eqref{eq:rhoeff}.

The distribution of $(p^{X}_{\rm K})_{kq}$ is shown in the bottom-left panel of Fig.~\ref{fig:histPhysicalQuantities}.
The volume-weighted mean is around \postRVcorr{5.9} and $-$\postRVcorr{0.6}\,$\times$\,$10^{-3}$\,${\rm M}_{\odot}\,{\rm km}\,{\rm s}^{-1}$/pc$^{3}$ for H{\sc i} and CO, respectively.
However, there are large excursions from these mean trends, characterized by standard deviations of roughly \postRVcorr{0.8} and \postRVcorr{0.2}\,${\rm M}_{\odot}\,{\rm km}\,{\rm s}^{-1}$/pc$^{3}$.
This implies that the local fluctuations are much more important than the mean values.

\begin{figure*}[ht]
\centerline{
\includegraphics[width=0.49\textwidth,angle=0,origin=c]{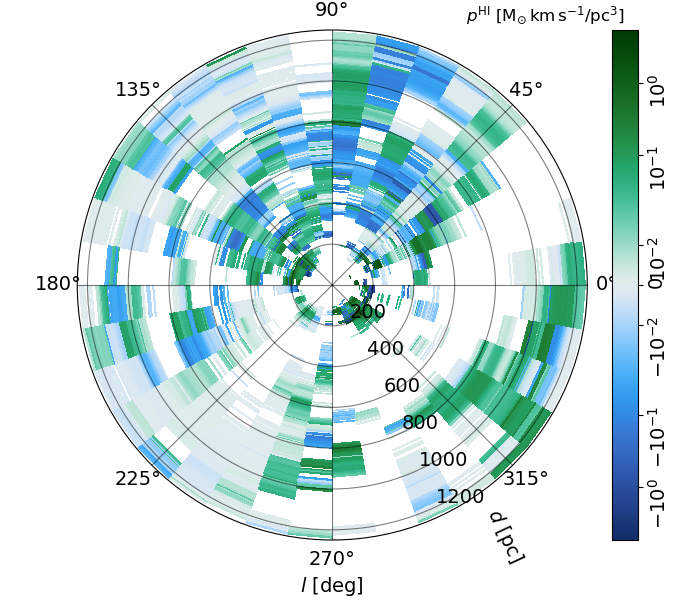}
\includegraphics[width=0.49\textwidth,angle=0,origin=c]{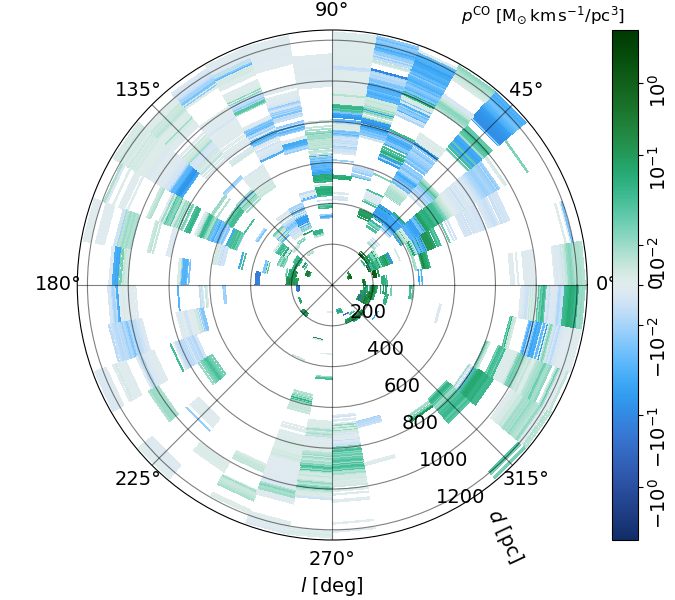}
}
\caption{Momentum density ($p$) obtained with the LOS velocity reconstructions in Fig.~\ref{fig:diffv} and Eq.~\eqref{eq:p}.}
\label{fig:p}
\end{figure*}

Figure~\ref{fig:p} shows the $p$ maps derived from the radial motions sampled by the two gas tracers.
\postreport{The peak momentum density in H{\sc i}, found toward $l$\,$\approx$\,345\deg\ at $d$\,$\approx$\,190\,pc, corresponds to a momentum around \postRVcorr{6.1}$\times$\,$10^{3}$\,${\rm M}_{\odot}\,{\rm km}\,{\rm s}^{-1}$.
This value can be interpreted by comparison with the radial momentum input from a single supernova remnant (SNR), identified as roughly between 1 and 15\,$\times$\,$10^{4}$\,${\rm M}_{\odot}\,{\rm km}\,{\rm s}^{-1}$ in numerical simulations of one SNR expansion in an inhomogeneous density field for different ambient medium's densities, metallicities \citep{martizzi2015}.
These values should be taken just as a reference; determining whether SNRs produce the reconstructed LOS momentum requires considering stochastic parameters such as initial conditions, efficiency in the conversion between thermal and kinetic energy, and the projection of momenta along the LOS.  
Our reconstruction aims to characterize the distribution of LOS momentum density and identify momentum sources in the local galactic plane.
Reconstruction of formation scenarios for individual regions is beyond the scope of this work.}

It is apparent from Fig.~\ref{fig:p} that there is a concentration of high $p$ values around $d$\,$\approx$\,200\,pc, both in H{\sc i} and CO. 
For regions beyond that heliocentric radius, the highest $|p|$ values appear toward 40\,$<$\,$l$\,$<$\,90\deg, roughly corresponding to the directions of the Split and the Radcliffe wave. 
There are also high $p$ values in H{\sc i} and CO toward $l$\,$\approx$\,270\deg\ for $800$\,$<$\,$d$\,$<$\,1100\,pc, roughly around the Vela C location indicated in Fig.~\ref{fig:polarDens}, and toward 300\,$<$\,$l$\,$<$\,360\deg\ for $d$\,$>$\,800\,pc.

\subsection{Mass flow rates}\label{sec:Mdot}

Finally, we considered the mass flow rates, $\dot{M}$, corresponding to the motion of the material in each distance channel across its radial width. 
For that purpose, we calculated 
\begin{equation}\label{eq:mdot}
\dot{M}_{kq}=\rho^{{\rm eff},X}_{kp^{*}q}\mathcal{V}_{kq}\left[(v^{X}_{\rm LOS})_{kp^{*}q}-(v^{R19}_{\rm LOS})_{kq}\right]/\Delta d_{q},
\end{equation}
where the radial velocities $(v^{X}_{\rm LOS})_{kp^{*}q}$ and $(v^{R19}_{\rm LOS})_{kq}$ are defined in the same way as in Eq.~\eqref{eq:Ek}, the effective density for tracer $X$, $\rho^{{\rm eff},X}_{kp^{*}q}$, is estimated using Eq.~\eqref{eq:rhoeff}, $\mathcal{V}_{kq}$ is the distance channel volume, and $\Delta d_{q}$ is the distance channel width.

The $\dot{M}$ distribution, shown on the bottom-right panel of Fig.~\ref{fig:histPhysicalQuantities}, is approximately centered around zero, implying that the LOS net mass flow for the studied region is close to zero.
The $\dot{M}$ standard deviations are around \postRVcorr{1.1}\,$\times$\,$10^{-3}$ and \postRVcorr{0.4}\,$\times$\,$10^{-3}$\,M$_{\odot}$/yr for H{\sc i} and CO, respectively.
The azimuthal $\dot{M}$ profiles are presented in Fig~\ref{fig:azprofileMdot}.
They correspond to the mean and the standard deviations of $\dot{M}$ along the LOS for H{\sc i} and CO.
The mean mass flows, displayed in Fig~\ref{fig:azprofileMdot}, show net matter displacements of roughly couple $10^{-3}$\,M$_{\odot}$/yr for lines of sight toward the first Galactic quadrant and 250\,$<$\,$l$\,$<$\,360\deg.
This could be interpreted as residual velocity from the Galactic rotation if not because it has the opposite sign of the expected LOS motions indicated on the right-hand-side panel of Fig.~\ref{fig:polarvLSRfromHIandCO}.
Further interpretation of these mass flows should consider that they may be a product of the LSR position and the boundaries of the 3D dust reconstruction. 

\begin{figure*}[ht]
\centerline{
\includegraphics[width=0.49\textwidth,angle=0,origin=c]{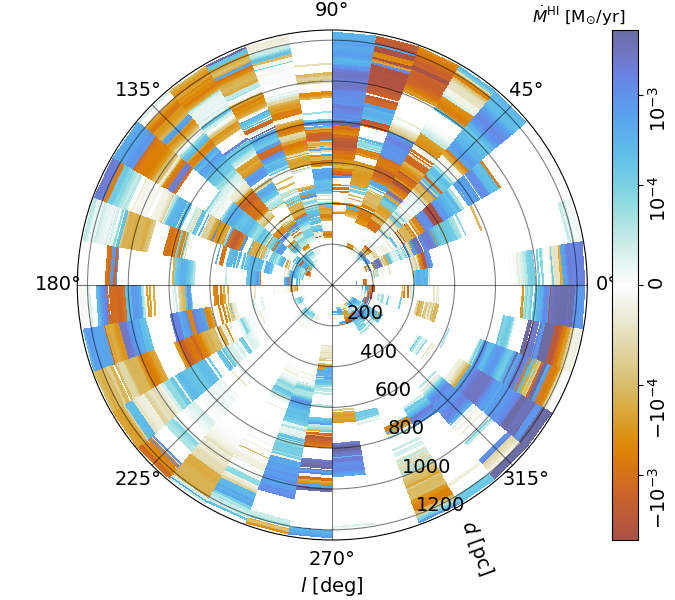}
\includegraphics[width=0.49\textwidth,angle=0,origin=c]{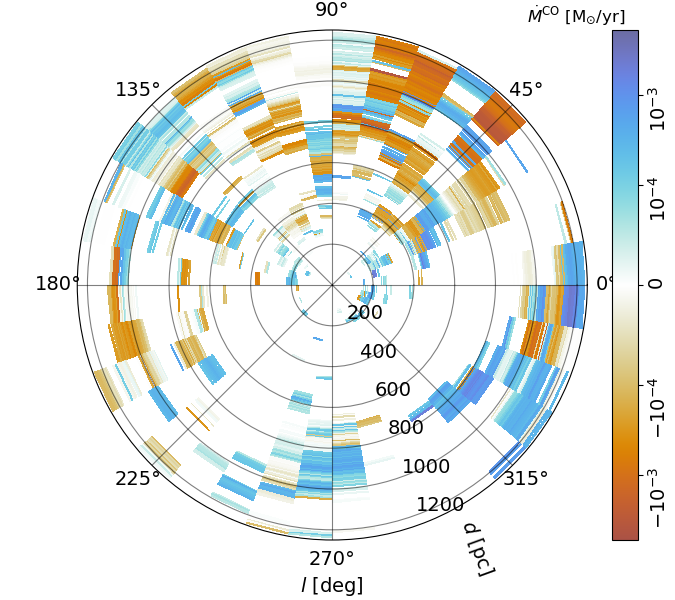}
}
\caption{Mass flow rates ($\dot{M}$) obtained with the LOS velocity reconstructions in Fig.~\ref{fig:diffv} and Eq.~\ref{eq:mdot}.}
\label{fig:Mdot}
\end{figure*}

The standard deviation of $\dot{M}$, shown in Fig~\ref{fig:azprofileMdot}, accounts for the amplitude of the mass flow fluctuations along the line of sight.
Two prominent peaks in this quantity traced by H{\sc i} appear toward $l$\,$\approx$\,40 and 80\deg, roughly the directions along the Split and the Radcliffe wave.
Additional peaks are found toward $l$\,$\approx$\,270 and 340\deg, coinciding with the locations of momentum overdensities in Fig.~\ref{fig:p}.
These peaks do not correspond to the $\dot{M}$ profiles derived from CO.
 
\begin{figure}[ht]
\centerline{\includegraphics[width=0.49\textwidth,angle=0,origin=c]{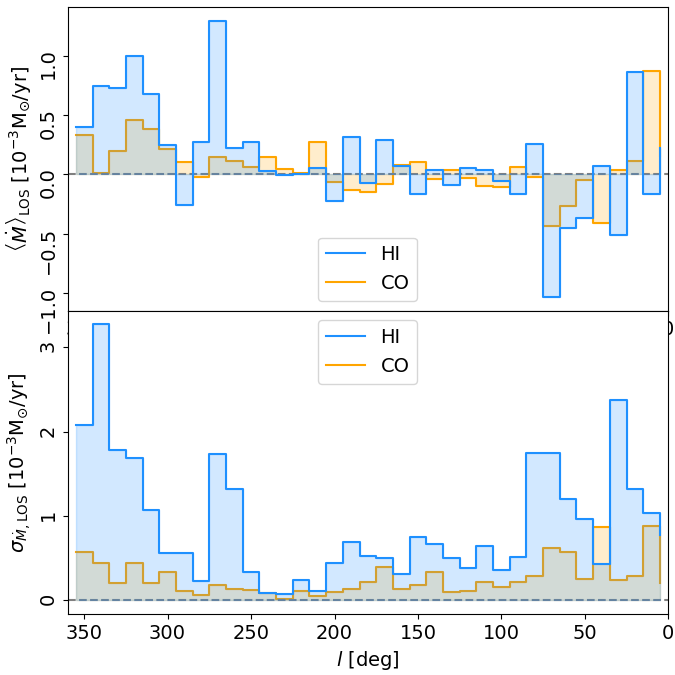}}
\caption{Azimuthal profiles of mean mass flow rates and its standard deviation.  
}
\label{fig:azprofileMdot}
\end{figure}

\section{Discussion}\label{sec:discussion}

\subsection{3D dust and line emission correlation}\label{sec:discHOGcorr}

Figure~\ref{fig:polarPRS} shows the significant similarity in the morphology of the structures traced by the 3D extinction modeling and the H{\sc i} and CO line emission.
Previous works demonstrated that dust and gas are well-mixed in the diffuse ISM, as inferred by the tight empirical linear relation between the H{\sc i} column density and dust extinction and emission \citep[see, for example,][]{bursteinANDheiles1978,boulanger1996}.
However, our results present the first quantitative estimation of the global likeness between the extinction structures and the line emission observations.

We found that roughly \postreport{67}\% and \postreport{15}\% of the 10\deg\,$\times$\,10\deg\ distance channels in Fig.~\ref{fig:polarPRS} show a significant morphological correlation, $\max(V_{\rm d})_{v}$\,$>$\,2.87, between the 3D dust and the H{\sc i} and the CO emission, respectively.
Around 9\% and 1\% of the distance channels present $\max(V_{\rm d})_{v}$\,$<$\,0.
Additional analysis of the $\max(V_{\rm d})_{v}$ distributions are presented in App.~\ref{app:physics}.

The lower correlation with CO can be explained by the lower volume filling factor of that tracer \citep{kwanANDsanders1986,kalberlaANDkerp2009}.
Clouds traced by CO are expected to be smaller and more concentrated than H{\sc i} structures.
Hence, CO has fewer independent CO gradients for the HOG morphological comparison on the same sky area and at the common angular scale set by the derivative kernel size. 

The lack of correlation in the rest of the channels can be attributed to several factors, which we separated into those related to the HOG method and those intrinsic to the dust and gas distribution in the local ISM.
Among the former factors, we can distinguish two cases.
First, there can be structures beyond 1.25\,kpc with a morphological match with the emission within the input range. 
But even if they become available with a future 3D dust reconstruction, including them would not affect the results in Fig.~\ref{fig:polarPRS} and only potentially fill a few gaps in the $lv$-diagrams presented in Fig.~\ref{fig:lvPRS}.
Second, there can be line emission structures in channels beyond the \vLOS\ input range with a morphological match with the dust within the 3D reconstruction domain. 
We considered this scenario in App.~\ref{app:vLOSrange}, where we reported the results for the input range $-120$\,$<$\,\vLOS\,$<$\,120\,\kps.
The expanded \vLOS\ range emphasizes higher streaming motion amplitudes at the inevitable price of an increased chance correlation.
However, it does not patch the low $V_{\rm d}$ channels in Fig.~\ref{fig:polarPRS}.

The low morphological correlation for certain portions of the studied volume can also be due to the intrinsic dust and gas distribution in the local ISM, which we consider in three main effects.
First, there is low extinction in local ISM volumes, such as found in the regions around 600\,$<$\,$d$\,$<$\,800\, pc toward $-45$\deg\,$<$\,$l$\,$<$\,30\deg\ and 1000\,$<$\,$d$\,$<$\,1200\, pc toward $250$\deg\,$<$\,$l$\,$<$\,290\deg, to mention a couple of examples.
Second, line emission is absent across some LOS velocity ranges toward particular lines of sight.
For example, toward the GSH~238+00+09 superbubble \citep{Heiles1998}, where there is very little CO, as illustrated in the right panel of  Fig.~\ref{fig:lvPRS}.
Identifying the exact physical conditions that produce the local ISM under-densities and a lack of emission toward particular regions is beyond the scope of this work.
However, the maps in Fig.~\ref{fig:polarPRS} and Fig.~\ref{fig:lvPRS} provide a global context for identifying the conditions toward specific regions and their implications for the ISM dynamics at smaller scales. 

There is a third global factor that reduces the correlation between the 3D dust and the line emission.
Observations of other ISM tracers, such as gamma-rays and the [C{\sc ii}] and OH line emission, reveal that not all of the ISM material is traced by H{\sc i} and CO emission \citep[see, for example,][]{grenier2005,langer2014,pineda2013,busch2021}.
Including this material in the distance-velocity correlation is not straightforward.
In the case of [C{\sc ii}], there is no map with enough angular resolution and coverage to obtain significant spatial correlations using HOG.
In the case of OH, the emission distribution involves the complex physical conditions that lead to the emission from that radical, which calls for a separate dedicated study \citep[see, for example,][]{rugel2018,dawson2022}.
We acknowledge that those and other line tracers should be accounted for in future reconstructions of the local ISM, adding richness to the results we present in this paper.
Yet, our analysis has shown we have enough distance channels where the correlation between the dust distribution and the line emission is significant enough to identify prevalent LOS velocities for different portions of the local ISM, which we discuss in the following section.

\subsection{Local ISM motions}

We used the morphological correlation across distances and LOS velocities to assign a \vLOS\ to the distance channels in each of the 10\deg\,$\times$\,10\deg\ regions in which we divided the studied volume.
This assignment simplifies the velocity field by identifying a single value to ISM parcels with volumes between 10$^2$ to 10$^5$\,pc$^{3}$\postreport{, corresponding to distance channels at the inner and outer edges of the 3D dust model}.
Thus, our velocity reconstruction represents the local ISM dynamics on spatial scales of tens of parsecs in the plane of the sky and around a parsec in the LOS direction.

The velocity assignment is unambiguous in most of the 10\deg\,$\times$\,10\deg\ regions, as illustrated in Fig.~\ref{fig:Vplanes} and Fig.~\ref{fig:VplanesROI1}.
There are distance ranges for which the assigned \vLOS\ shows an oscillatory pattern.
We do not interpret these as actual oscillations since they can arise from the prevalence of the correlation measured by $V_{\rm d}$ alternating between different structures in the same sky area. 
We do not discard the possibility of oscillations, such as the ones recently identified in works such as \cite{henshaw2020}. Still, its identification requires a dedicated analysis beyond this work's scope. 

The maps of reconstructed LOS velocities shown in Fig.~\ref{fig:polarvLSRfromHIandCO} indicate that the local LOS velocity field carries the imprint of the Galactic rotation pattern in both tracers, as inferred from the comparison with the expected \vLOS\ in the \cite{reid2019} Galactic rotation model, shown in Fig.~\ref{fig:polarReid2019vLSR}.
The large-scale pattern of H{\sc i} and CO emission is deeply influenced by Galactic rotation, as shown in figure~B.2 of \cite{hi4pi2016} and figure~3 of \cite{dame2001}.
However, the peculiar velocities of MCs within 3\,kpc cannot be explained purely by Galactic rotation \citep[see, for example,][]{starkANDbrand1989}.

The radial velocity of stars from the Gaia DR2 catalog within 1\,kpc of the Galactic mid-plane and roughly 4\,kpc from the Sun, reported by \cite{katz2019}, show a global pattern consistent with the LOS-projected differential rotation of the stars around the Galactic center, as observed from the Sun.
The HOG method enabled mapping this rotational pattern in the gas in the Solar neighborhood by anchoring the LOS velocity to a distance using the morphological similarity between the dust and gas tracers.
Figure~\ref{fig:polarvLSRfromHIandCO} strongly suggests that despite its collisional nature and being exposed to the influence of local conditions, such as the energy and momentum input from stellar feedback, the gas within 1.25\,kpc follows the pattern of Galactic rotation.

We subtracted the \cite{reid2019} model of Galactic rotation from the LOS velocity pattern in Figure~\ref{fig:polarvLSRfromHIandCO} to characterize the deviations from pure Galactic rotation in the local ISM \citep{burton1971}.
We refer to the resulting LOS motions as streaming motions to avoid attributing them to fluctuations in the large-scale turbulent velocity field or local fluctuations introduced, for example, by SN explosions.
Separating the mean and the turbulent velocity fields is not straightforward, even when all velocity field components are available.
Thus, our characterization of the streaming motions reported in Figure~\ref{fig:diffv} aims to relate them to the known structure of the local ISM rather than a generic separation that may be hindered by the limitations of our reconstruction.

Figure~\ref{fig:histdiffv} shows that the distribution of LOS streaming motions roughly centered around zero, within the uncertainties imposed by the velocity channel widths in the input data.
The standard deviations of the rotation-subtracted LOS velocities are \sigmavHI\ and \sigmavCO\,\kps\ for H{\sc i} and CO, respectively.
Previous studies have identified similar deviations from Galactic rotation. 
\cite{clemens1985} used the Massachusetts-Stony Brook Galactic plane CO survey to identify streaming motions with an amplitude of around 5\,\kps\ on length scales larger than $0.22R_{\rm 0}$, where $R_{\rm 0}$ is the radius of the solar orbit around the Galactic center, so that $0.22R_{\rm 0}$ corresponds to a scale of between 1.87 and 2.2\,kpc for the two values considered in that work. \cite{brandANDblitz1993} reported streaming motions with mean values of 12\,\kps\ in their modeling of the spectro-photometric distances for a sample H{\sc ii} regions and the LOS velocities for their associated MCs. 
The authors found a quadrupolar LOS velocity pattern, qualitatively comparable to those in Fig.~\ref{fig:polarvLSRfromHIandCO}, and report that the molecular gas is streaming past the LSR at about 3.8\,\kps, which are within a factor two we obtained for CO in the Solar vicinity.
Our reported velocity dispersions in H{\sc i} are also comparable to those found in The H{\sc i} Nearby Galaxy Survey \citep[THINGS;][]{walter2008,schmidt2016}. 
However, the physical conditions in their sample are not in general equivalent to those in the solar vicinity. 

The difference in the amplitude of the streaming motions estimated with the HOG method for H{\sc i} and CO can be related to the properties of the global turbulent motion in the media dominated by each tracer.
Dense, CO-dominated clouds form in regions where the velocity field is convergent, corresponding to regions where shocks dissipate turbulent kinetic energy in a supersonically turbulent flow \citep[see, for example,][]{koyama2000,audit2005,heitsch2006}. 
Therefore, it is natural that the velocity dispersions in those regions should be smaller.
Moreover, the observed linear relation between size and velocity dispersion implies that smaller regions sampled by CO have lower dispersion than the more extended H{\sc i} clouds \citep{larson1981,heyer2009}.

\subsection{Comparison to previous works}

The most direct precedent of our global reconstruction of \vLOS\ as a function of distance is that presented in \cite{tchernyshyov2017}, from here on \citetalias{tchernyshyov2017}.
Other previous reconstructions have focused on individual spiral arms, using ISM flow models to invert emission line observations, \citep[see, for example,][]{shane1972,foster2006} or individual regions \citep[see, for example,][]{traficante2014}.
\cite{reid2016} used the Very Long Baseline Interferometry (VLBI) observations of maser sources associated with young massive stars \citep[see][and references therein]{reid2009,honma2012} in combination with what the authors considered spiral arm signatures in CO and H{\sc i} surveys to produce distance probability density functions as a function of \vLOS.
None of these approaches uses the information available in extinction-based 3D dust density reconstructions.

\begin{figure}[ht]
\centerline{\includegraphics[width=0.49\textwidth,angle=0,origin=c]{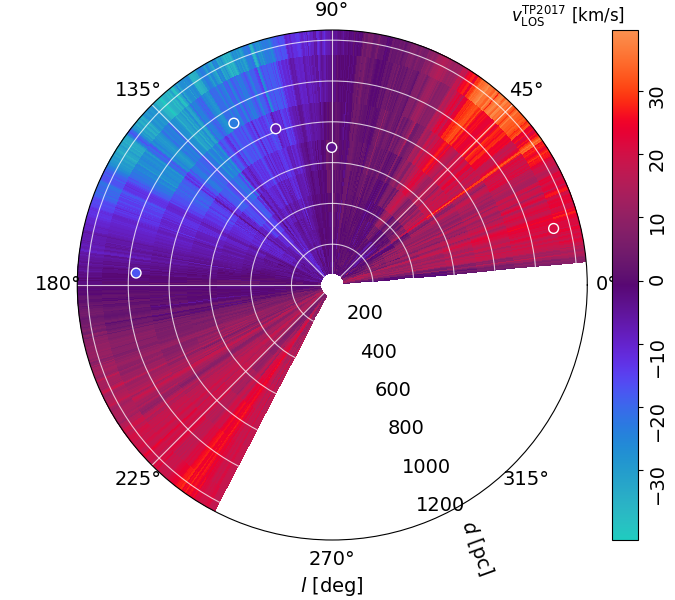}}
\caption{Line-of-sight velocity reconstruction from \cite{tchernyshyov2017} for the distance range considered in this paper.}
\label{fig:polarTP2017vlos}
\end{figure}

\citetalias{tchernyshyov2017} used as input maps of reddening as a function of $l$, $b$, and $d$ (PPP) and maps of H{\sc i} and CO line emission as a function of $l$, $b$, and $v_{\rm}$ (PPV) to produce the ISM distribution as a function of $l$, $b$, $d$, and $v_{\rm}$ (PPDV).
Their method is based on minimizing the difference between the observed PPV cube, from H{\sc i} or CO line emission, and a model PPV cube, obtained by modeling the density distribution in PPDV as the product of a density function and a Gaussian distribution of $v_{\rm}$ centered on the LOS velocity from a flat Galactic rotation curve, and projecting into PPV by summing across distance channels.
The product of this minimization is a map of the LOS velocity residual, which is regularized by adding a term penalizing significant differences between the LOS velocities of contiguous pixels.

The HOG-based kinetic tomography differs from that in \citetalias{tchernyshyov2017} in three main aspects.
First, the \citetalias{tchernyshyov2017} input is the radial difference of the 3D interstellar dust reddening presented in \cite{green2015}.
These maps cover the distances between 63\,pc and 63\,kpc in steps of half a distance modulus (roughly 12.6\,pc), but are limited to the Galactic longitude range 5\deg\,$<$\,$l$\,$<$\,243\deg. 
We employed the 3D dust reconstruction from \cite{edenhofer2024}, which is limited to distances between 69\,pc and 1.25\,kpc but uses an improved source selection based on observations not available at the time of \cite{green2015}.

Second, the method in \citetalias{tchernyshyov2017} can be considered as an inversion of the usual kinematic distance method, assuming rotational motions around the Galactic center as a prior. 
The HOG does not assume a prior on the velocity distribution.
This means that the two methods would converge where \vLOS\ is close to the expectation from rotation around the Galactic center, within the 5\,\kps\ LOS velocity dispersion assumed for the regularization in \citetalias{tchernyshyov2017}, but diverge where it is not.
Thus, the HOG-based kinetic tomography favors the identification of unusual velocity offsets, assuming that the plane-of-the-sky morphological matching represents the relation between the dust and gas distributions.

Third, \citetalias{tchernyshyov2017} is a pixel-by-pixel analysis where additional information is introduced in the model by assuming similar \vLOS\ for adjacent pixels.
The HOG method uses the gradient orientation, a non-local quantity calculated over the angular scale set by the Gaussian derivative kernel size.
Then, it accumulates the statistics over the 10\deg\,$\times$\,10\deg\ region to obtain the relation between $d$ and \vLOS.
Thus, \citetalias{tchernyshyov2017} obtains a higher-resolution reconstruction of PPDV than the HOG method at the expense of having to assume priors for the LOS velocity and spatial and velocity coherence. 
The higher angular resolution in \citetalias{tchernyshyov2017} favors the comparison with the VLBI observations. 
Still, it can be biased toward identifying the most massive object along a \postreport{LOS} and associating it with the most massive object for the same position in the line emission data. 
This would correctly assign a distance and velocity to a dense object like a giant molecular cloud (GMC), but produce a less accurate solution for \postreport{less-concentrated neutral gas, which is more exposed to the acceleration by SN feedback and deviations from pure Galactic rotation \citep{iffrigANDhennebelle2015,walch2015}.}

Figure~\ref{fig:polarTP2017vlos} \postreport{presents} the LOS velocity reconstruction obtained in \citetalias{tchernyshyov2017} for the region within 1250\,pc from the Sun. 
From visual comparison with Fig.~\ref{fig:polarvLSRfromHIandCO}, it is evident that \citetalias{tchernyshyov2017} also recovers the quadrupolar \vLOS\ pattern resulting in circular motions around the Galactic center. 
However, the amplitude of the reconstructed \vLOS\ in \citetalias{tchernyshyov2017} is much higher than the values we report in Fig.~\ref{fig:polarvLSRfromHIandCO}, reaching maxima and minima of up to $\pm$40\,\kps.
These large residual velocities are beyond the range of our HOG-based reconstruction, given our input range $-25$\,$<$\,\vLOS$<$,25\,\kps.
They are recovered using the broader input \vLOS\ range considered in App.~\ref{app:vLOSrange}, but at the expense of a higher chance correlation that biases the HOG results toward higher \vLOS. 

\begin{figure}[ht]
\centerline{\includegraphics[width=0.49\textwidth,angle=0,origin=c]{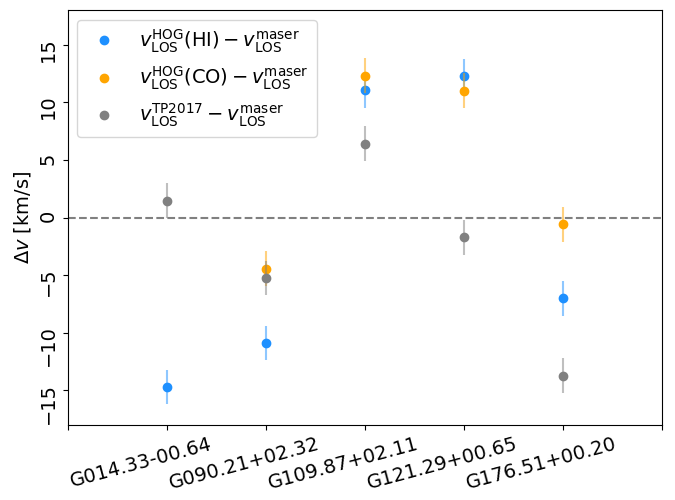}}
\caption{Line-of-sight velocity reconstruction from \cite{tchernyshyov2017} for the distance range considered in this paper.}
\label{fig:maserComparison}
\end{figure}

There are only five VLBI maser observations within the ISM volume considered in this paper.
Figure~\ref{fig:maserComparison} shows the difference between the \vLOS\ reconstructions presented in Fig.~\ref{fig:polarvLSRfromHIandCO} and \citetalias{tchernyshyov2017} and the observations for those five sources.
For our reconstruction, we considered the difference between the nearest distance channel with an assigned \vLOS\ and the position of the maser.
The \vLOS\ differences between the masers and the \citetalias{tchernyshyov2017} reconstruction are positive and negative and characterized by a standard deviation of roughly 6.7\,\kps. 
The most significant discrepancy is for the source G176.51+00.20, where the difference is around $-$14\,\kps. 
The \vLOS\ differences between the masers and our reconstructions are also in either direction.
They are characterized by standard deviations around \postreport{13.5} and \postreport{11.3}\,\kps for H{\sc i} and CO, respectively.
The most considerable difference is \postreport{for source G14.33-00.64, where the lack of morphological correlation between CO and 3D dust limits the \vLOS\ reconstruction}.

The HOG-based \vLOS\ reconstruction requires the averaging over large sky areas. 
Thus, its results are expected to show less agreement with the masers results than the \citetalias{tchernyshyov2017} reconstruction, which has an angular resolution much closer to the VLBI observations.
Yet, Fig.~\ref{fig:maserComparison} shows that the discrepancy between the maser and HOG \vLOS\ for CO is less than \postreport{a few} \kps\ for most of the sources, and in the case of G090.21+02.11 it is closer than \citetalias{tchernyshyov2017}.
The fact that the HOG \vLOS\ for H{\sc i} is systematically further from the maser values reflects that it samples more diffuse gas, which is not necessarily anchored to the denser gas producing the maser emission.
Given the low number of maser sources to anchor both reconstructions, the comparison with the VLBI results is limited.
Thus, we compare our results to those in \citetalias{tchernyshyov2017}.
We reserve further comparison between the HOG and the masers for Appendix \ref{app:masers}.

The differences in the HOG and \citetalias{tchernyshyov2017} \vLOS\ reconstruction methods provide several potential causes for the discrepancies in the results reported in Fig.~\ref{fig:polarvLSRfromHIandCO} and Fig.~\ref{fig:polarTP2017vlos}.
It is plausible that the high \vLOS\ values in \citetalias{tchernyshyov2017} for the range $d$\,$<$\,1250\,kpc are due to features not resolved in the \cite{edenhofer2024} 3D extinction models or washed away by the spatial averaging required to compute the HOG.
\cite{peek2022} presented a detailed study of a collection of CO clouds in the range 135\deg\,$<$\,$l$\,$<$\,160\deg\ using the 3D dust reconstructions from \cite{green2019} and identified a few objects within $d$\,$<$\,1250\,kpc with \vLOS\,$<$\,$-30$\,\kps; namely the clouds identified with numbers 3397, 3430, 4338, 5935, 6078, and 6142 in the \cite{miville-deschenes2017} catalog.
The angular size of these clouds ranges between 0\pdeg23 and 0\pdeg67, so it is unlikely that they are the dominant feature of the average across each 10\deg\,$\times$\,10\deg\ distance channels in which the HOG is computed.
In that case, the discrepancy is introduced by the different scales sampled by each method.

The HOG-based kinetic tomography employs the information contained in the plane-of-the-sky distribution of the tracers to distinguish the motions in H{\sc i} and CO.
It provides additional information to break the degeneracies in the inversion problem tackled with the \citetalias{tchernyshyov2017}.
Yet, it is limited by the resolution of the 3D dust reconstruction and the spatial averaging.
Thus, an optimal approach is the combination of both methods, using the pixel-by-pixel inversion in \citetalias{tchernyshyov2017} and the HOG-based spatial correlation to regularize the solutions and solve the LOS\ reconstruction ambiguities.
This merging of techniques is not straightforward and falls beyond the scope of this paper, but should bridge between our results and those in \citetalias{tchernyshyov2017}.

\begin{figure}[ht]
\centerline{\includegraphics[width=0.50\textwidth,angle=0,origin=c]{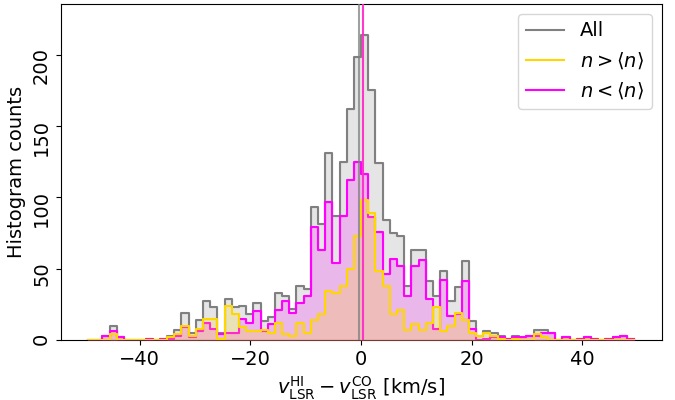}}
\caption{Histogram of the difference between LOS velocities across distance channels reconstructed using H{\sc i} and CO emission.}
\label{fig:histvHIminusCO}
\end{figure}


One of the advantages of HOG-based kinetic tomography is the separate treatment of H{\sc i} and CO as tracers of different ISM dynamics.
Although the global LOS velocity reconstructions reported in Fig.~\ref{fig:polarvLSRfromHIandCO} are qualitatively similar, the streaming motions obtained are different, as illustrated in Fig.~\ref{fig:diffv}.
The differences between the streaming motions from H{\sc i} and CO, reported in Fig.~\ref{fig:histvHIminusCO}, are roughly around zero, indicating that, on average, the two tracers are co-moving in the studied volume. 
The dispersion of the LOS velocity difference distribution is around 15\,\kps, suggesting that there are significant local velocity offsets in the material sampled by each tracer.

Several physical conditions can lead to a velocity offset between H{\sc i} and CO.
Large-scale shocks, from spiral density waves or SNe, would couple more efficiently to the diffuse material sampled by H{\sc i}, accelerating it with respect to the denser material, traced by CO \citep[see, for example,][]{gatto2015,iffrigANDhennebelle2015,izquierdo2021}.

\begin{figure}[ht]
\centerline{\includegraphics[width=0.5\textwidth,angle=0,origin=c]{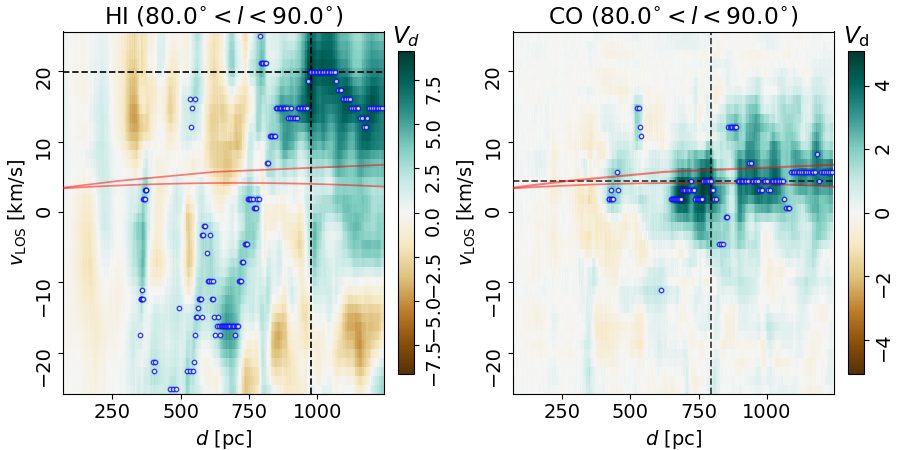}}
\caption{Same as Fig.~\ref{fig:Vplanes}, but for the tile 80\,$<$\,$l$\,$<$\,90\deg.}
\label{fig:VplanesROI1}
\end{figure}

There are some tiles where the H{\sc i} shows diverging velocities that are not matched by the CO motions.
A significant example of this behavior is found at $d$\,$\approx$\,800\,pc toward the region 80\,$<$\,$l$\,$<$\,90\deg.
In this particular example, which is located right at the axis of the Radcliffe wave, the distance-LOS velocity mapping produced with the HOG, shown in Fig.~\ref{fig:VplanesROI1}, indicates that the structure traced by CO closely follows Galactic rotation.
In contrast, the structures traced by H{\sc i} appear offset from Galactic rotation by roughly 10\,\kps\ in either direction.
The center of these diverging motions appears to be close to the location of the North America nebula \citep[NGC 7000; 81\pdeg7\,$<$\,$l$\,$<$\,86\pdeg6, $-2$\pdeg2\,$<$\,$b$\,$<$\,0\pdeg1, $d$\,$\approx$\,731 to 878\,pc;][]{zucker2020}.
Previous observations have reported expanding motions in the CO MCs around H{\sc ii} region W80 \citep{bally1980}.
Clustering of the young stellar objects (YSOs) in this region reveals six spatio-kinematic groups, three expanding with velocity gradient between 0.3 and 0.5\,\kps\,pc$^{-1}$. 
However, on a global scale, the relative motions of the groups do not appear either divergent or convergent \citep{kuhn2020}.
Determining whether or not this and other star-forming regions and their stellar content are responsible for the velocity offset in H{\sc i} is further considered in the context of the energy and momentum estimates discussed in the following section.

\subsection{Energy and momentum in the Solar neighborhood}

\subsubsection{Kinetic energy density}

We used our streaming motion estimates to calculate the kinetic energy densities, $E_{\rm k}$, across tiles in the studied volume.
These correspond to just one component of the velocity field.
However, until new revolutionary techniques or observational methods are invented, this is the only component of the gas velocity we can measure. 

The mean $E_{\rm k}$ values of \meanEkHI\ and \meanEkCO\,eV/cm$^{3}$ obtained for H{\sc i} and CO are comparable to the 0.22\,eV/cm$^{3}$ obtained by assuming standard conditions for the Solar neighborhood \citep{draine2011}.
Assuming equipartition between the three components of the velocity vector, these estimates should be roughly within a factor of three from the total values.

It is a known fact that previous generic estimates of the kinetic energy density are close to the thermal energy density, estimated from the distribution of thermal pressure throughout the ISM, 0.49\,eV/cm$^{3}$ \citep{jenkins2011}.
They are also within a factor of a few from the cosmic ray energy density, 1.39\,eV/cm$^{3}$ \citep{webber1983}, and the magnetic energy density derived from the H{\sc i} Zeeman effect observations, 0.89\,eV/cm$^{3}$ \citep{heilesANDcrutcher2005}.
Finally, they are also close to the energy densities in starlight and far-infrared radiation from dust, 0.54 and 0.31 \citep[see, chapter 12 in][]{draine2011}.
Our analysis provides a new and independent estimate of the kinetic energy density, demonstrating that this near equipartition is also found in the local ISM.

Still, a novelty introduced by our analysis is the mapping of the distribution of this quantity, which shows maxima a factor of ten and more above the mean value for particular regions, as illustrated in Fig.~\ref{fig:polarEk}.
The distribution of $E_{\rm k}$ across heliocentric distances, presented in the top panel of Fig.~\ref{fig:profilesEk}, \postreport{displays} overdensities concentrated at lower distances.
This is most likely produced by the sampling of different angular scales across distances.
That is, nearby distance channels have higher spatial resolutions, which resolve dense structures on spatial scales that are not sampled in the most distance channels.
This results in a concentration of higher densities at a lower heliocentric radius that is reflected in our $E_{\rm k}$ estimates.

The potential distance bias is marginalized in the $E_{\rm k}$ azimuthal profiles, presented in the bottom panel of Fig.~\ref{fig:profilesEk}.
There it is evident that $E_{\rm k}$ overdensities of a factor of a few are found toward 40\deg\,$<$\,$l$\,$<$\,90\deg\ in H{\sc i} and CO, around $l$\,$\approx$\,170\deg\ and 270\deg\ in CO, and 320\deg\,$<$\,$l$\,$<$\,340\deg\ in H{\sc i}.
The first $l$ range coincides with the position of the Radcliffe wave and the Split, as illustrated in Fig.~\ref{fig:polarDens}.

\postreport{
The Radcliffe wave comprises the majority of nearby star-forming regions and contains about three million solar masses of gas \citep{alves2020}.
Its existence was not among the premises of our study, but it is easily distinguishable in the H{\sc i} $E_{\rm k}$ distribution presented in Fig.~\ref{fig:polarEk}.
\cite{konietzka2024} used the motion of young stellar clusters to identify oscillatory motions with an amplitude of 14\,\kps\ along the Radcliffe wave.
The energy overdensities we identified toward this structure are likely related to the oscillatory motions.
\cite{konietzka2024} argued that ``a superposition of feedback-driven structures could reproduce the observed wavelength and amplitude of the wave''. 
The expanding motions in H{\sc i} identified in Fig.~\ref{fig:VplanesROI1} potentially be remnants of such an input from stellar feedback, but further comparison calls for a study of the Radcliffe wave in it full extent and not just in the region within $|b|$\,$\leq$\,5\deg.
}

The $l$\,$\approx$\,270\deg\ \postreport{energy-overdensity} corresponds to the location of the Vela C MC, also shown in Fig.~\ref{fig:polarDens}, that is representative of the Vela GMC complexes, which host intermediate-mass star formation, up to early B-type, late O-type stars \citep{murphy1991,netterfield2009,massi2019,hottier2021}.
The region around 320\deg\,$<$\,$l$\,$<$\,340\deg\ covers the position of the Ara OB1 association ($l$\,$\approx$\,336\pdeg7; $b$\,$\approx$\,$-$1\pdeg6), a stellar group extended across 1\deg\ on the plane of the sky, where the velocity distribution of H{\sc i} has been suggested to correspond to an expanding structure \citep{arnal1987}.
It is unlikely that these individual star-forming regions would produce the energy overdensities we find in H{\sc i}. 
However, they may be guideposts to larger-scale energy input into the ISM.

Mechanical and radiative energy input from high-mass stars is one of the primary sources of kinetic energy and ISM turbulence \citep[see, for example,][]{hennebelleANDfalgarone2012,klessenANDglover2016}.
The largest contribution from high-mass stars to interstellar turbulence most likely comes from SN explosions, although recent studies indicate that the mechanical energy injected by the supernovae alone is not sufficient to explain the kinetic energy of the ionized gas in nearby galaxies \citep{egorov2023}.
Numerical simulations of a single supernova remnant (SNR) expanding in an inhomogeneous density field for different ambient medium's densities, metallicities, and structures indicate thermal energy inputs of up to $10^{51}$\,erg \citep[][]{martizzi2015,iffrigANDhennebelle2015}.
The efficiency of energy transfer from supernova blast waves to the interstellar gas depends on many factors, but it has been estimated to be around 10\% \citep{normanANDFerrara1996,thorton1998}.

\postreport{The total amount of kinetic energy in the studied volume is around 1.7\,$\times$\,$10^{52}$\,erg.
The total amount of kinetic energy excess relative to a background level, calculated by subtracting the mean from each reconstruction element, is 
1.2\,$\times$\,$10^{52}$\,erg.}
Given the energy input and energy transfer efficiency described above, it would take roughly \postreport{120} SNe to introduce that energy into \postreport{that} volume.
Recently, \cite{swiggum2024} used {\it Gaia} astrometry and other large-scale spectroscopic surveys to show that around 57\% of the young \postreport{stellar clusters} within one kiloparsec from the Sun arise from three distinct spatial volumes and these families produced more than 200 SNe in the past 30\,Myr, including the sources of the Local Bubble and GSH~238+00+09.
This number of SNe is roughly consistent with the 15\,Myr$^{-1}$ SNe for gas at surface densities compared to the local ISM, around 10\,M$_{\odot}$/pc$^{2}$ \citep{walch2015}.
Given the \sigmavHI\,\kps\ velocity dispersion, turbulent mixing would have spread this input energy covering scales of roughly 400\,pc in \postreport{36}\,Myr.
Thus, SN feedback \postreport{is} a plausible candidate for the source of the energy overdensities reported in Fig.~\ref{fig:polarEk}, but it does not entirely explain its distribution.

\begin{figure*}[ht]
\centerline{
\includegraphics[width=0.49\textwidth,angle=0,origin=c]{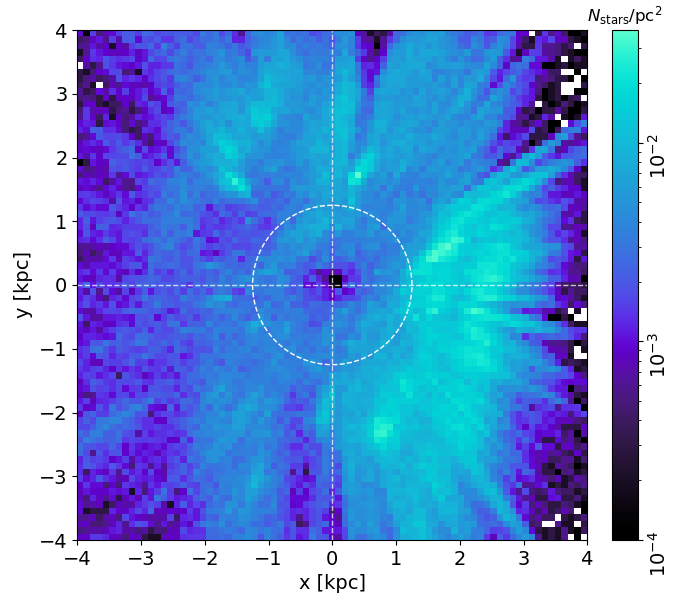}
\includegraphics[width=0.49\textwidth,angle=0,origin=c]{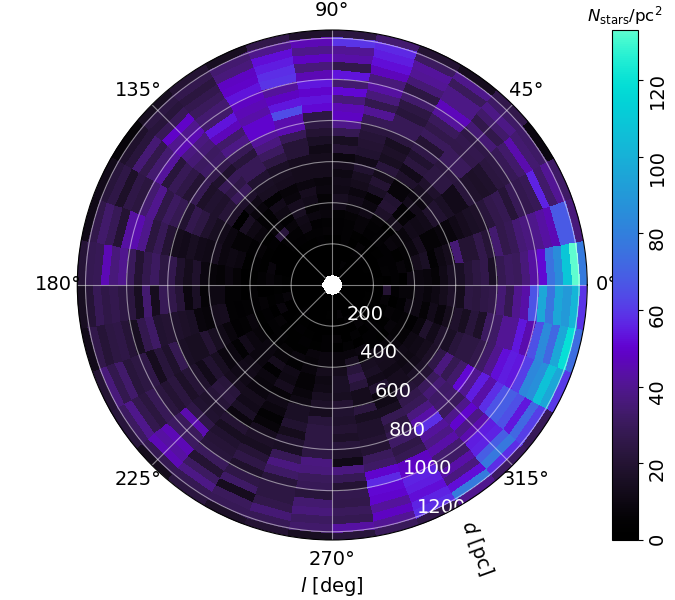}
}
\caption{Distribution of O-, B-, and A-type stars in the \cite{zari2021} catalog limited to the range $b$\,$<$\,5\deg\ presented in Cartesian coordinates ({\it left}) and projected into the polar grid introduced in Fig.~\ref{fig:polarDens} ({\it right}).}
\label{fig:OBAstars}
\end{figure*}

In the absence of a direct tracer to locate the SN energy input in the local ISM, our best proxy for the potential location of SNe is the distribution of the current population of high-mass stars.
One of the state-of-the-art catalogs of O-, B-, and A-type (OBA) stars in the Solar vicinity is presented in \cite{zari2021}.
Although this catalog is not complete, due, for example, to the saturation of {\it Gaia} for bright sources, it provides a global view of the potential stellar energy input into the ISM.
Figure~\ref{fig:OBAstars} shows the distribution of OBA stars projected into the same grid used for the streaming motion reconstructions.
It is evident that the energy overdensities found toward 320\deg\,$<$\,$l$\,$<$\,340\deg\ in Fig.\ref{fig:polarEk} are in the vicinity of an OBA-type stellar overdensity, making plausible the argument of SNe or other types of stellar feedback as the source of the kinetic energy accumulation.
However, this argument does not hold in the region toward 40\deg\,$<$\,$l$\,$<$\,90\deg.

If the kinetic energy overdensities reported in Fig.~\ref{fig:polarEk} do not correspond to overdensities in the population of high-mass stars, we can consider three alternative scenarios.
The first is that the stellar population producing the SN feedback is not currently located at the position of the energy overdensities.
This would mean we see the imprint of a past SNe not traced by the current distribution of high-mass stars.
However, it is improbable that the cluster families identified in \cite{swiggum2024} originate from the Radcliffe Wave. 
Most likely, they primarily formed in compact regions 30 to 60\,Myr ago, before what we call the Radcliffe Wave existed. 
There are signs that the cluster families were responsible for forming the Radcliffe Wave and the Split by driving superbubbles from the Alpha Per and Cr135 families, as illustrated in figure~2 of \cite{swiggum2024}.

A second scenario is that the extinction within the Split and the Radcliffe Wave hides the stellar population from which the SN feedback originated.
Redding by dust extinction complicates the identification of OBA stars, making catalogs incomplete toward high-extinction regions.
\cite{zari2021} identified in their figure~14 that the dust density in the range 40\deg\,$<$\,$l$\,$<$\,90\deg\ is higher than the stellar density, a fact that they interpreted as the result of the decoupling of gas and stars after the passage of the spiral arm. 
Testing whether this is the case is problematic because it would involve measuring the plane-of-the-sky gas motions with the same accuracy as we do for stellar motions, which is beyond the reach of current observational techniques.
Yet, it is unlikely that the dust in the Split and the Radcliffe Wave are hiding a prominent overdensity of high-mass stars such as that shown in Fig.~\ref{fig:OBAstars} for distances between 900 and 1250\,kpc toward the fourth Galactic quadrant.

A third scenario is one in which the energy overdensities are not exclusively produced by SNe but by large-scale driving \citep[see, for example,][]{elmegreen2024}. 
Large-scale energy sources such as spiral arms, magneto-rotational instability (MRI), gravity-driven turbulence, tidal interactions, and inflows can drive turbulence on scales larger than one kiloparsec \citep[see, for example,][]{wada2008,bournaud2010,colman2022}.
Although the properties of turbulence in the inertial range do not depend on the injection mechanism, large-scale turbulence implies that an additional factor drives the distribution of energy beyond the domain of our reconstruction.

\cite{tchernyshyov2018} presents a kinematic-tomography-based study of Galactic-scale numerical simulations with a fixed background potential, corresponding to the spiral density wave model \citep{linANDshu1964}, and a dynamically evolving background potential, corresponding to a transient and recurrent spiral structure model \citep{sellwoodANDcarlberg1984}, without feedback.
The results of the comparison with their Milky Way kinetic tomography favor the dynamic spiral structure simulation but show that both spiral arm models produce streaming motion structures that are coherent over a few kiloparsecs \citep[see figure~5 in][]{tchernyshyov2018}. 
Whether or not these structures are compatible with the Split or the Radcliffe way is a matter of discussion. 
Still, they show that large-scale motions can produce spatially coherent energy overdensities without the necessity of supernova feedback.
Yet, as our momentum estimates further indicate, separating a purely SN-driven scenario from a large-scale forcing scenario is challenging.

\subsubsection{Momentum density}

We reported our estimates of radial momentum density, $p$, in Fig.~\ref{fig:p}.
As in the streaming motions maps, $p^{\rm HI}$ reveals multiple reversals along the line of sight, the most prominent of which is found around the location of the North America MC, along the Radcliffe wave. 
However, the are significant concentration of relatively high $|p^{\rm HI}|$ for $d$\,$<$\,700\,pc toward 290\deg\,$<$\,$l$\,$<$\,360\deg and 700\deg\,$<$\,$d$\,$<$\,1100\,pc toward $l$\,$\approx$\,270\deg.
The accumulations of relatively high $|p^{\rm CO}|$ values are found toward 40\deg\,$<$\,$l$\,$<$\,80\deg for $d$\,$<$\,900\,pc and for $d$\,$<$\,700\,pc toward 290\deg\,$<$\,$l$\,$<$\,360\deg.

Matter accumulation along the Radcliffe Wave may produce star formation that injects energy and momentum into the ISM.
However, the initial accumulation of matter that formed the Radcliffe Wave is probably the product of the large-scale Galactic dynamics.
Thus, we are again faced with the dichotomy between an SN-dominated and large-scale-forcing-dominated scenario, as we were in the case of the energy distribution over the whole studied volume. 
Without further evidence, we must conclude that our reconstruction of the momentum density distribution is most likely a combination of both effects.

\subsubsection{Mass flow rates}

Finally, we considered the mass flows related to the reconstructed streaming motions.
The distributions of $\dot{M}$, presented in Fig.~\ref{fig:Mdot}, indicate a series of converging and diverging mass flows along the line of sight with standard deviations around \postreport{3.0}\,$\times$\,$10^{-3}$ and \postreport{1.2}\,$\times$\,$10^{-3}$\,M$_{\odot}$/yr.
It is beyond the scope of this work to determine whether there is a known MC complex at the interphase of all the H{\sc i} converging flows, as suggested by some of the MC formation scenarios \citep[see, for example,][]{koyama2000}.
Reconstructing the specific flows in and around MC calls for dedicated studies of each region, as done with the HOG method for the Taurus MC in \cite{soler2023a}.
The map in Fig.~\ref{fig:Mdot} and the azimuthal profiles in Fig.~\ref{fig:azprofileMdot} provide a framework to distinguish specific flows from the global fluctuations and establish a statistical comparison with the results of numerical simulation of the Galactic environment. 

A rough interpretation of the $\dot{M}$ standard deviation for H{\sc i}, 3.0\,$\times$\,$10^{-3}$\,M$_{\odot}$/yr, can be made considering that these mass displacements correspond to stochastic fluctuations in the density and velocity fields.
If we assume that the mass displacements continue stochastically with an amplitude equal to the $\dot{M}$ standard deviation, they lead to an accumulation or dissipation of a mass comparable to that of a low-mass star formation region such as the Taurus MC \citep[$M$\,$\approx$\,2.4\,$\times$\,10$^{4}$\,$M_{\odot}$;][]{goldsmith2008} in about 8\,Myr.
This oversimplification ignores the effect of gravity, stellar feedback, and the bias introduced by exclusively measuring radial motions with respect to the LSR.
Still, it provides a rough estimate of the timescales implied by the derived $\dot{M}$.

\postreport{Following the same line of argumentation, the $\dot{M}$ standard deviation for H{\sc i}} implies that a GMC like Orion A \cite[$M$\,$\approx$\,7.57\,$\times$\,$10^{5}$\,$M_{\odot}$;][]{lombardi2011} would assemble or disperse in roughly 250\,Myr.
This is considerably larger than the cloud dispersal times of a few Myr estimated from various observations and statistical considerations \citep[see, for example,][]{elmegreen2000,kruijssen2019}.
If we assume that the $\dot{M}$ standard deviation for H{\sc i} exclusively represents converging flows, the times are equally large compared to the estimated cloud formation times \citep[see, for example]{fukui2009,maclow2017}.
Thus, we conclude that the typical $\dot{M}$ fluctuation\postreport{s} we identified in our analysis \postreport{do not directly correspond to GMC-forming or dispersing flows but to fluctuations in the ISM density and velocity fields}.
However, identifying the flows that lead to the formation and dissipation of star-forming regions like Taurus or Orion requires a focused study beyond the scope of the global averages considered in this paper.
\section{Conclusions}\label{sec:conclusions}

In this paper, we studied the morphological relation between a model of three-dimensional dust extinction and the H{\sc i} and CO line emission. 
We identified a significant correlation between dust and gas that we employed to study the ISM velocities across the region within $|b|$\,$<$\,5\deg\ and $d$\,$<$\,1250\,pc.
We produced maps of the LOS motions, streaming motions, kinetic energy and moment densities, and mass flow rate distributions across the studied volume.

We found that the material within 1250\,pc from the Sun carries the global kinematic imprint of circular motions around the Galactic center.
Yet, we also identified average departures from Galactic rotation, streaming motions, of around \sigmavHI\ and \sigmavCO\,\kps\ for the material sampled by H{\sc i} and CO, respectively.

We estimated the kinetic energy density corresponding to streaming motions. 
Even if we only have access to one component of the gas velocity field, we obtained mean values compatible with previous estimates, which confirms the near-equipartition with other forms of energy in the ISM. 
However, our result is novel because it maps that quantity across the studied volume, producing a reference point for reconstructions and deeper comparison with the results of numerical experiments.

We found energy overdensities of up to a factor of ten above the mean concentrated in the Radcliffe Wave, the Split, and regions around the Vela C and Ara star-forming regions. 
We provided evidence that these overdensities are most probably produced by the combined effect of large-scale forcing and SN explosions.
Marginalizing the impact of one effect in favor of the other is unlikely to fully describe the physical conditions in the local 1.25-kpc radius.
Yet, we did not find evidence that the local spiral arm affects the studied region's energy or momentum density distributions.

We found that the energy overdensities correspond to regions with momentum densities and mass flow rates above the mean.
It is difficult to assign these concentrations to individual feedback events or accumulations of material, as establishing those processes would require modeling the gas's past configuration with the only velocity component available to observations. 
However, the synergy of this kind of analysis with the reconstruction of the stellar dynamics enabled by {\it Gaia} and other observations and the novel techniques of data modeling promises to produce a complete picture of the ISM evolution in the Solar neighborhood and beyond.

\begin{acknowledgements}
We thank the anonymous referee for carefully reading our manuscript and providing insightful comments and suggestions.
JDS thanks the following people who helped with their encouragement and conversation: Jo\~{a}o Alves, Bruce Elmegreen, Torsten E{\ss}lin, \'Alvaro Hacar, Sebastian Hutschenreuter, Claire Murray, Martin Piecka, Eugenio Schisano, Mark Heyer, and \postreport{Oleg Egorov}. 
JDS also thanks Daniel Seifried and Stefanie Walch-Gassner for providing the numerical simulations and synthetic observations to test the HOG method.
RSK also thanks the 2024/25 Class of Radcliffe Fellows for the stimulating discussions.
Some of the crucial discussions that led to this work took place under the program Milky-Way-Gaia of the PSI2 project, which is funded by the IDEX Paris-Saclay, ANR-11-IDEX-0003-02.
This project was funded by the European Research Council via the ERC Synergy Grant ``ECOGAL'' (project ID 855130). 
RSK also acknowledges financial support from the German Excellence Strategy via the Heidelberg Cluster of Excellence (EXC 2181 - 390900948) ``STRUCTURES'', and from the German Ministry for Economic Affairs and Climate Action in project ``MAINN'' (funding ID 50OO2206).
The computations for this work were performed at the Max-Planck Institute for Astronomy (MPIA) {\tt astro-node} servers.
RSK is also grateful for computing resources provided by the Ministry of Science, Research and the Arts (MWK) of the State of Baden-W\"{u}rttemberg through bwHPC and the German Science Foundation (DFG) through grants INST 35/1134-1 FUGG and 35/1597-1 FUGG, and also for data storage at SDS@hd funded through grants INST 35/1314-1 FUGG and INST 35/1503-1 FUGG. 

{\it Software}: 
{\tt astropy} \citep{astropy2018}, 
{\tt SciPy} \citep{2020SciPy-NMeth}, 
{\tt astroHOG} \citep{astrohog2020}.
\end{acknowledgements}

\bibliographystyle{aa}
\bibliography{KineticTomography.bbl}

\appendix

\section{Insights into the HOG method}\label{app:hog}

\subsection{Direction- and orientation-sensitive Rayleigh statistic}\label{app:VandVd}

The original implementation of the HOG method presented by \cite{soler2019} used the projected Rayleigh statistic, defined as
\begin{equation}\label{eq:V}
V_{lm} = \frac{\sum_{ij}w_{ij,lm}\cos(2\theta_{ij,lm})}{\left(\sum_{ij}w_{ij,lm}/2\right)^{1/2}}.
\end{equation}
In this definition, the angles between the gradient vectors, $\theta_{ij,lm}$, are doubled, effectively comparing their orientation and not their direction.
This means that the result of Eq.~\eqref{eq:V} is the same for parallel or antiparallel gradient vectors.
This selection was based on the applications of this metric to the studies of magnetic field orientations, where the directions of interest are 0\deg\ and 90\deg\ \citep[see][and references therein]{jow2018}.
However, in our application there is additional information in the direction of the gradient vectors.

The orientation between the dust density gradients and CO emission is expected to be parallel in matching structures.
The same expectation applies to the H{\sc i} emission, with the exception of H{\sc i}SA, where the dust density gradients and HI emission are antiparallel.
Thus, the directions of interest are 0\deg\ and 180\deg.
For that reason, we applied a direction-sensitive projected Rayleigh statistic ($V_{\rm d}$; Eq.~\ref{eq:Vd}), which corresponds to Eq.~\eqref{eq:V} without the doubling of the angle.

\begin{figure}[ht]
\centerline{\includegraphics[width=0.5\textwidth,angle=0,origin=c]{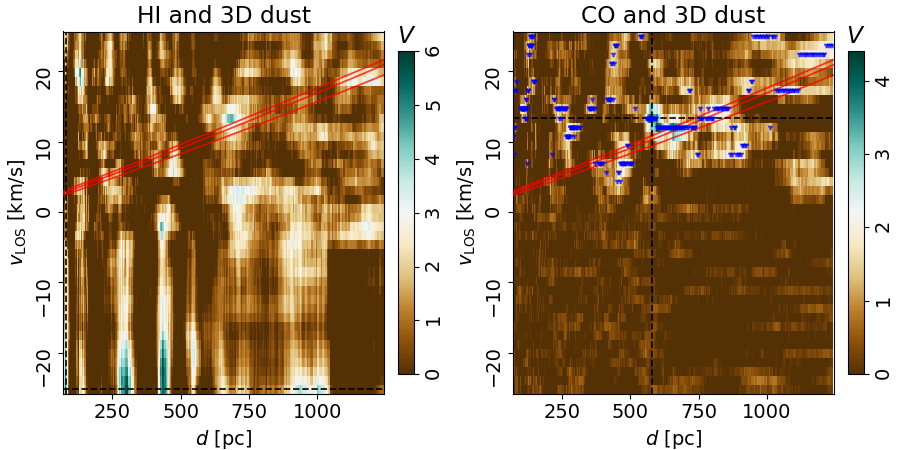}}
\centerline{\includegraphics[width=0.5\textwidth,angle=0,origin=c]{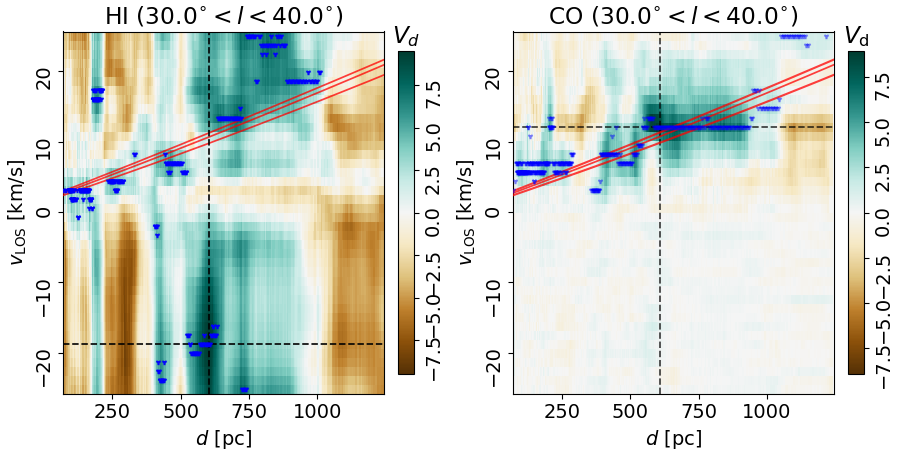}}
\caption{Comparison of the results obtained with $V$ and $V_{\rm d}$ for the mean 3D dust and the line emission toward the tile covering the Galactic longitude range 30\deg\,$<$\,$l$\,$<$\,40\deg.
}
\label{fig:comparisonVandVd}
\end{figure}

Fig.~\ref{fig:comparisonVandVd} compares $V$ and $V_{\rm d}$ for a test region.
The distinction between parallel and antiparallel gradients increases the significance of the distance-\vLOS\ pairs with morphological correlation, remarkably increasing the contrast in comparing the 3D dust density and the CO emission.
It also improves the comparison between the 3D dust density and the H{\sc i} emission by excluding outlying distance-\vLOS\ pairs that are antiparallel, as shown by their $V_{\rm d}$\,$<$\,$0$ values.
These outliers may be produced by H{\sc i}SA structures or chance correlation.
The detailed study of these features requires further H{\sc i}SA identification using spectral methods \citep[see, for example,][]{gibson2005,syed2023}, which are beyond the scope of this work.

Using $V_{\rm d}$ also reduces the fluctuations in the distance-\vLOS\ mapping for the different regions in which we split the Galactic plane, as illustrated in the example shown in Fig.~\ref{fig:comparisonVandVd}. 
This regularity enabled the assignment of \vLOS\ to the dust density channels that are at the core of our study, which we discuss in App.~\ref{app:physics}
In what follows, we present studies of the effects of noise and segmentation assuming $V_{\rm d}$ as the core metric of our analysis.

\subsection{Error estimation}\label{app:errors}

There are two primary sources of error in the input data used in the HOG analysis.
First, the uncertainties in the observations of the H{\sc} and CO line emission.
Second, the uncertainties in the reconstruction of the 3D dust.
In what follows, we describe how these uncertainties were propagated through the HOG analysis.

\subsubsection{Line emission uncertainties}\label{sec:sigmaLineEmission}

We used Monte Carlo sampling to propagate the uncertainties in the line emission observations following the procedure described in appendix~B of \cite{soler2019}.
For each gas tracer, we produced $N_{\rm MC}$ velocity channel maps $I^{*}_{ij,l}$ generated from random draws of Gaussian probability distribution centered on the observed values, $I_{ij,l}$, and standard deviation equal to the noise level of the observation, $\sigma_I$.
This data model assumes that the noise is constant throughout the line emission observations, which is adequate for the combination of surveys in H{\sc i}4PI but less so for the CO data.
However, given that we assume conservative values for the noise level in the \cite{dame2001} observations, we are most likely overestimating the observational uncertainties in the CO in exchange for a homogeneous treatment of the data.

\begin{figure}[ht]
\centerline{\includegraphics[width=0.5\textwidth,angle=0,origin=c]{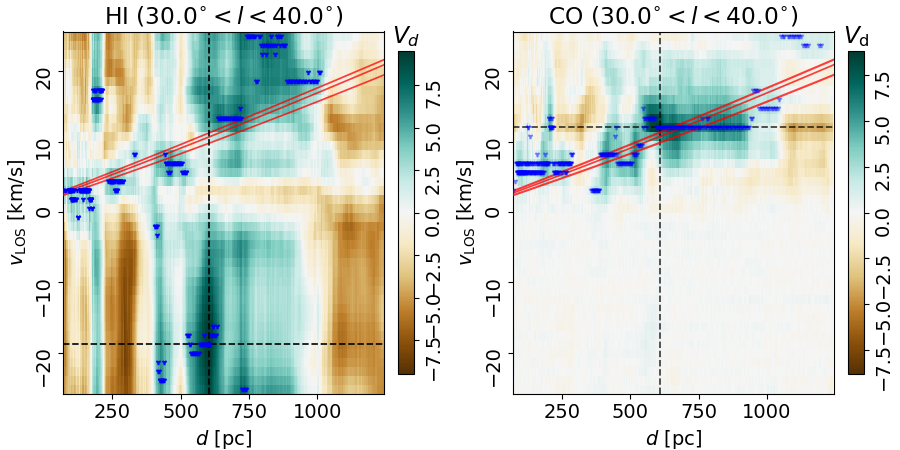}}
\centerline{\includegraphics[width=0.5\textwidth,angle=0,origin=c]{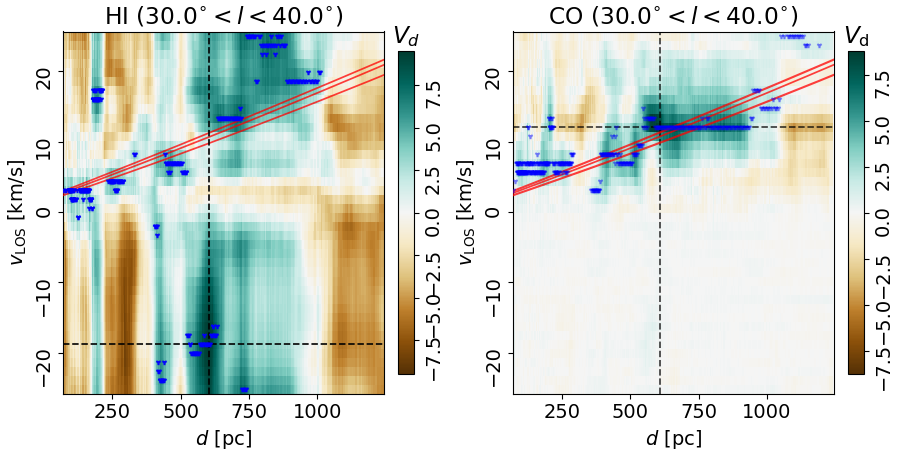}}
\caption{Same as the bottom Fig.~\ref{fig:comparisonVandVd}, but for $N_{\rm MC}$\,$=$\,10 and 100, shown in the top and bottom panels respectively.}
\label{fig:PRSmclines}
\end{figure}

Figure~\ref{fig:PRSmclines} shows an example of the line emission and 3D dust density correlation for $N_{\rm MC}$\,$=$\,10 and 100.
For comparison, we used the mean 3D dust density cube.
The negligible differences between the bottom panel of Fig.~\ref{fig:comparisonVandVd} and Fig.~\ref{fig:PRSmclines} suggest that the noise in the line emission observations is not a crucial uncertainty factor in the HOG method results.
This is expected if we consider that the noise in the line emission observations produces a spurious gradient orientation that is unlikely to be correlated with the structure in the 3D dust maps.

\begin{figure}[ht]
\centerline{\includegraphics[width=0.5\textwidth,angle=0,origin=c]{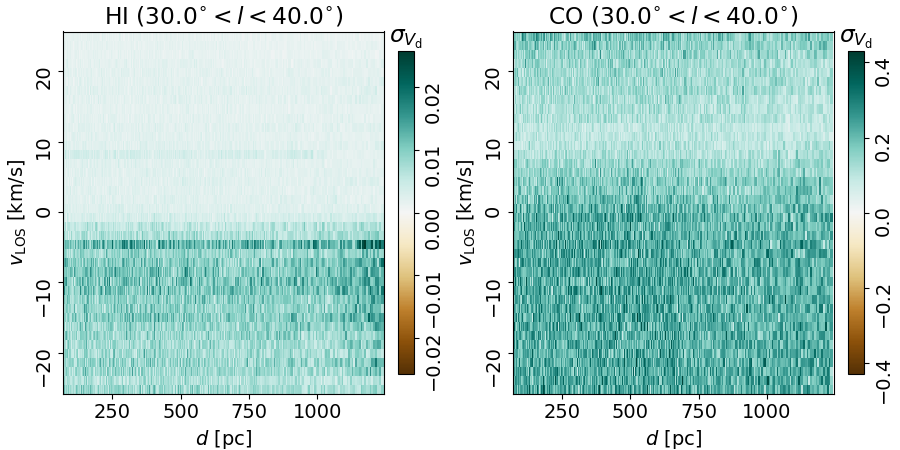}}
\centerline{\includegraphics[width=0.5\textwidth,angle=0,origin=c]{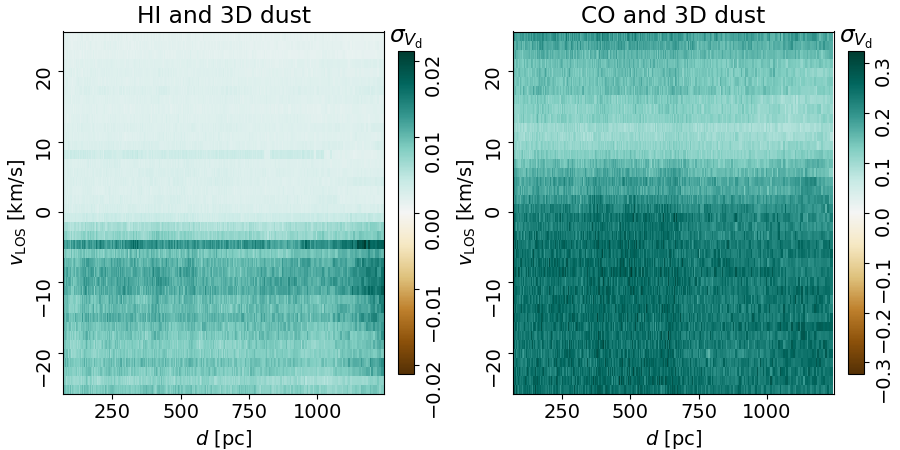}}
\caption{Standard deviation of $V_{\rm d}$ obtained for the comparison between the mean 3D dust cube and the Monte Carlo sampling of the line emission observations with for $N_{\rm MC}$\,$=$\,10 and 100, shown in the top and bottom panels respectively.}
\label{fig:s_PRSmclines}
\end{figure}

Figure~\ref{fig:s_PRSmclines} \postreport{presents} the $V_{\rm d}$ standard deviation for the example presented in Fig.~\ref{fig:PRSmclines}.
As expected from the negligible variations in the results obtained with $N_{\rm MC}$\,$=$\,0, 10, and 100, the values of $\sigma_{V_{\rm d}}$ are less than 1\% below the mean values obtained with the Monte Carlo sampling.  
In essence, the maps in Fig.~\ref{fig:s_PRSmclines} reflect the significance of the emission detection across \vLOS.
Thus, the lower $\sigma_{V_{\rm d}}$ for the H{\sc i} observations are the result of the lower noise level in those observations, in comparison with the CO composite survey.

\begin{figure*}[ht]
\centerline{
\includegraphics[width=0.5\textwidth,angle=0,origin=c]{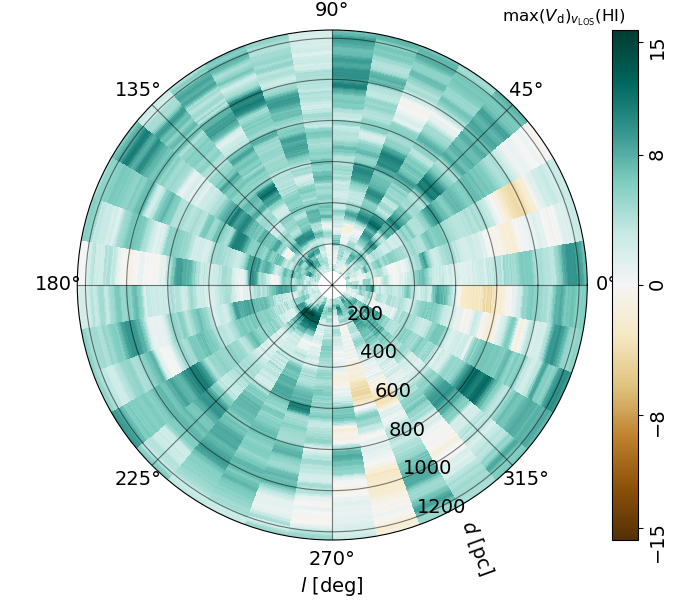}
\includegraphics[width=0.5\textwidth,angle=0,origin=c]{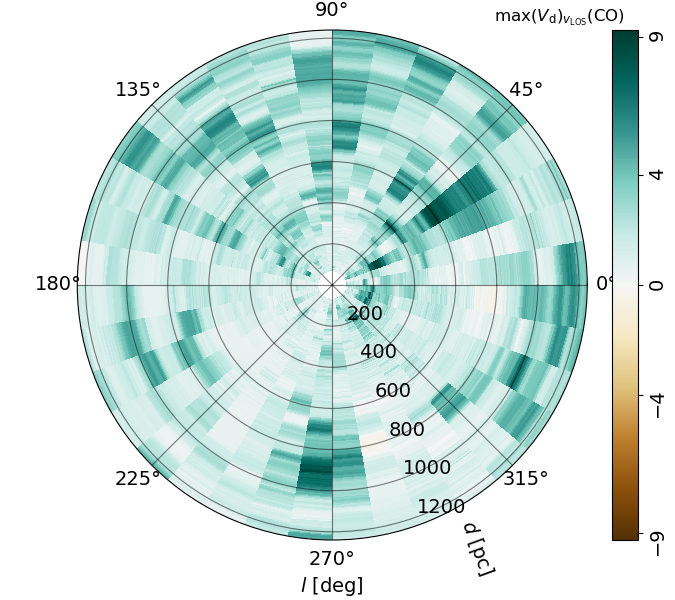}
}
\centerline{
\includegraphics[width=0.5\textwidth,angle=0,origin=c]{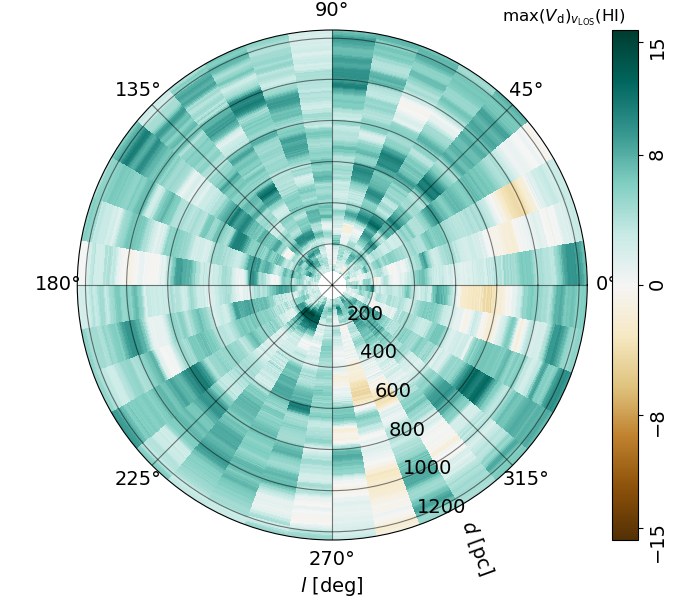}
\includegraphics[width=0.5\textwidth,angle=0,origin=c]{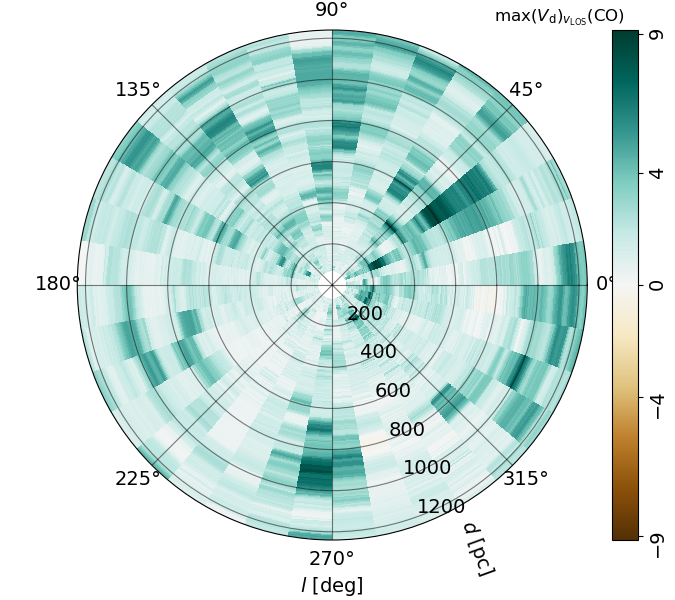}
}
\caption{Same as Fig.~\ref{fig:polarPRS}, but for the mean 3D dust density and the line emission data without and with Monte Carlo sampling, shown in the top and bottom panels, respectively.}
\label{fig:polarPRSmctest}
\end{figure*}

Figure~\ref{fig:polarPRSmctest} presents the global results of the HOG method applied to the mean 3D dust maps and the line emission observations with and without Monte Carlo sampling.
The negligible effect of the noise level in the line emission observations reported for a particular example in Fig.~\ref{fig:s_PRSmclines} can be extended to the whole Galactic longitude range.
This indicates that the noise in the line emission surveys does not produce any features that introduce a spurious morphological correlation in the HOG comparison with the 3D dust.
The mapping technique and instrumental variations undoubtedly introduce spatial features related to the noise level, as can be seen in figure~B.1 of \cite{riener2020} for the CO observations of the Galactic plane.
However, given that these noise features are unlikely to be found in the 3D dust reconstruction, they do not acutely affect the HOG analysis results.

\subsubsection{3D dust reconstruction uncertainties}\label{sec:sigma3Ddust}

The dust extinction modeling result presented in \cite{edenhofer2024} is a set of 12 samples from the variational posterior of the 3D dust distribution.
We computed the correlation with the line emission data for each sample to propagate the reconstruction uncertainties into the HOG analysis. 
We reported the mean value of $V_{\rm d}$.
This data model assumes that each sample is a plausible reconstruction of the 3D dust distribution and that the variance across the set represents the reconstruction's uncertainties.

\begin{figure}[ht]
\centerline{\includegraphics[width=0.5\textwidth,angle=0,origin=c]{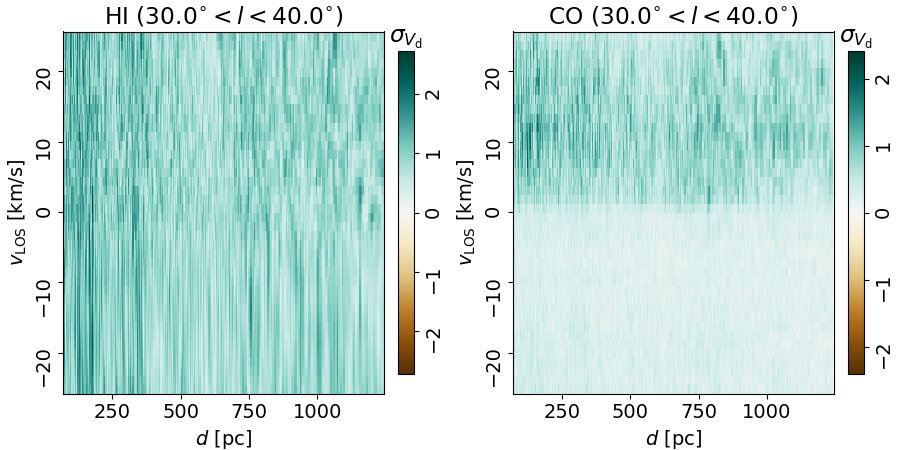}}
\caption{Standard deviation of $V_{\rm d}$ obtained for comparing the 12 posterior samples of the 3D dust cube and line emission observations without Monte Carlo sampling.}
\label{fig:s_PRS3DdustSamples}
\end{figure}

Fig.~\ref{fig:s_PRS3DdustSamples} presents the $V_{\rm d}$ standard deviation calculated from the 12 samples of the 3D dust reconstruction toward the test region used in Fig.~\ref{fig:comparisonVandVd}.
The values of $\sigma_{V_{\rm d}}$ are significantly larger than those found with the Monte Carlo sampling of the uncertainties in the line emission maps, thus indicating that the variance in the 3D dust reconstructions is the dominant source of error in the HOG results. 
To deal with this effect, we report in the main body of this paper the mean values of $V_{\rm d}$ obtained from the set of 3D dust samples.
We establish the significance of the HOG results for each distance-\vLOS\ pair by comparing the mean $V_{\rm d}$ with the standard deviation $\sigma_{V_{\rm d}}$.
We consider different levels of contrast to $\sigma_{V_{\rm d}}$ when assigning a prevalent velocity to a 3D dust channel and estimating other physical quantities from the HOG results, as we detail in App.~\ref{app:physics}.

\subsubsection{Chance correlation}\label{app:chancecorrelation}

\postreport{A significant effect to consider in the morphological matching using the HOG method is the chance correlation, which is the increase in $V_{\rm d}$ due to the accidental similarity between images.
For example, most landscape photographs taken by humans on the surface of the Earth are correlated by chance due to the presence of the horizon. 
In the same way, the vertical symmetry of the Galactic plane imposes a general level of morphological correlation in all tracers.
This effect can be minimized by employing the highest possible number of independent gradient vectors to resolve individual characteristics in the two images and produce contrast above the chance correlation level.
In our application, the number of independent gradient vectors is fixed by the 3D dust reconstruction's angular resolution.
Thus, we employ additional tests to marginalize the effect of change correlation.} 

\begin{figure}[ht]
\centerline{\includegraphics[width=0.5\textwidth,angle=0,origin=c]{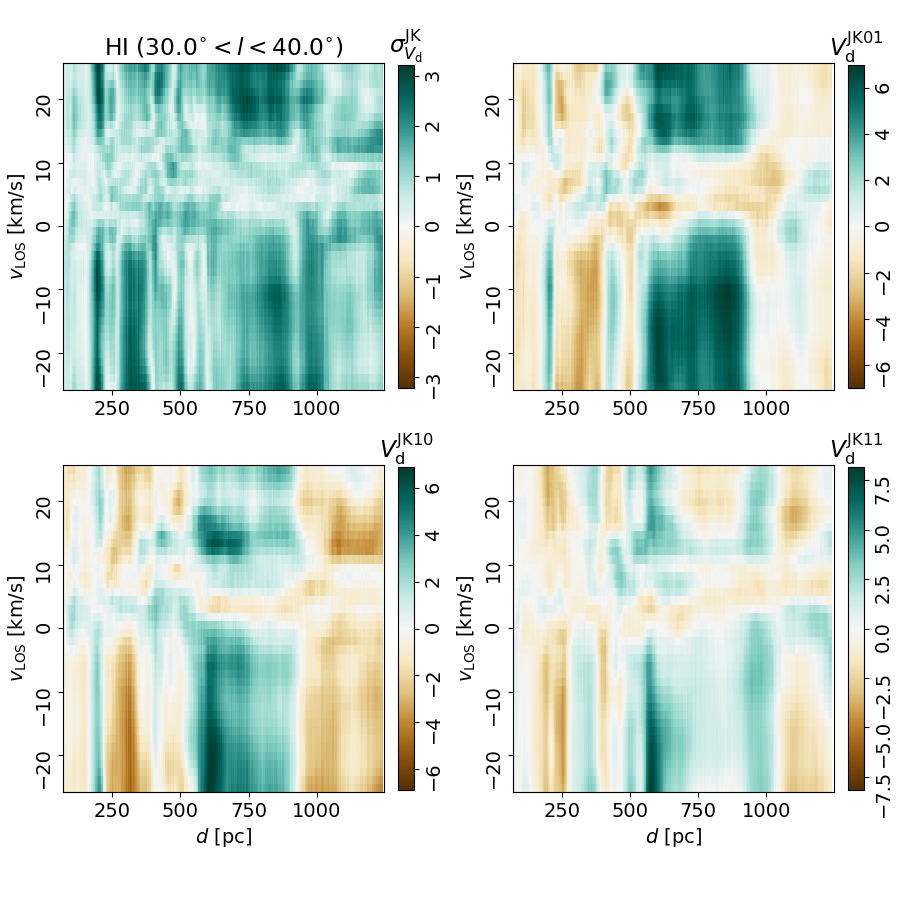}}
\caption{Standard deviation of $V_{\rm d}$ obtained for comparing the 12 posterior samples of the 3D dust cube and line emission observations without Monte Carlo sampling.}
\label{fig:s_PRSjkHI}
\end{figure}

\begin{figure}[ht]
\centerline{\includegraphics[width=0.5\textwidth,angle=0,origin=c]{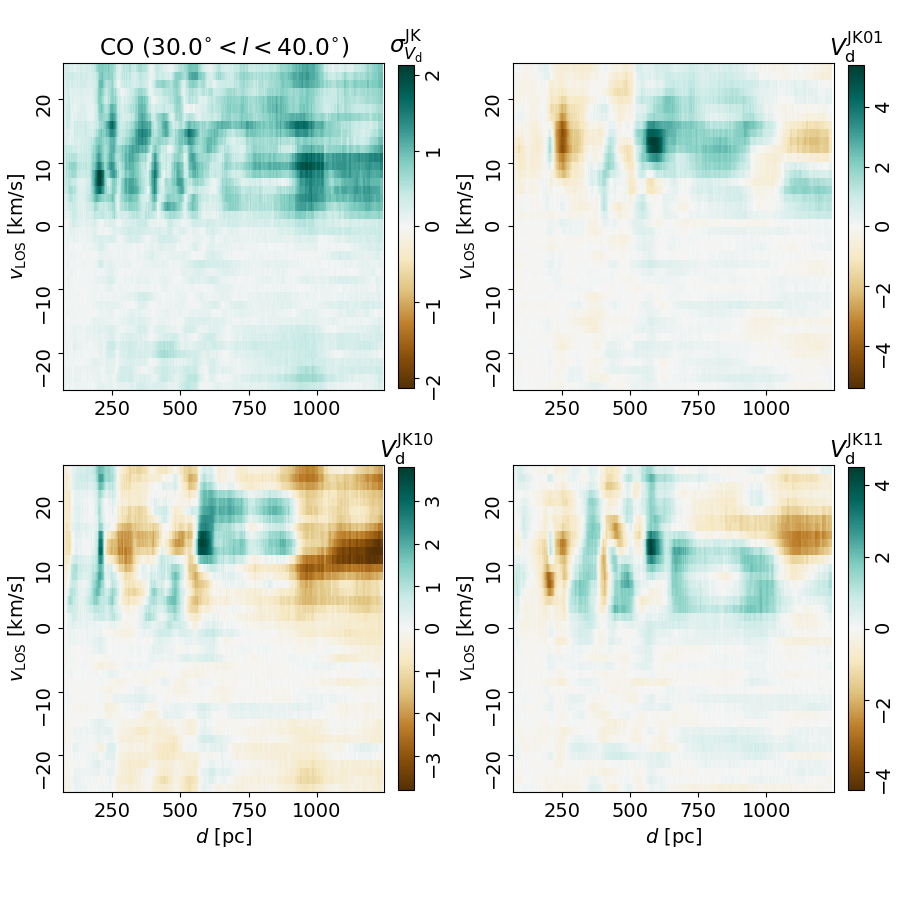}}
\caption{Same as Fig.~\ref{fig:s_PRSjkHI}, but for CO.}
\label{fig:s_PRSjkCO}
\end{figure}

\postreport{Figures~\ref{fig:s_PRSjkHI} and \ref{fig:s_PRSjkCO} present the $V_{\rm d}$ values obtained for the test tile in Fig.~\ref{fig:Vplanes} when flipping the emission PPV cubes with respect to the 3D dust PPD cube in the vertical (JK01), horizontal (JK10), and diagonal (JK11) directions.
The JK label comes from \cite{soler2019}, where this operation was introduced as ``jackknife'' tests.
The flipped cubes have the same global statistical properties, but their morphology should differ from the originals.
That difference level is quantified by the $V_{d}$ values obtained with the flipped sets.
For example, if the line emission is distributed like a large symmetric blob, it would produce a similar $V_{d}$ in the three flip tests, thus suggesting that it has insufficient unique features to produce an unambiguous morphological match.}

\postreport{The root-mean-square of the jackknife tests, $\sigma^{\rm JK}_{V_{\rm d}}$, reveals a significant number of $d$-\vLOS\ pairs where the chance correlation can be acute, as illustrated in Fig.~\ref{fig:s_PRSjkHI} and Fig.~\ref{fig:s_PRSjkCO}.
This is expected, given the multiscale correlations in the ISM that can mimic similarity when observed or reconstructed at low angular resolution.
We account for this effect in \vLOS\ reconstruction by evaluating the analysis results against the jackknife tests' amplitude, as described in Sec.~\ref{sec:methods}.}

\subsection{Selection of analysis parameters}

The sky portion chosen for our study of the morphological correlation between the 3D dust and the H{\sc i} and CO line emission is mainly driven by the coverage of the \cite{dame2001} composite survey, $|b|$\,$<$\,5\deg.
However, we were flexible in choosing the spatial scale at which we made the comparison, which was set by the derivative kernel size and the segmentation of the Galactic plane.
We discuss these two analysis choices in the following sections.

\subsubsection{Kernel size}

The $\Delta$\,$=$\,30\parcm0 FWHM derivative kernel used for the HOG results in the main body of this paper was mainly driven by the angular resolution of the data.
That kernel size is roughly twice the resolution of the H{\sc i}4PI data and covers four telescope beams in the \cite{dame2001} composite survey, which was only beam-width sampled in many regions.
Introducing a larger derivative kernel implies a loss in $V_{\rm d}$ significance, given the lower number of independent gradient vectors for the same sky area.
However, we tested $\Delta$\,$=$\,60\parcm0, and 90\parcm0 kernels to determine the effect of that choice in our results. 

\begin{figure}[ht]
\centerline{\includegraphics[width=0.5\textwidth,angle=0,origin=c]{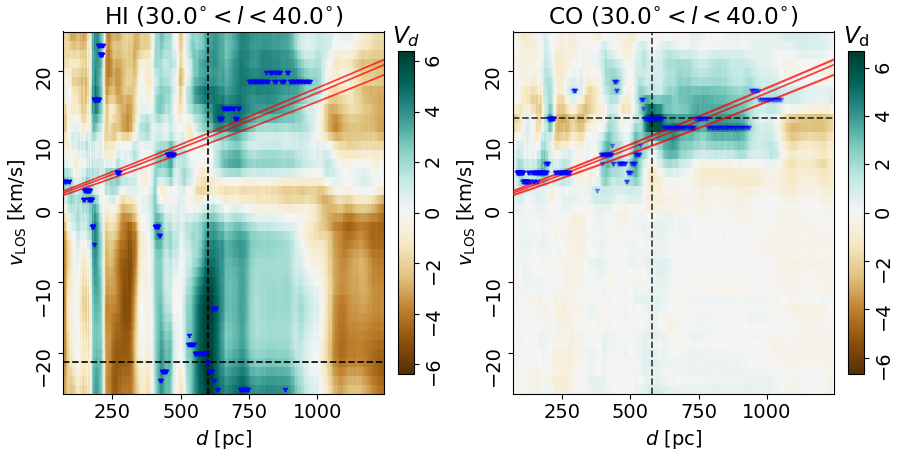}}
\centerline{\includegraphics[width=0.5\textwidth,angle=0,origin=c]{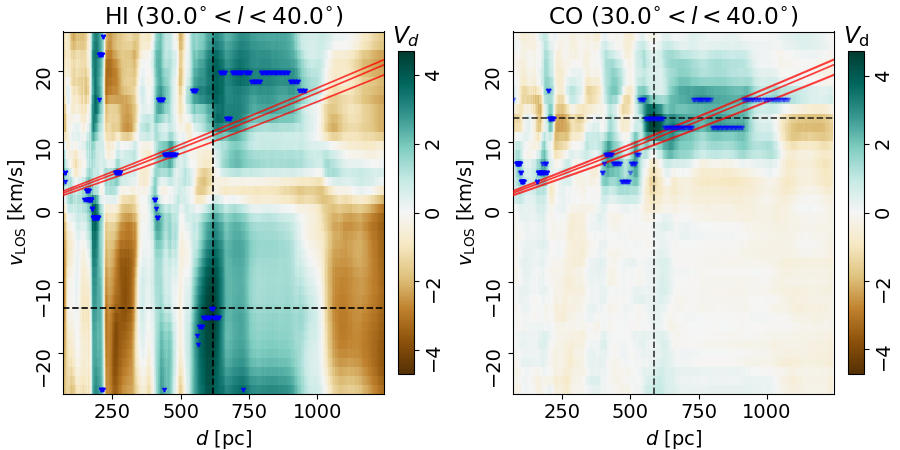}}
\caption{Same as bottom panel of Fig.~\ref{fig:comparisonVandVd}, but for derivative kernel sizes $\Delta$\,$=$60\pdeg0 and 90\pdeg0, shown in the top and bottom panels, respectively}
\label{fig:VplaneMultiksz}
\end{figure}

Figure~\ref{fig:VplaneMultiksz} shows the effect of the two kernel sizes for the same example region considered in Fig.~\ref{fig:comparisonVandVd}.
The distance-\vLOS\ pairs for which we obtain the highest correlation seem unaffected by the kernel size.
This can be explained by the persistence of the 3D dust density and line emission structures producing the highest $V_{\rm d}$ across these angular scales; for example, a blob producing high $V_{\rm d}$ at the 30\parcm0 scale is unlikely to disappear at 60\parcm0 and 90\parcm0.
However, the increase in the kernel size leads to a decrease in the maximum values of $V_{\rm d}$.
Thus, we set for $\Delta$\,$=$\,30\parcm0 FWHM to maintain a reasonable high $V_{\rm d}$ level in the 10\deg\,$\times$\,10\deg\ tiles throughout the $|b|$\,$<$\,5\deg\ band. 

\subsubsection{Galactic plane segmentation}\label{app:glondivision}

We chose to report the result of a segmentation of the $|b|$\,$<$\,5\deg\ band into 10\deg\,$\times$\,10\deg\ tiles starting at $l_{0}$\,$=$\,0\deg.
This segmentation is convenient in the Galactic coordinates reference frame but arbitrary.
Thus, we considered two additional segmentations starting at the reference Galactic longitudes $l_{0}$\,$=$\,$-$2\pdeg5 and $-$5\pdeg0, which are also arbitrary but we use to illustrate and estimate the effect of our division of the Galactic plane in the quantities derived with the HOG method.

\begin{figure*}[ht]
\centerline{
\includegraphics[width=0.5\textwidth,angle=0,origin=c]{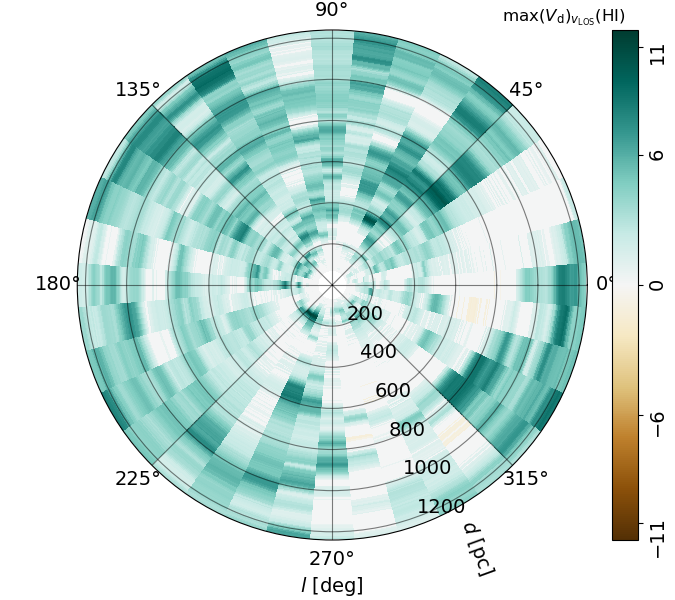}
\includegraphics[width=0.5\textwidth,angle=0,origin=c]{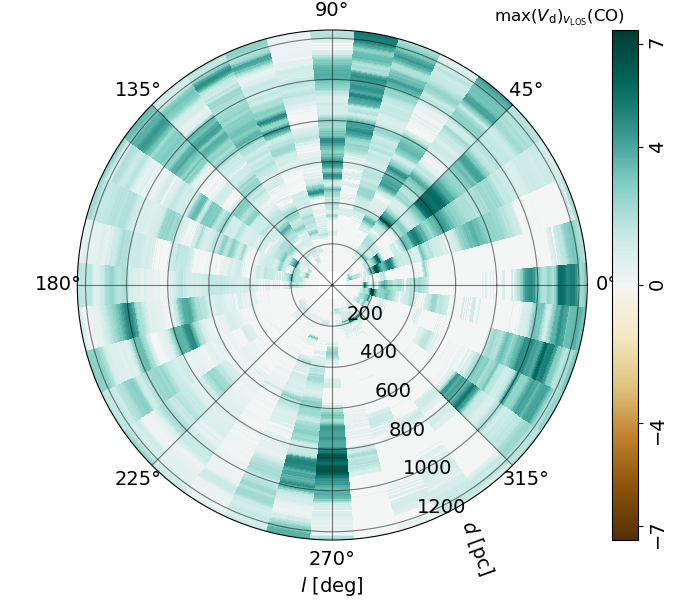}
}
\centerline{
\includegraphics[width=0.5\textwidth,angle=0,origin=c]{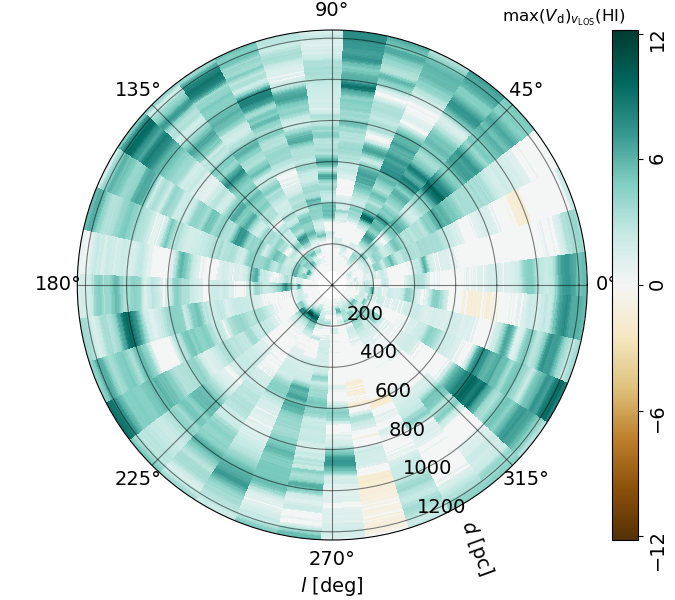}
\includegraphics[width=0.5\textwidth,angle=0,origin=c]{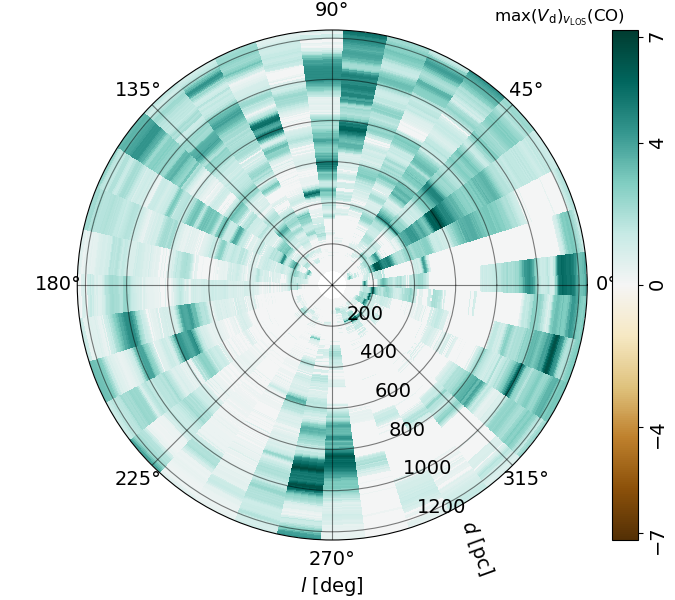}
}
\caption{Same as the top panel of \postreport{Fig.~\ref{fig:polarPRS}}, but Galactic longitude segmentations starting on $l$\,$=$\,$-$2\pdeg5 and $-$5\pdeg0, as shown in the top and bottom panels respectively.}
\label{fig:polarPRSgrid2and3}
\end{figure*}

Figure~\ref{fig:polarPRSgrid2and3} shows the $V_{\rm d}$ values for the two additional segmentations.
The values obtained for individual regions differ, which is expected given that a high-correlation feature may be split between tiles in one segmentation or another.
However, there is a considerable similarity in the global correlation pattern.
We used the variations among these segmentations to quantify the variance in the physical quantities estimated in App.~\ref{app:physics}

\begin{figure}[ht]
\centerline{\includegraphics[width=0.5\textwidth,angle=0,origin=c]{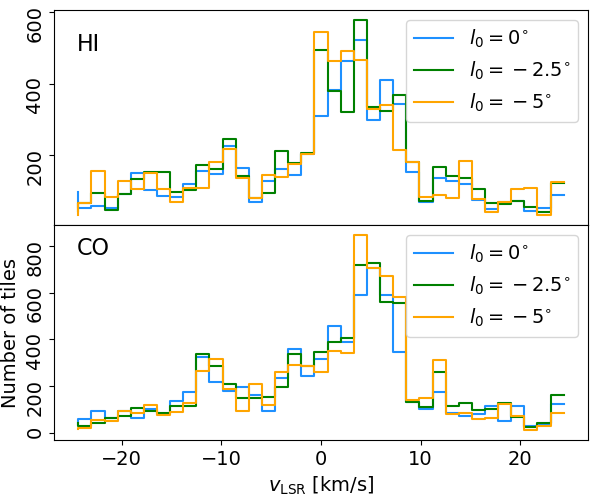}}
\caption{Histogram of the \vLOS\ assigned to the 3D dust tiles in three Galactic longitude segmentations, characterized by the position of the right edge of the first tile, $l_{0}$.}
\label{fig:gridcompareVlsrHIandCO}
\end{figure}

\postreport{
We evaluated the effect of the Galactic plane segmentation in the \vLOS\ assignment by considering the results obtained for three $l$ divisions. 
Figure~\ref{fig:gridcompareVlsrHIandCO} \postreport{presents} the distribution of the \vLOS\ assigned to the dust tiles using the H{\sc i} and CO emission.
We found that the reconstructed \vLOS\ have a global distribution primarily unaffected by the sky segmentation.
The discrepancies between the three \vLOS\ distributions indicate the suppression of some velocity features toward particular lines of sight, which are worth exploring in a focused study of specific regions but do not significantly change the conclusions of our work.}

\subsubsection{Line-of-sight velocity range}\label{app:vLOSrange}

In the main body of this paper, we chose the range $-25$\,$<$\,\vLOS\,$<$\,25\,\kps\ for the line emission data input.
This selection was \postreport{initially} motivated by the maximum LOS velocity expected from the \cite{reid2019} Galactic rotation model within the heliocentric distances $d$\,$<$\,1.25\,kpc, which is 21.6\,\kps.
\postreport{This restricted input \vLOS\ range aims to mitigate the effect of chance correlation in the morphological comparison since, in principle, line emission at higher \vLOS\ is not necessarily related to the local material.
This section tests that assumption by considering the results obtained with broader \vLOS\ input ranges..}

\begin{figure*}[ht]
\centerline{
\includegraphics[width=0.5\textwidth,angle=0,origin=c]{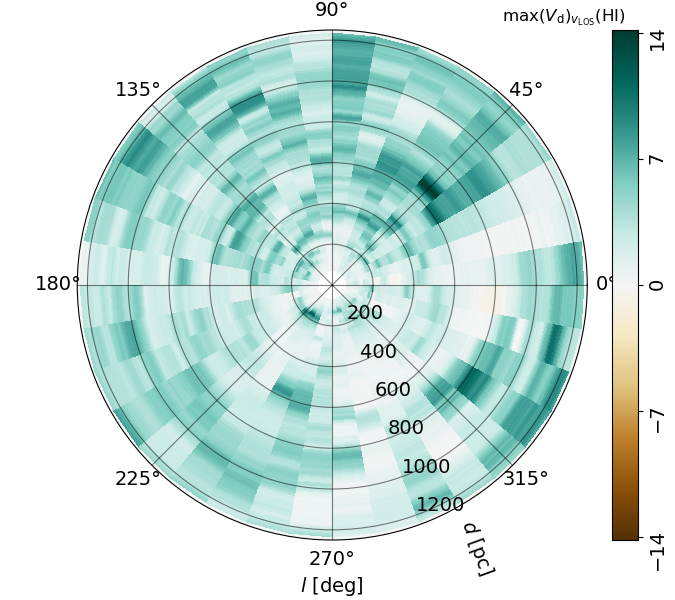}
\includegraphics[width=0.5\textwidth,angle=0,origin=c]{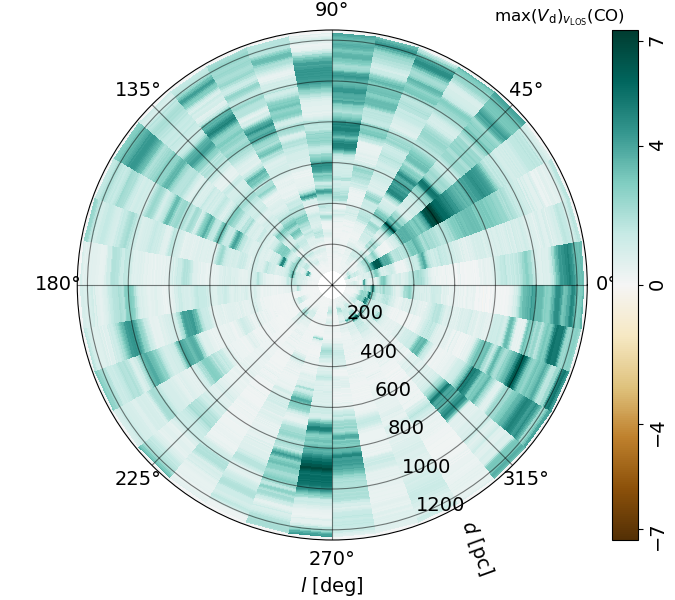}
}
\caption{Same as Fig.~\ref{fig:polarPRS}, but for the input LOS velocity ranges $-120$\,$<$\,\vLOS\,$<$\,120\,\kps.}
\label{fig:polarPRSvlos-40to40}
\end{figure*}

\postreport{Figure~\ref{fig:polarPRSvlos-40to40} presents the distribution of the maximum $V_{\rm d}$ for the comparison between the 3D dust and the H{\sc i} and CO line emission in the range $-80$\,$<$\,\vLOS\,$<$\,80\,\kps.
Visual comparison between Fig.~\ref{fig:polarPRS} and Fig.~\ref{fig:polarPRSvlos-40to40} demonstrates that the extension of the \vLOS\ input range does not increase the morphological correlation, quantified by $\max(V_{\rm d})_{V_{\rm LOS}}$. 
This result suggests that the low-correlation regions, $\max(V_{\rm d})_{V_{\rm LOS}}$\,$\approx$\,0, are not the consequences of the limited \vLOS-input but a result of the 3D dust distribution across distance channels.
}

\begin{figure}[ht]
\centerline{\includegraphics[width=0.5\textwidth,angle=0,origin=c]{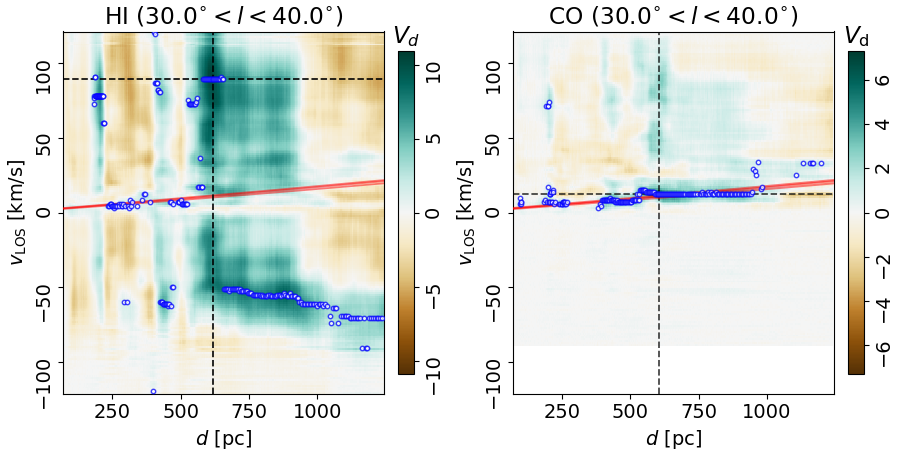}
}
\caption{\postreport{Same as Fig.~\ref{fig:Vplanes}, but for the input LOS velocity range $-120$\,$<$\,\vLOS\,$<$\,120\,\kps.}}
\label{fig:VplaneGLON30to40vLOS-40to40}
\end{figure}

\postreport{Figure~\ref{fig:VplaneGLON30to40vLOS-40to40} shows the results of expanded LOS velocity input ranges in the distance-\vLOS\ diagrams introduced in Fig.~\ref{fig:Vplanes}.
For the region in that example, a significant morphological correlation exists for H{\sc i} beyond the $|v_{\rm LOS}|$\,$<$\,25\,\kps limit.
Except for a few outliers, the most significant morphological correlation for CO is within the range $|v_{\rm LOS}|$\,$<$\,25\,\kps.
This result implies that the \vLOS\ input limit excludes some significant departures from circular motions, effectively limiting the amplitude of the reconstructed streaming motions and the energy and momentum estimates.
However, the { \tt HOG} method also indicates contamination by chance correlation when employing a broader \vLOS\ range.}

\begin{figure}[ht]
\centerline{\includegraphics[width=0.49\textwidth,angle=0,origin=c]{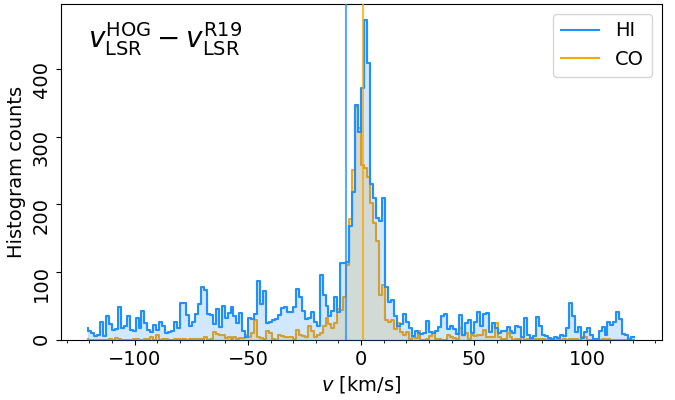}}
\caption{Same as Fig.~\ref{fig:histdiffv}, but for the input LOS velocity range $-120$\,$<$\,\vLOS\,$<$\,120\,\kps.}\label{fig:histdiffv40to40kps}
\end{figure}

\postreport{Figure~\ref{fig:histdiffv40to40kps} presents the distribution of the streaming motions derived with the extended input range $-120$\,$<$\,\vLOS\,$<$\,120\,\kps.
The histogram shows a background signal in H{\sc i} that extends beyond the range $|v_{\rm LOS}|$\,$<$\,25\,\kps, particularly for negative \vLOS.
This signal is unlikely to have a natural origin, as it would imply a significant velocity gradient in the solar vicinity, in tension with the observations of other tracers and the conditions for the Galactic disk stability.
It is more plausible that this signal results from accidental correlation remanent after the selection described in App.~\ref{app:chancecorrelation}. 
The CO emission appears less affected by this additional chance correlation, so we use it as a reference to set the input \vLOS\ limits of the reconstruction.
For the main body of the paper, we chose an input \vLOS\ range centered on zero and spanning across three times the standard deviation of the CO streaming motions estimated with the extended for the range $-120$\,$<$\,\vLOS\,$<$\,$120$\,\kps, which is 52.1\,\kps.
Hence the resulting $|v_{\rm LOS}|$\,$<$\,25\,\kps\ range.
We reserve the study of broader departures from circular motion to a subsequent, less general study. 
}

\section{H{\sc i} self absorption and the HOG results}\label{app:hisa}

\begin{figure}[ht]
\centerline{\includegraphics[width=0.5\textwidth,angle=0,origin=c]{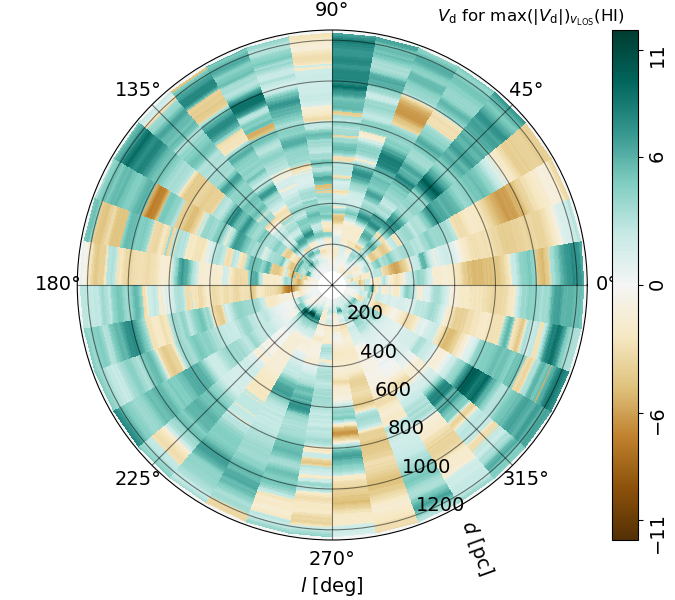}}
\caption{Same as Fig.~\ref{fig:polarPRS}, but reporting the $V_{\rm d}$ values corresponding to the maximum $|V_{\rm d}|$.}
\label{fig:polarmaxabsPRS}
\end{figure}

The plots in Fig.~\ref{fig:polarPRS} and Fig.~\ref{fig:lvPRS} present the maximum values of $V_{\rm d}$ identified across \vLOS\ channels for the volume cells in which we divided the studied region.
This selection highlights structures where the density and line emission gradients are parallel, thus \postreport{highlighting a} positive correlation between emission intensity and density.
However, there may be information in the minimum values of $V_{\rm d}$\postreport{, which in the case of significant negative values} may highlight a HISA feature, where the gradients for the dust density and H{\sc i} emission are antiparallel.
To evaluate this possibility, we considered the values of $V_{\rm d}$ corresponding to the maximum $|V_{\rm d}|$ across distance channels, as we report in Fig.~\ref{fig:polarmaxabsPRS}.
That selection \postreport{accentuates} distance channels dominated by antiparallel density and emission gradients.

\postreport{Figure~\ref{fig:polarmaxabsPRS} presents the $V_{\rm d}$ values corresponding to the maximum $|V_{\rm d}|$ across distance channels in the studied volume.
Visual comparison with Fig~\ref{fig:polarPRS} shows that, in general, the negative $V_{\rm d}$ are only dominant in regions with relatively low $\max(V_{\rm d})$.
That result implies the positive correlation between 3D dust density and H{\sc i} emission dominates throughout the studied volume.
However, we performed additional tests to determine whether dust structures are associated with HISA in the distance channels with significant $V_{\rm d}$\,$<$\,0. 
}

\begin{figure}[ht]
\centerline{\includegraphics[width=0.5\textwidth,angle=0,origin=c]{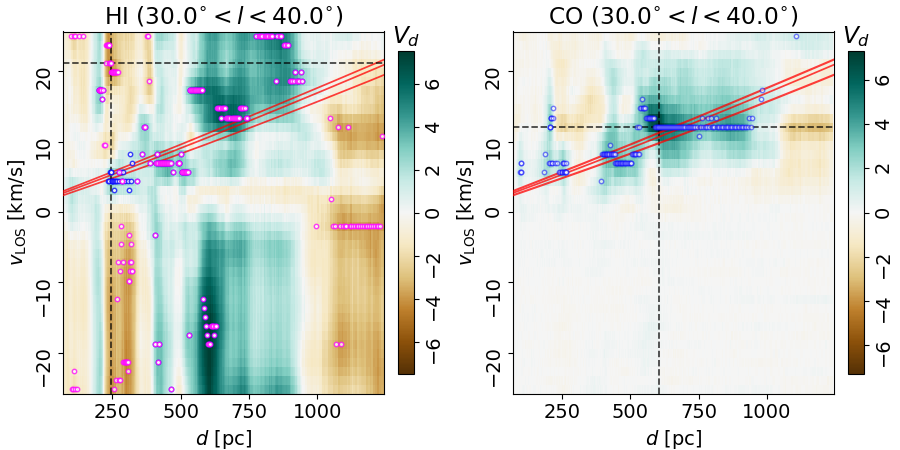}
}
\caption{\postreport{Same as Fig.~\ref{fig:Vplanes}, but also indicating the maximum $|V_{\rm d}|$ for H{\sc i} in each distance channel (magenta markers).
The dashed vertical and horizontal lines indicate the distance and \vLOS\ with the minimum $V_{\rm d}$ toward this region.}}
\label{fig:VplaneHISA}
\end{figure}

\postreport{Figure~\ref{fig:VplaneHISA} show the morphological correlation between 3D dust distance channels and line-emission \vLOS\ channels, indicating the \vLOS\ with the maximum $|V_{\rm d}|$ at each distance after the selection criteria described in Sec.~\ref{sec:methods}.
As expected from the results in Fig.~\ref{fig:polarmaxabsPRS}, in most distance channels, the maximum $|V_{\rm d}|$ coincides with the maximum $V_{\rm d}$.
However, in this example, the distance channel with the minimum $V_{\rm d}$ corresponds to a \vLOS\ that would have been excluded in selecting positive correlation by maximum $V_{\rm d}$.
}

\begin{figure*}[ht]
\centerline{\includegraphics[width=0.98\textwidth,angle=0,origin=c]{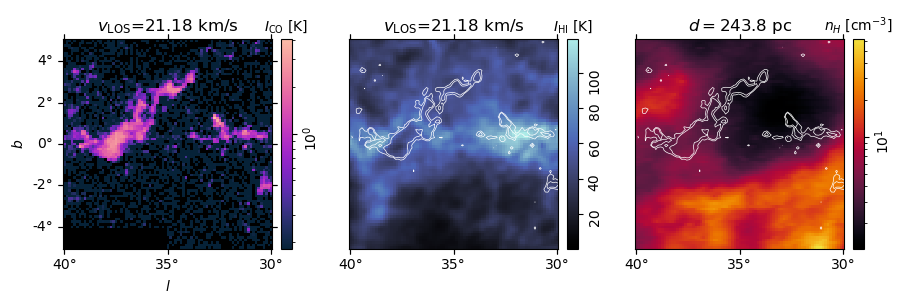}}
\vspace{-1.0mm}
\centerline{\includegraphics[width=0.98\textwidth,angle=0,origin=c]{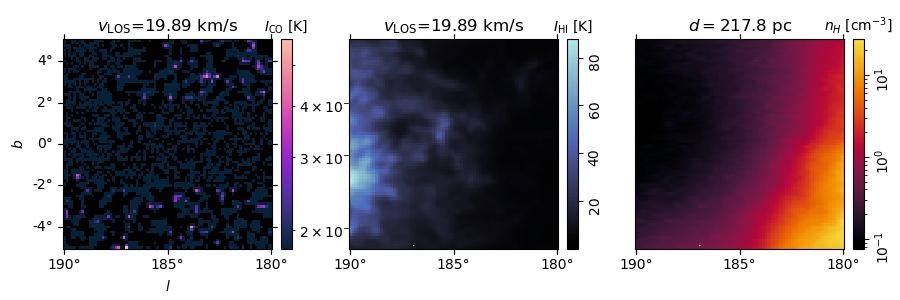}}
\caption{Two examples of regions where the H{\sc i} emission gradients are mainly antiparallel to the 3D dust density gradients, as identified by the minimum values of $V_{d}$.}
\label{fig:mapsHISAtest}
\end{figure*}

\postreport{The top panel of Fig.~\ref{fig:mapsHISAtest} presents the line emission and 3D dust density for the channels with the minimum $V_{\rm d}$ identified in Fig.~\ref{fig:VplaneHISA}.
It is apparent from the images that there is a decrease in the H{\sc i} intensity toward the region with high CO emission on the central-left part of the maps.
Inspection of the H{\sc i} spectrum confirms the presence of a few-\kps-wide dip that usually characterizes HISA \citep{gibson2000}.
A large fraction of the $V_{\rm d}$ signal comes from the morphological coincidence between the depression in the H{\sc i} emission and the 3D structures toward the lower portion of the map.
The spectra toward that portion of the map are not straightforward to interpret but show some promising indications of HISA. 
However, a generalization of this result is complicated.
}

\postreport{The bottom panel of Fig.~\ref{fig:mapsHISAtest} displays the line emission and 3D dust density for the distance-\vLOS\ pair with the lowest $V_{\rm d}$ in the studied volume.
The gradient in 3D dust density coincides with the H{\sc i} emission decrease.
In this case, however, the depression in the H{\sc i} emission is extended tens of \kps, thus it is unlikely to come from HISA.
Our results can be used to indicate the potential HISA locations, but their confirmation requires additional spectral analysis beyond the scope of this work. 
Consequently, we report only the results for the positive correlation between H{\sc i} emission and 3D dust in the main body of the paper.}


\section{Effects of fixed angular resolution and distance}\label{app:physics}

\postreport{In this section, we consider the effect of the fixed angular resolution introduced by the derivative kernel size in the $V_{\rm d}$ distribution and its potential effect in the results of the HOG method.
Our selection of a 0.5\deg\ derivative kernel implies that we are sampling scales around 0.6\,pc at the front of the 3D dust reconstruction, $d$\,$=$\,69\,pc, and 10.9\,pc at the back, $d$\,$=$\,1250\,pc.
We evaluate the potential effect of this feature in the reported $V_{\rm d}$ values by considering its distribution across distance channels.}

\begin{figure}[ht]
\centerline{\includegraphics[width=0.5\textwidth,angle=0,origin=c]{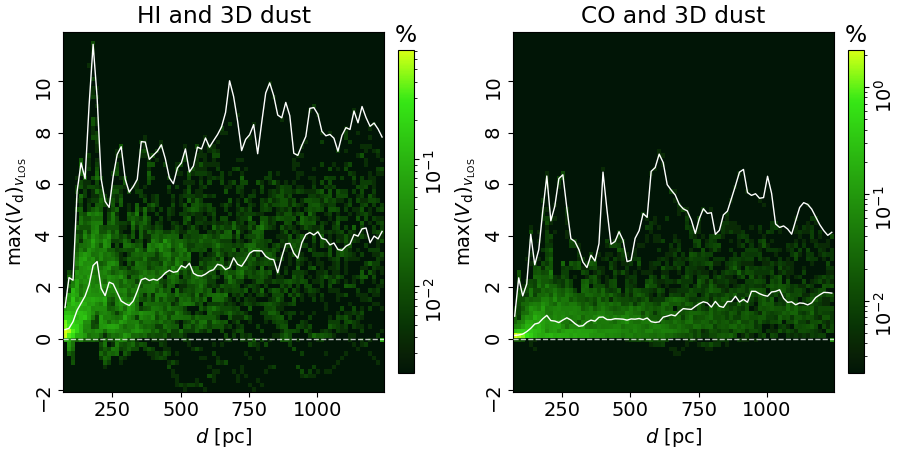}}
\caption{Two-dimensional histograms showing the distribution of $V_{\rm d}$ and distances for the data presented in Fig.~\ref{fig:polarPRS}.
\postreport{The solid white lines correspond to the 50th and 99th percentile of the $\max(V_{\rm d})$ values in each distance bin.}}
\label{fig:histPRSHIandCOvsDistance}
\end{figure}

\postreport{Figure~\ref{fig:histPRSHIandCOvsDistance} presents the distribution of maximum $V_{\rm d}$ values  reported in Fig.~\ref{fig:polarPRS}, but divided in distance bins.
The trends obtained for the 50th and 99th percentile of this quantity indicate that the fixed angular resolution does not produce a systematic fall in the $\max(V_{\rm d})$ values, but that these rather increase with distance.
This increasing trend is most likely the result of the inclusion of additional ISM structures in the most distant channels compensating for the lack of spatial resolution.
These results are not unexpected, considering the self-similar nature of the ISM, which implies that the correlation between H{\sc i}, CO, and 3D dust is extended across the scales considered in this study.
Figure~\ref{fig:histPRSHIandCOvsDistance} indicates that the highest $\max(V_{\rm d})$ values are not clustered around particular distances but are distributed along the line of sight, thus indicating that HOG is not biased by an effect related to the distance in the 3D dust model.}

\begin{figure}[ht]
\centerline{\includegraphics[width=0.5\textwidth,angle=0,origin=c]{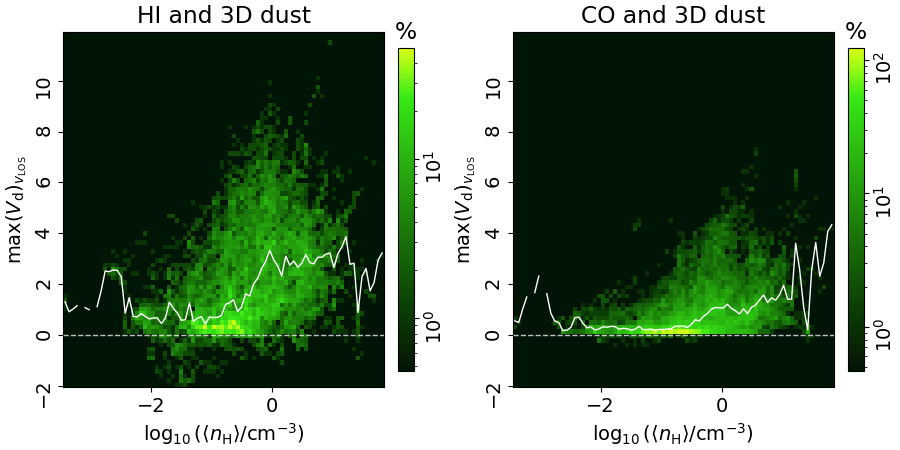}}
\caption{Two-dimensional histograms showing the distribution of $V_{\rm d}$ and $\left<n_{\rm H}\right>$ for the data presented in Fig.~\ref{fig:polarPRS}.
\postreport{The solid white lines correspond to the 50th percentile of the $\max(V_{\rm d})$ values in each density bin.}}
\label{fig:histPRSHIandCOvsDensity}
\end{figure}

We also considered the potential bias introduced by the mean density, $\left<n_{\rm H}\right>$, in a tile in the HOG results.
It is expected that $\max(V_{\rm d})$ should have higher values toward tiles with higher densities due to the larger number of significant density gradients for the same area.
Yet, a morphological correlation should not be exclusively found toward high-$\left<n_{\rm H}\right>$ tiles but distributed across $\left<n_{\rm H}\right>$.
This is what we found in the distribution of $V_{\rm d}$ as a function of $\left<n_{\rm H}\right>$ reported in Fig.~\ref{fig:histPRSHIandCOvsDensity}.
We found that the highest $V_{\rm d}$ are not exclusively found in high-$\left<n_{\rm H}\right>$ tiles, thus suggesting the HOG is finding a correlation that is not only related to the amount of dust in a tile but to the coincidence between its distribution and that of the line emission tracer in the plane of the sky.

The definition in Eq.~\eqref{eq:Vd} implies that $V_{\rm d}$ depends on the number of gradient pairs.
This is useful in our application because $V_{\rm d}$ highlights regions with a high number of parallel gradients but does not provide information about the portion of the maps that is producing the morphological correlation.
To estimate the percentage of gradients producing the morphological correlation identified with $V_{\rm d}$, we used the normalized projected Rayleigh statistic introduced in \cite{mininni2024},
\begin{equation}\label{eq:VoverVmax}
\chi_{kpq}=\frac{(V_{\rm d})_{kpq}}{(V_{\rm d, max})_{kpq}}=\frac{(V_{\rm d})_{kpq}}{\left(2\sum_{ij}w_{ijkpq}\right)^{1/2}}.
\end{equation}

The quantity of $V_{d, {\rm max}}$ represents the maximum value of Eq.~\eqref{eq:Vd}, which corresponds to all gradient pairs being parallel, $\theta_{ijkpq}$\,$=$\,0\deg, or the comparison of identical images.
In the particular case $w_{ijkpq}$\,$=$\,1, $\left(V_{d, {\rm max}}\right)_{kpq}$\,$=$\,$(2N_{kpq})^{1/2}$, where $N_{kpq}$ is the number of relative orientation angles between gradients in distance channel $p$ and velocity channel $q$.
The values of $\chi$ are bound to the range $-1$ and $1$ and roughly correspond to the percentage of the gradient pairs representing the parallel or antiparallel direction, for $V_{\rm d}$\,$>$\,0 and $V_{\rm d}$\,$<$\,0, respectively.

\begin{figure*}[ht]
\centerline{
\includegraphics[width=0.5\textwidth,angle=0,origin=c]{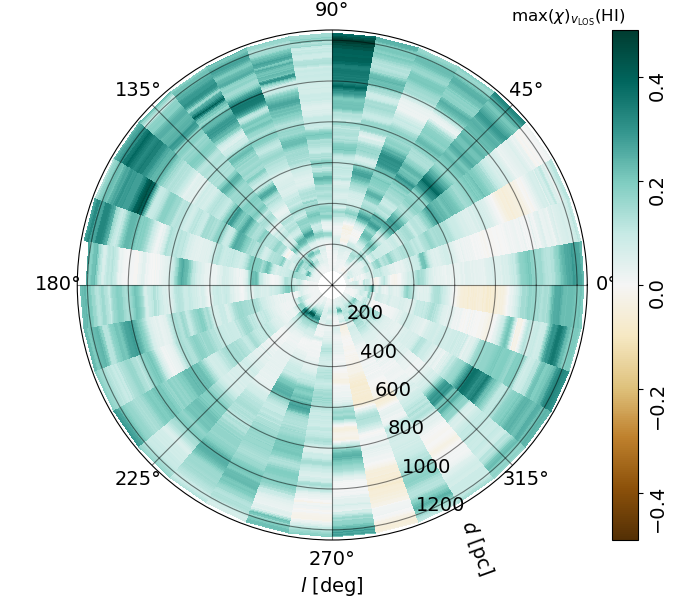}
\includegraphics[width=0.5\textwidth,angle=0,origin=c]{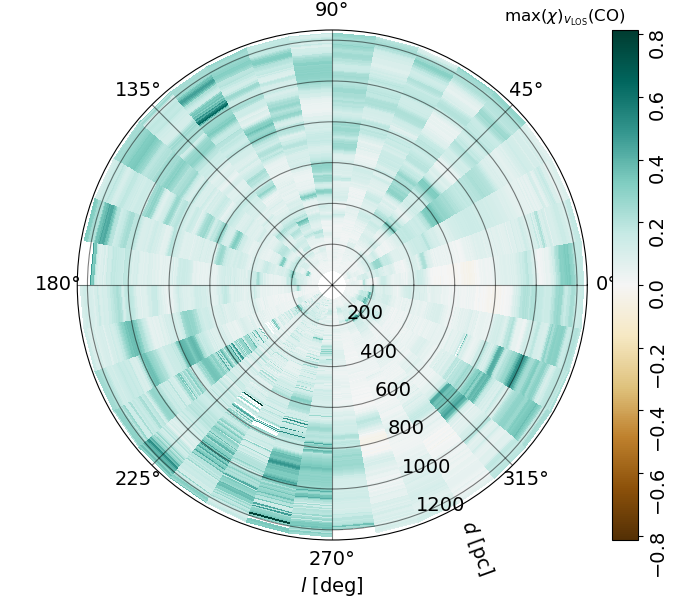}
}
\caption{Same as Fig.~\ref{fig:polarPRS}, but for the normalized projected Rayleigh statistic, $\chi$, Eq.~\eqref{eq:VoverVmax}.}
\label{fig:polarvovervmax}
\end{figure*}

\postreport{Figure~\ref{fig:polarvovervmax} shows the maximum $\chi$ across \vLOS\ channels for each distance channel in our reconstruction.
This quantity can be considered an analog of $\max(V_{\rm d})$ in Fig.~\ref{fig:polarPRS}, but marginalizing over the difference in the number of independent gradients across tiles.
Visual comparison between Fig.~\ref{fig:polarPRS} and Fig.~\ref{fig:polarvovervmax} indicates that the $\max(V_{\rm d})$ and $\max(\chi)$ distributions are very similar in H{\sc i}, as expected from the homogeneous number of independent gradients across distance and \vLOS\ channels.
However, in CO, some portions of the $b$\,$\leq$\,5\deg range are not uniform, producing a dissimilar number of independent gradients across Galactic longitudes and \vLOS.
This effect is evident in the differences between $\max(V_{\rm d})$ and $\max(\chi)$ for that tracer toward the third Galactic quadrant and other low CO emission regions. 
In those regions, the normalization implied in Eq.~\ref{eq:VoverVmax} may emphasize a morphological correlation represented by just a few independent gradients. 
Thus, we settle for $V_{\rm d}$ as a morphological metric rather than $\chi$. 
}

\section{The HOG method in synthetic observations}\label{app:synthobs}

\postreport{We conducted a HOG study of the synthetic H{\sc i} and CO observations of a simulated MC in the Simulating the life-Cycle of molecular clouds (SILCC) Zoom project \citep{seifried2017,seifried2020}.
It corresponds to a 125-pc-side cube extracted for the kiloparsec-sized stratified boxes in the SILCC project \citep{walch2015,girichidis2016}, modeled at 0.06-pc resolution using adaptive mesh refinement in combination with a chemical network to follow heating, cooling and the chemistry of H{\sc i} and the formation of H$_{2}$ and CO including (self-) shielding. 
We used their MC1-MHD model, one of the four clouds studied through H{\sc i} synthetic observations in \cite{seifried2022}.
We refer to the aforementioned publications for further details on the simulations and synthetic observations.
.}

\begin{figure*}[ht]
\centerline{\includegraphics[width=0.98\textwidth,angle=0,origin=c]{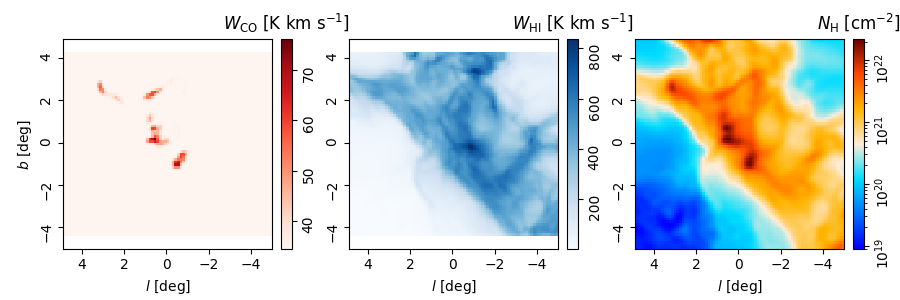}}
\caption{Projected synthetic $^{12}$CO(1-0) and H{\sc i} line emission and 3D dust reconstruction from the SILCC Zoom MC1-MHD simulation placed at 709\,pc from the LSR.}
\label{fig:synthObs}
\end{figure*}

\postreport{We used the synthetic H{\sc i} and $^{12}$CO(1-0) from the SILCC-Zoom MC1-MHD simulation to mimic observations comparable to those employed in this paper.
We placed the simulated cube at a distance of 709\,pc parsec to cover the range $l$\,$<$\,$\pm 5$\deg.
We smoothed each observation using a 2D-Gaussian point spread function matching the corresponding beam and used an identical grid to that employed in the analysis of observational data, as described in Sec.~\ref{sec:data}.
As a proxy for the 3D dust PPD cube, we used the 3D density cube from the simulation but smoothed employing a 2D-Gaussian kernel to match the angular resolution of the \cite{edenhofer2024} reconstruction and projected it into a regular distance grid with a channel width of 0.24\,pc.
The resulting maps are presented in Fig.~\ref{fig:synthObs}.
}

\begin{figure}[ht]
\centerline{\includegraphics[width=0.5\textwidth,angle=0,origin=c]{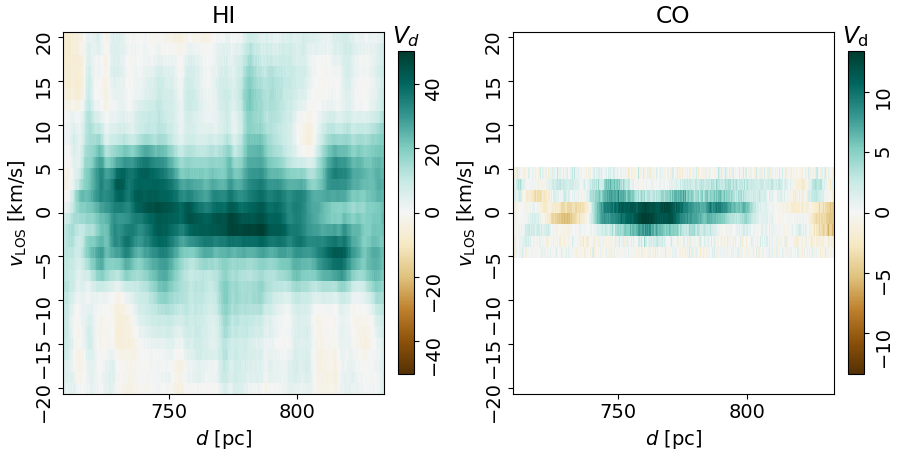}}
\caption{Same as Fig.~\ref{fig:Vplanes}, but for the synthetic observations of the SILCC-Zoom MC1-MHD simulation.}
\label{fig:hogSynthObs}
\end{figure}

\postreport{We applied the {\tt HOG} analysis to the MC1-MHD synthetic observations using the same parameters described in Sec.~\ref{sec:methods}.
The results, shown in Fig.~\ref{fig:hogSynthObs}, confirm the capabilities of {\tt HOG} to identify the line emission from an object in the PPD density cube. 
We employed this result to evaluate the density and \vLOS\ reconstructions presented in Sec.~\ref{sec:physics}.
}

\begin{figure}[ht]
\centerline{\includegraphics[width=0.5\textwidth,angle=0,origin=c]{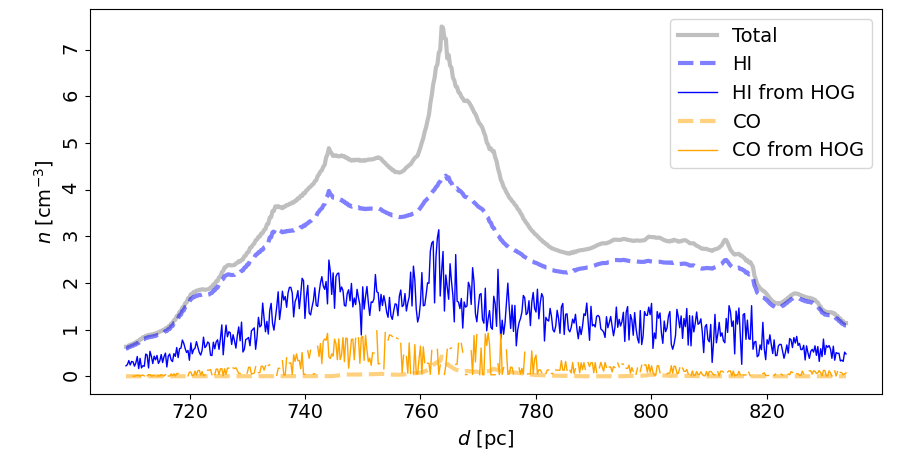}}
\caption{Comparison between the total, H{\sc i}, and CO densities across distance slices obtained directly from the simulation cube and reconstructed with HOG using Eq.~\eqref{eq:neff}.}
\label{fig:hogSynthObsDensities}
\end{figure}

Figure~\ref{fig:hogSynthObsDensities} shows a comparison of the densities across LOS distance channels obtained directly from the simulation and reconstructed with the {\tt HOG} using Eq.~\eqref{eq:neff}.
The reconstructed densities for H{\sc i} are roughly within a factor of two from their values in the simulation and recover the global density profile along the LOS.
The values reconstructed with the {\tt HOG} are smaller than the actual values, which is an expected consequence of estimating the densities using block averaging.
Reconstructions and line emission observations with higher angular resolutions enable finer segmentation and better estimates.
For CO, the {\tt HOG} method tends to overestimate the density, possibly by smearing the concentrated emission along the block area. 
However, this comparison is limited by the amount of extended, diffuse CO emission in this model, which is still a subject of active discussion \citep[e.g.,][]{gloverANDclark2012COinGMCs,levrier2012,godard2023,beitia-antero2024}

\postreport{Regarding \vLOS\ reconstruction, the {\tt HOG} method correctly reproduces the global LOS motion in the MC1-MHD simulation, which is 0\,\kps.
The width of the $V_{\rm d}$ signal in the \vLOS\ axis of Fig.~\ref{fig:hogSynthObs} corresponds to the emission linewidth coming from this isolated object 
\citep{seifried2022}.
Excursions from the central \vLOS\ are most likely related to the internal motions of the cloud and its interaction with the environment.
Further identification of the internal MC dynamics using the {\tt HOG} is beyond the scope of this work. 
However, this numerical experiment indicates that the recovered \vLOS\ dispersion is related to the H{\sc i} and CO emission linewidths.}

\section{Comparison with parallax and proper motions of star-forming regions}\label{app:masers}

The best estimates of the connection between the line of sight distances and velocities throughout the Galaxy come from VLBI observations of maser sources associated with young massive stars \citep[see, for example,][and references therein]{reid2009,honma2012}.
We considered the compilation of parallaxes and proper motions for about 200 masers presented by \cite{reid2019}, which is the result of observations from the National Radio Astronomy Observatory's Very Long Baseline Array (VLBA), the Japanese VLBI Exploration of Radio Astrometry (VERA) project, the European VLBI Network, and the Australian Long Baseline Array.
Only five observations in this catalog are within the Galactic volume considered in this paper.
We summarize their properties in Table~\ref{table:masers}.

\begin{table}
\caption{Parallaxes and proper motions of high-mass star-forming regions within the studied volume}              
\label{table:masers}      
\centering                                      
\begin{tabular}{c c c c}          
\hline\hline                        
Source & Distance\footnote{Estimated from 1/parallax} & \vLOS & Reference \\    
            & [pc] & [\kps] &  \\ 
\hline                                   
 G014.33$-$00.64   & 1119.8\,$\pm$126  & 22\,$\pm$\,5 & \cite{sato2010} \\      
 G090.21+02.32      & 674.3\,$\pm$17   & $-3$\,$\pm$\,5 & \cite{xu2013} \\
 G109.87+02.11      & 814.3\,$\pm$16   & $-7$\,$\pm$\,5 & \cite{moscadelli2009} \\
 & & & \cite{xu2016}  \\
 G121.29+00.65      & 928.5\,$\pm$34   & $-23$\,$\pm$\,5 & \cite{rygl2010} \\
 G176.51+00.20      & 963.4\,$\pm$19   & $-17$\,$\pm$\,5 & \cite{xu2013} \\
\hline                                             
\end{tabular}
\end{table}

\begin{figure}[ht]
\includegraphics[width=0.49\textwidth,angle=0,origin=c]{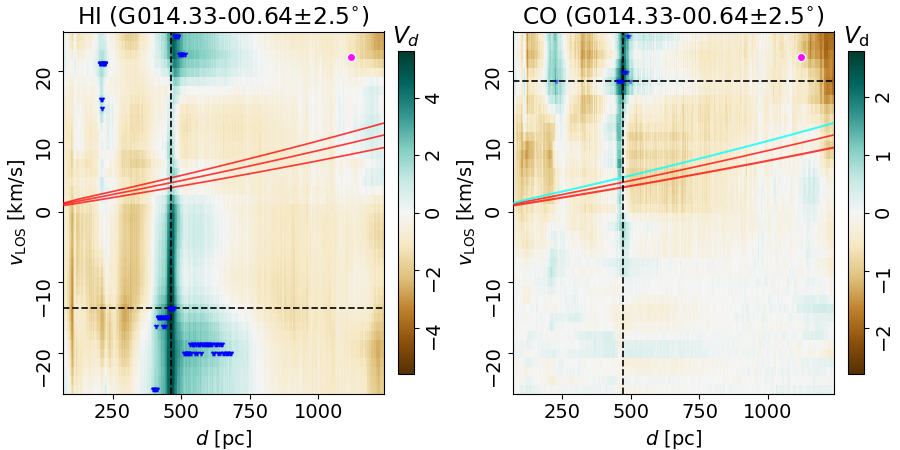}
\includegraphics[width=0.49\textwidth,angle=0,origin=c]{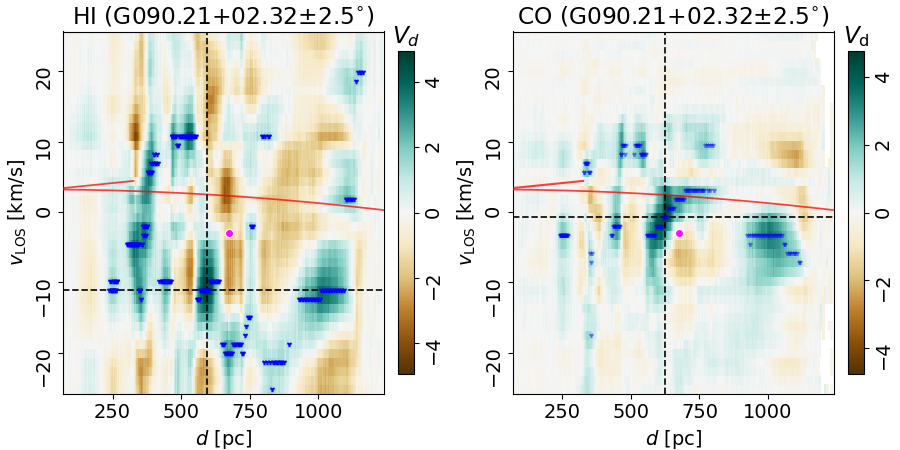}
\includegraphics[width=0.49\textwidth,angle=0,origin=c]{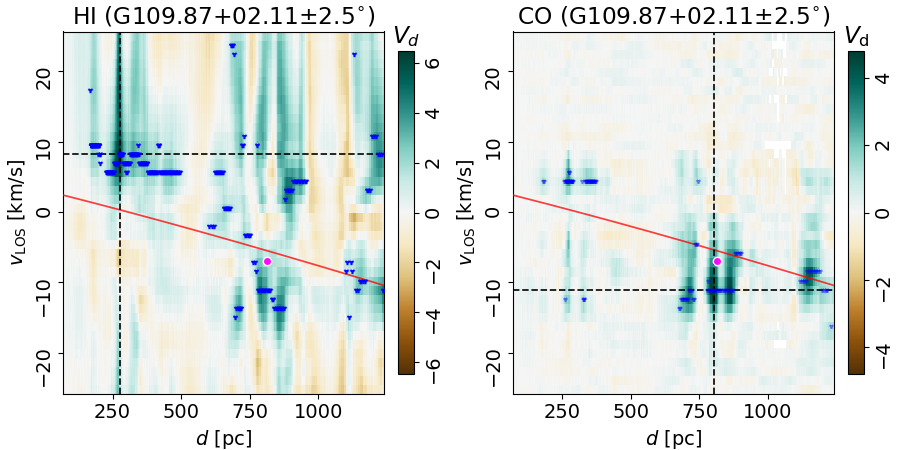}
\includegraphics[width=0.49\textwidth,angle=0,origin=c]{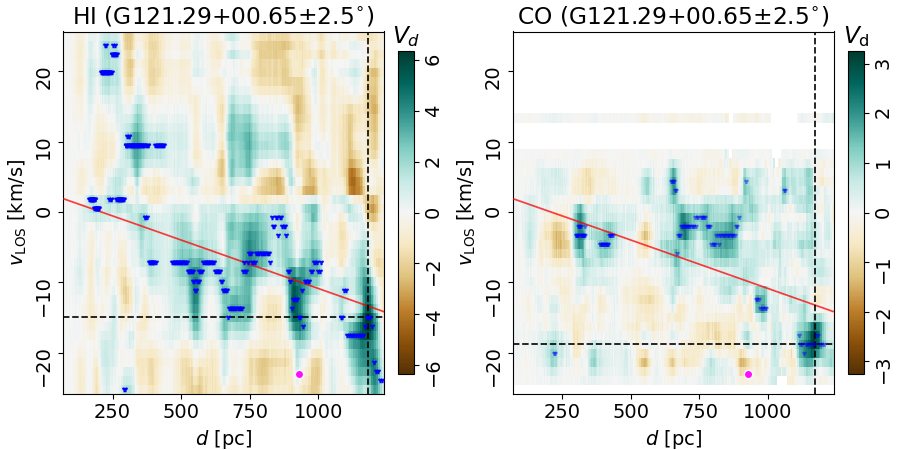}
\includegraphics[width=0.49\textwidth,angle=0,origin=c]{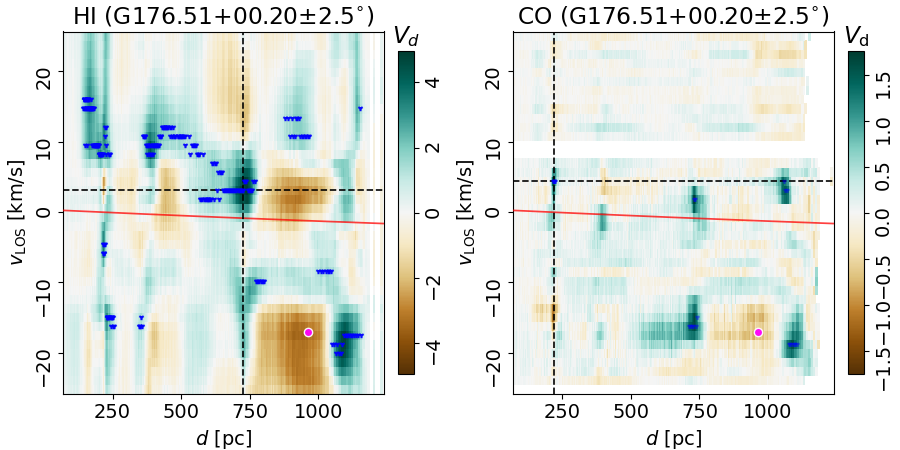}
\caption{Same as Fig.~\ref{fig:Vplanes}, but for a 5\deg\,$\times$\,5\deg\ area centered on the position of maser sources within $|b|$\,$<$\,5\deg.   
They are, from top to bottom, G014.33$-$00.64, G090.21$+$02.32, G109.87$+$02.11, G121.29$+$00.65, and G176.51$+$00.20.
The magenta disks indicate the parallax-derived distance and \vLOS\ for the maser sources.}
\label{fig:VplanesMaser1}
\end{figure}

Figure~\ref{fig:VplanesMaser1} shows the result analysis for H{\sc i} and CO emission and the 3D dust in 5\deg\,$\times$\,5\deg\ regions centered on the position of each source in Table~{table:masers}.
The region's size is selected to achieve minimum significance levels in $V_{\rm d}$. 
However, it implies that the volume sampled with the HOG can include ISM volumes not associated with the maser. 
We found a coincidence between the HOG analysis results and the VLBI observations for sources G090.21+02.32 and G109.87+02.11, both for CO.
The HOG-derived distance and LOS velocity for G121.29+00.65 in CO are within 200\,pc and a few kilometers per second from the VLBI observations.
For G$176.51+00.20$ there are $d$-\vLOS\ pairs with high $V_{\rm d}$ in the vicinity of the maser, but they do not correspond to the highest $V_{\rm d}$ toward that region.
Finally, G014.33$-$00.64 does not show a significant correlation close to the $d$ and \vLOS\ for the maser source.

More than revealing a profound truth about the ISM, the general lack of agreement between the HOG and VLBI results indicates that the two methods measure different ISM features.
The VLBI observes high-density regions where the physical conditions favor the maser emission while the HOG matches features in the emission across larger scales.
The two methods coincide in some cases, for example, in the CO emission toward G090.21+02.32 and G109.87+02.11. 
The cases in which the two methods disagree reveal regions where the dynamics sampled by the diffuse, extended gas emission differ from those sampled by the denser gas.
Thus, the HOG reveals a velocity structure across distance channels inaccessible to the maser observations throughout the Galaxy.

\section{Distance-LOS velocity correlation toward the Galactic center and anticenter}\label{app:anticenter}

\begin{figure}[ht]
\centerline{\includegraphics[width=0.5\textwidth,angle=0,origin=c]{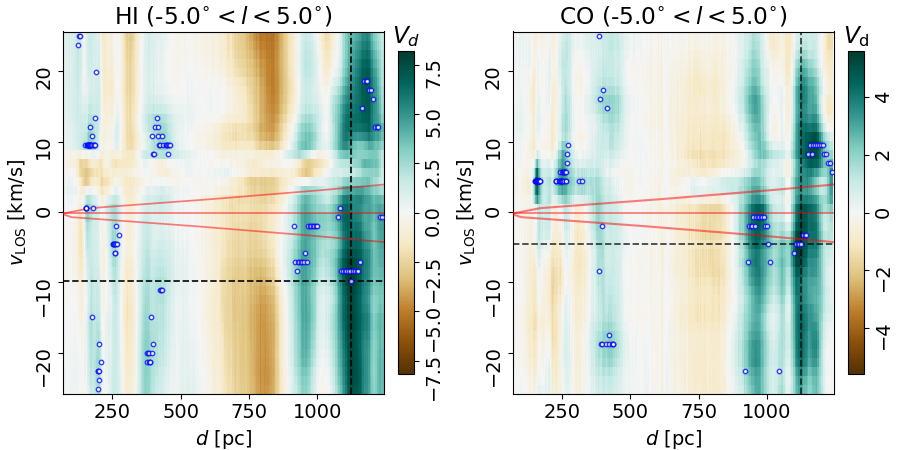}}
\centerline{\includegraphics[width=0.5\textwidth,angle=0,origin=c]{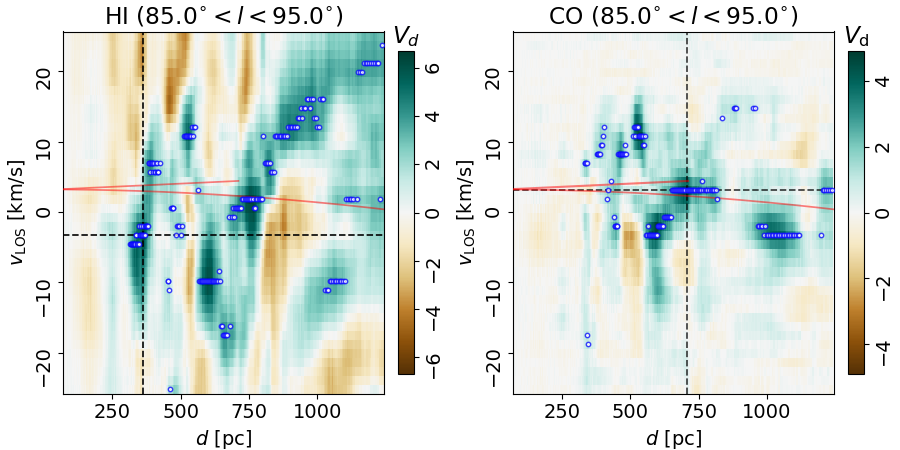}}
\centerline{\includegraphics[width=0.5\textwidth,angle=0,origin=c]{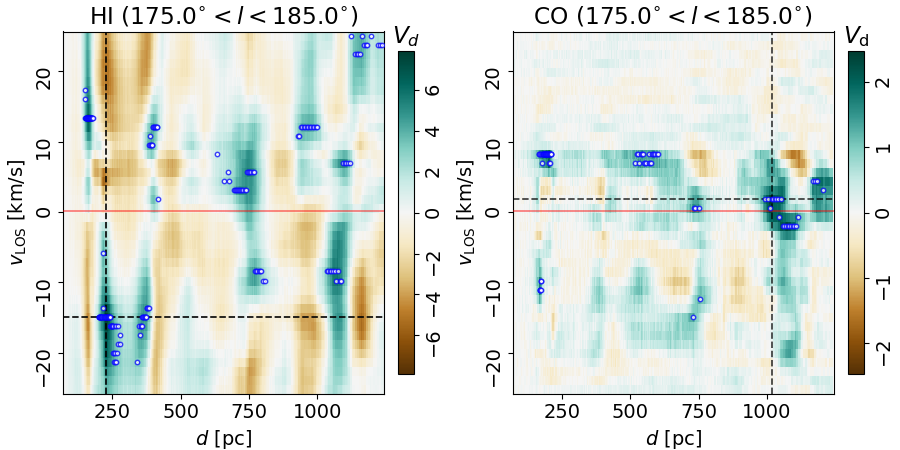}}
\centerline{\includegraphics[width=0.5\textwidth,angle=0,origin=c]{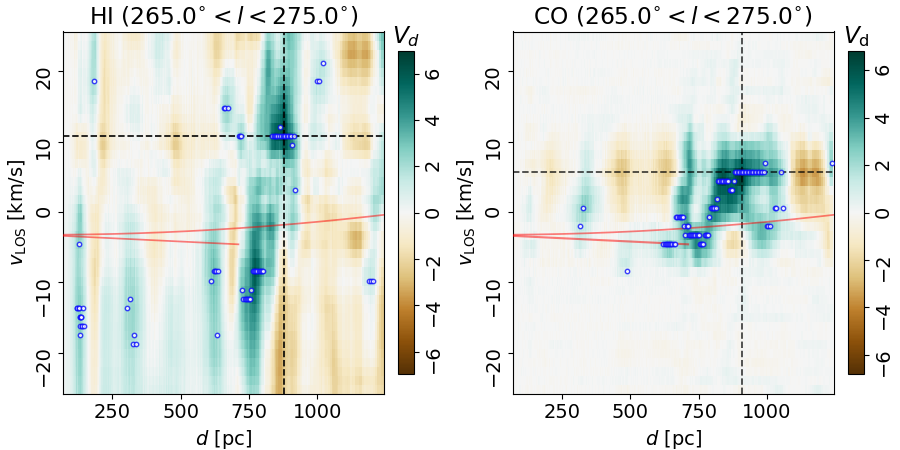}}
\caption{Same as Fig.~\ref{fig:Vplanes}, but for four 10\deg\,$\times$\,10\deg\ regions of interest centered on $b$\,$=$\,0\deg, including the Galactic center ($|l|$\,$<$\,5\deg) and anticenter (175\deg$<$\,$|l|$\,$<$\,185\deg).}
\label{fig:testL-5to5}
\end{figure}

The assumption of circular motions around the Galactic center to derive distances from the observed emission across \vLOS\ is particularly catastrophic in directions where the LOS component of the velocity is small, for example, toward the Galactic center and anticenter \citep[see, for example,][and references \postreport{therein}]{hunter2024}.
Therefore, we report the HOG-derived distance-\vLOS\ mapping for four reference regions in Fig.~\ref{fig:testL-5to5}.

Toward the Galactic center, the analysis shows that the highest $V_{\rm d}$ are distributed in dust parcels located at $d$\,$\approx$\,950 and 1125\,pc 
\postreport{It is apparent that the signal from the dust clouds in that distance range is smeared across a broad \vLOS\ range.  
This is unlikely the result of the cloud dynamics but rather a manifestation of the mapping of one density structure into multiple velocities in line emission due to the large velocity gradients along the LOS \citep[see, for example,][]{beaumont2013}.
The presence of substantial \vLOS\ gradients and the high densities sampled across this LOS also implies that a significant portion of the line emission may originate from distances beyond the range of the reconstruction.
However, testing this hypothesis requires deeper LOS 3D dust reconstructions, which are not yet available.
}

Toward the Galactic anticenter, the highest $V_{\rm d}$ values appear distributed across multiple distances \postreport{in a limited \vLOS\ range.
This is an archetypal example of the effect known as ``velocity crowding'', which is the superposition of multiple PPD structures into a few channels in PPV  \citep[see, for example,][]{ballesteros-paredes2002}.}
For CO, the highest $V_{\rm d}$ appears around \vLOS\,$\approx$\,10 and $-10$\,\kps\ across distances between 69 and 600\,pc, although the radial velocity expectation is close to zero.
\postreport{There are also considerable offsets from the expected \vLOS\ in H{\sc i}.
For both gas tracers, the amplitude of the departure from the expected \vLOS\ is comparable to the global velocity offsets reported in Fig.~\ref{fig:histvHIminusCO}.
}

\raggedright

\end{document}